\magnification=1200
\baselineskip=16truept
\hsize 15.5 truecm \vsize 21.7truecm  
\parindent=18truept
\parskip=3truept plus 2truept minus 3truept
\tolerance=10000
\hoffset= 0.25truecm  
\font\text=cmr10 at 12 truept
 at 11 truept
\font\note=cmr10 at 10 truept
\font\bf=cmbx10
\font\it=cmti10

\def\pni{\par \noindent}
\def\vsh{\smallskip}
\def\vs{\medskip}
\def\vvs{\bigskip}
\def\vsp{\par}
\def\vsn{\vsh\pni}
\def\cen{\centerline}
    
\def\eg{{\it e.g.}\ } \def\ie{{\it i.e.}\ }

    \def\rp{\item{}}
    \def\rf#1{\item{{#1}.\ }}  
\pageno=-1
\cen{CISM  \ LECTURE \ NOTES}
\cen{International Centre for Mechanical Sciences}
\cen{Palazzo del Torso, Piazza Garibaldi, Udine, Italy}
\vs\hrule\vvs
\vskip 0.5 truecm
\cen{{\bf FRACTIONAL \ CALCULUS :}}
\vsh
\cen{{\bf Integral and Differential Equations of Fractional Order}}
\vskip 0.5 truecm
\line{\bf Rudolf GORENFLO \hfill and \hfill Francesco MAINARDI}
\vsh
\line{Department of Mathematics and Informatics \hfill
      Department of Physics}
\line{Free University of Berlin \hfill University of Bologna}
\line{Arnimallee  3   \hfill Via Irnerio 46}
\line{D-14195 Berlin, Germany \hfill I-40126 Bologna, Italy}
\line{{\tt gorenflo@mi.fu-berlin.de} \hfill
      {\tt francesco.mainardi@unibo.it}}
\vskip 0.5 truecm
\cen{{\bf URL: www.fracalmo.org}}
\vs
\cen{{\bf  FRACALMO \ PRE-PRINT \ 54 pages : pp. 223-276}}
\vskip 0.5truecm
\def\indice{\leaders\hbox to 1 em {\hss.\hss}\hfill}
\def\hb#1{\hbox to 1.25 truecm{p. \hss#1}}
\line{\ \ \ \ ABSTRACT\indice  \hb{223}}
\line{1. INTRODUCTION TO FRACTIONAL CALCULUS
   \indice \hb{224}}
\line{2. FRACTIONAL INTEGRAL EQUATIONS
 \indice \hb{235}}
\line{3. FRACTIONAL DIFFERENTIAL EQUATIONS: 1-st PART \indice	\hb{241}}
\line{4. FRACTIONAL DIFFERENTIAL EQUATIONS: 2-nd PART \indice	\hb{253}}
\line{\ \ \ \ CONCLUSIONS    \indice   \hb{261}}
\line{\ \ \ \ APPENDIX : THE MITTAG-LEFFLER TYPE FUNCTIONS
    \indice   \hb{263}}
\line{\ \ \ \ REFERENCES \indice \hb{271}}
\vskip 0.5truecm
\hrule\vs
The paper is based on the lectures  delivered by the authors
at the CISM Course
{\it Scaling Laws and Fractality in Continuum Mechanics: A Survey of the
Methods based on Renormalization Group and Fractional Calculus},
held at the seat of CISM, Udine, from 23 to 27 September 1996,
under the direction of Professors A. Carpinteri  and F. Mainardi.
\pni
This {\TeX} pre-print is a revised version (December 2000)
of the chapter published in
 \pni
 {\bf A. Carpinteri and F. Mainardi (Editors):}
{\bf Fractals and Fractional Calculus in Continuum Mechanics,}
{\bf Springer Verlag, Wien and New York 1997, pp. 223-276.}
\pni
Such book is the volume No. 378 of the series CISM COURSES AND LECTURES
[ISBN 3-211-82913-X]
\vfill\eject
$\null$  \vskip 18.5 truecm
\cen{\copyright\ 1997, 2000 ~Prof. ~Rudolf ~Gorenflo - Berlin - Germany}
\cen{\copyright\ 1997, 2000 ~Prof. ~Francesco ~Mainardi - Bologna - Italy}
\vs
\vfill\eject
\input epsf
\baselineskip=16truept
\hsize 15.5 truecm \vsize 21.7truecm  
\parindent=18truept
\parskip=3truept plus 2truept minus 3truept
\tolerance=10000
\hoffset= 0.25truecm  
\font\text=cmr10 at 12 truept
 at 11 truept
\font\note=cmr10 at 10 truept
\font\bf=cmbx10
\font\it=cmti10

\def\pni{\par \noindent}
\def\vsh{\smallskip}
\def\vs{\medskip}
\def\vvs{\bigskip}
\def\vsp{\par}
\def\vsn{\vsh\pni}
\def\cen{\centerline}
    
\def\eg{{\it e.g.}\ } \def\ie{{\it i.e.}\ }

    \def\rp{\item{}}
    \def\rf#1{\item{{#1}.\ }}  
\def\e{{\rm e}}
\def\exp{{\rm exp}}
\def\ds{\displaystyle}

\def\q{\quad}	 \def\qq{\qquad}

\def\l{\left} \def\r{\right}

\def\rec#1{{1\over{#1}}}
\def\bar{\tilde}

\def\NN{{\rm I\hskip-2pt N}}
\def\RR{\vbox {\hbox to 8.9pt {I\hskip-2.1pt R\hfil}}}
\def\CC{{\rm C\hskip-4.8pt \vrule height 6pt width 12000sp\hskip 5pt}}
\def\II{{\rm I\hskip-2pt I}}
 
\def\u{\widetilde{u}}	\def\v{\widetilde{v}}
\def\ql{\widetilde{q}}

\def\riga{\vskip .1  truecm   \hrule \vskip .2	    truecm \noindent }
\def\HeadLinea#1#2{	\headline={\vbox to 0pt{\vss\noindent
{\ifnum \pageno=1  \hfill {\bf \folio}		 
\else {\ifodd \pageno			   
{\noindent     \hfill  {\it     #2} \quad {\bf \folio}
 }\riga
\else				 
{\noindent {\bf\folio} \quad  {\it  #1} \hfill 	}
\riga
\fi } \fi }			}}
}
\nopagenumbers
\voffset= 2 truecm

\HeadLinea{Fractional Calculus: Integral and Differential Equations of
 Fractional Order} {R. Gorenflo and F. Mainardi}
\pageno=223
$\null$
\vskip 1.8truecm
\cen{{\bf FRACTIONAL \ CALCULUS :}}
\vsh
\cen{{\bf Integral and Differential Equations of
   Fractional Order}}
\vskip 1.8 truecm
\line{Rudolf GORENFLO \hfill and \hfill Francesco MAINARDI}
\footnote {} {This research was
partially supported by	Research Grants of the Free University
of Berlin and the University of Bologna.
The authors also appreciate the support given by the National Research
Councils of Italy (CNR-GNFM) and by the International
Centre of Mechanical Sciences (CISM).}
\vsh
\line{Department of Mathematics and Informatics\hfill
      Department of Physics}
\line{Free University of Berlin \hfill University of Bologna}
\line{Arnimallee  3   \hfill Via Irnerio 46}
\line{D-14195 Berlin, Germany \hfill I-40126 Bologna, Italy}
\line{{\tt gorenflo@math.fu-berlin.de} \hfill
      {\tt francesco.mainardi@unibo.it}}
\vskip 0.5 truecm
\cen{{\bf URL: www.fracalmo.org}}
\vskip 1.5 truecm
\line{ABSTRACT \hfill}
\vsn
In these 
lectures we introduce  the linear operators of fractional
integration and fractional differentiation in the framework
of the {Riemann-Liouville fractional calculus}.
Particular attention is devoted to the technique of Laplace
transforms  for treating these operators in a way
accessible to applied scientists, avoiding unproductive
generalities and excessive  mathematical rigor.
By applying this technique  we shall derive the analytical  solutions
of the most simple linear integral and differential equations of
fractional order. We shall
show the fundamental role of
the Mittag-Leffler function, whose properties are reported
in an {\it ad hoc} Appendix.
The topics discussed here will be:
(a) essentials	of Riemann-Liouville fractional calculus
  with basic formulas of Laplace transforms,
(b) Abel type integral equations of first and  second kind,
(c) relaxation and oscillation type differential equations
    of fractional order.
\vs
\line{{\it 2000 Math. Subj. Class.}: 26A33,
 33E12, 
 33E20, 44A20,	45E10, 45J05.  \hfill}
 \vfill\eject
\line{1. INTRODUCTION TO FRACTIONAL CALCULUS \hfill} \vsn
\line{{\it 1.1 Historical Foreword \hfill}}
\vsp
Fractional calculus is the field of mathematical analysis which deals
with the investigation and applications of integrals and
derivatives of arbitrary order.
The term {\it fractional} is a misnomer, but it is
retained 
following the  prevailing use.
\vsp
The fractional calculus may be considered an {\it old} and
yet {\it novel}  topic.
It is an {\it old} topic since, starting from some speculations of
G.W. Leibniz
(1695, 1697) and L. Euler (1730), it has been developed up to nowadays.
A list of 
mathematicians,
who have provided important contributions
up to the middle of our century, includes
P.S. Laplace (1812),
J.B.J. Fourier (1822),	N.H. Abel (1823-1826),
J. Liouville (1832-1873),
B. Riemann (1847),
H. Holmgren (1865-67),
A.K. Gr\"unwald (1867-1872), A.V. Letnikov (1868-1872),
H. Laurent (1884), P.A. Nekrassov (1888), A. Krug (1890),
J. Hadamard (1892), O. Heaviside (1892-1912), S. Pincherle (1902),
G.H. Hardy  and  J.E. Littlewood (1917-1928), H. Weyl (1917), P. L\'evy
(1923), A. Marchaud (1927), H.T. Davis (1924-1936),
A. Zygmund (1935-1945),
E.R. Love (1938-1996),
A. Erd\'elyi (1939-1965), H. Kober (1940),
D.V. Widder (1941),
M. Riesz (1949).
\vsp
However, it may be considered a {\it novel} topic as well, since
only from a little more than twenty years it has been
object of specialized conferences
and treatises. For the first conference the merit is ascribed to
B. Ross who
organized the {\it First Conference on Fractional Calculus and its
Applications}  at the University of New Haven  in June 1974,
and edited the proceedings, see [1]. For the first monograph
the merit is ascribed to K.B. Oldham and J. Spanier, see [2],
who, after a joint collaboration started in 1968, published
a book devoted to fractional calculus in 1974.
Nowadays, the list of texts and proceedings devoted solely or partly
to fractional calculus and its applications includes about
a dozen of titles [1-14], among which the encyclopaedic treatise
by Samko, Kilbas \& Marichev [5]  is the most prominent.
Furthermore, we   recall
the attention to
the treatises by Davis [15],
Erd\'elyi [16], Gel'fand \&  Shilov [17], Djrbashian [18, 22], Caputo [19],
Babenko [20],  Gorenflo  \&  Vessella [21],
which contain a detailed analysis   of
some mathematical aspects and/or physical applications	of
fractional calculus,
although without explicit mention in their titles.
\vsp
In recent years considerable interest in fractional calculus has been
stimulated by the applications that this calculus finds in
numerical analysis and different
areas of physics and engineering, possibly including fractal phenomena.
In this respect  A. Carpinteri and F. Mainardi
have edited  the present book	of lecture notes  and entitled it
as {\it Fractals and Fractional Calculus in Continuum Mechanics}.
For the topic of fractional calculus, in addition to this  joint article
of introduction, we have contributed  also with  two
single articles, one  by Gorenflo [23],
devoted to  numerical methods,	and one by Mainardi [24],
concerning  applications in mechanics.
\vfill\eject
\line{1.2 {\it The Fractional Integral}\hfill}
\vsp
According to the Riemann-Liouville approach  to fractional calculus
the notion of fractional integral of order $\alpha$
($\alpha >0$)
 is a natural consequence
of the well known formula (usually attributed to Cauchy),
that reduces the calculation of the $n-$fold primitive of a function
$f(t)$ to a single integral of convolution type.
In our notation the Cauchy formula reads
$$
    J^n f(t) := f_{n}(t)=
 \rec{(n-1)!}\, \int_0^t \!\!  (t-\tau )^{n-1}\,f(\tau) \, d\tau\,,
    \q t > 0\,,\q n \in \NN   \,, \eqno(1.1) $$
where $\NN$ is the set of positive integers.
From this definition we note that $f_{n}(t)$
vanishes at $t=0$ with its
derivatives of order $1,2, \dots, n-1\,. $
For convention we require that	$f(t)$ and henceforth
$f_{n}(t)$ be a {\it causal} function, \ie identically
vanishing for $t<0\,. $
\vsp
In a natural way  one is  led
to extend the above formula
from positive integer values of the index to any positive real values
by using the Gamma function.
Indeed, noting that $(n-1)!= \Gamma(n)\,, $
and introducing the arbitrary {\it positive} real number
 $\alpha\,, $
one defines  the
 \ \underbar{{\it Fractional Integral of order} $\alpha >0 $} :
$$
J^\alpha \,f(t) :=
      \rec{\Gamma(\alpha )}\,
 \int_0^t (t-\tau )^{\alpha -1}\, f(\tau )\,d\tau \,,
   \q t > 0\, \q \alpha  \in \RR^+
 \,,  \eqno(1.2) $$
where $\RR^+$ is the set of positive real numbers.
For complementation we define
$J^0 := I\, $ ({Identity operator)}, \ie we mean
$J^0\, f(t) = f(t)\,. $ Furthermore,
by $J^\alpha f(0^+)$ we mean the limit (if it exists)
of $J^\alpha f(t)$ for $t\to 0^+\,;$ this limit may be infinite.
\vsp
\line{\underbar{Remark 1} : \hfill} \pni
Here, and in all our following treatment, the integrals are intended
in the {\it generalized} Riemann sense, so that any function
is required to be {\it locally} absolutely integrable
in $\RR^+$.
However, we will not bother to give  descriptions of sets
of admissible func\-tions and will not hesitate, when necessary,
to use {\it formal} expressions with generalized
functions ({\it distributions}),
which,	as far as possible,  will be
re-interpreted in the framework of classical functions.
The reader interested in the strict mathematical rigor is referred
to [5], where the fractional calculus is treated
in the framework of Lebesgue spaces of summable functions
and Sobolev spaces of generalized functions.
\vsp
\line{\underbar{Remark 2} : \hfill} \pni
In order to  remain in accordance with the
standard notation $I$ for the Identity operator
we use the character $J$ for the integral operator and its power
$J^\alpha$. If one likes to denote by $I^\alpha $  the integral
operators, he would adopt a different notation for the	Identity,
\eg $\II\,, $ to avoid a possible confusion.
\vfill\eject
We note the {\it semigroup property}
$$
J^\alpha J^\beta = J^{\alpha +\beta}\,,
   \q
\alpha\,,\;\beta  \ge 0\,,\eqno(1.3)$$
which implies the {\it commutative property}
$J^\beta  J^\alpha= J^\alpha J^\beta\,,$
and  the effect of our operators $J^\alpha$
on the power functions
$$
J^\alpha t^\gamma ={\Gamma (\gamma +1)\over \Gamma(\gamma +1+\alpha)}\,
		   t^{\gamma+\alpha}\,, \q \alpha >0\,,
  \q \gamma >-1\,, \q t>0\,.
\eqno (1.4)
$$
The properties (1.3-4) are of course a natural generalization
of those known when the order is a positive integer.
The proofs, see \eg  [2], [5] or [10], are based on the
properties of the two Eulerian integrals, \ie
the {\it Gamma} and  {\it Beta} functions, 
$$ \Gamma(z) := \int_0^\infty\!\! \e^{-u}\, u^{z-1}\, du\,,
   \q \Gamma(z+1)=z\, \Gamma(z) \,, \q {\rm Re}\, \{z\} >0\,,\eqno(1.5)$$
$$ B(p,q) := \int_0^1\!\! (1-u)^{p-1}\, u^{q-1}\, du\  =
    { \Gamma(p)\, \Gamma(q)\over \Gamma(p+q)}
    =B(q,p)\,,
 \q {\rm Re}\, \{p\,,\,q\} >0\,.
  \eqno(1.6)$$
\vsp
It may be convenient to introduce the following causal
function
$$ \Phi_\alpha (t) := { t_+^{\alpha -1}\over \Gamma(\alpha)}\,,
   \q \alpha  >0\,, \eqno(1.7)$$
where the suffix $+$ is just denoting that the function is
vanishing for $t<0\,. $  Being $\alpha >0\,, $
this function turns out to be {\it locally} absolutely
integrable  in $\RR^+\,. $
Let us now recall the notion of {\it Laplace convolution}, \ie the
convolution integral with two causal functions,
which reads in a standard notation
$\, f(t) * g(t) := \int_0^t f(t-\tau )\, g(\tau )\,d\tau
   = g(t)* f(t) \,. $
\vsp
Then we  note from (1.2) and (1.7) that the fractional integral
of order $\alpha >0$
 can be considered as the Laplace convolution between
 $\Phi_\alpha (t)$ and $f(t)\,, $ \ie
$$ J^\alpha \, f(t) = \Phi_\alpha (t) \,*\, f(t)\,,
  \q \alpha >0\,.  \eqno(1.8)$$
Furthermore, based on the Eulerian integrals, one
proves the {\it composition rule}
$$ \Phi_\alpha (t) \,*\, \Phi_\beta (t) = \Phi_{\alpha +\beta }(t)\,,
 \q
\alpha\,,\;\beta  > 0\,,\eqno(1.9)$$
which can be used to re-obtain (1.3) and (1.4).
\vsp 
Introducing
the Laplace transform by the notation
$ {\cal{L}}\, \l\{  f(t) \r\}  := \int_0^\infty \!\!
   \e^{-st}\, f(t)\, dt = \widetilde f(s)\,, \; s \in \CC\,,$
and  using the sign $\div$ to denote a Laplace transform pair,
\ie
$ f(t) \div  \widetilde f(s) \,, $
we note the following rule for the   Laplace transform of
the fractional integral,
$$	   J^\alpha \,f(t) \div
     {\widetilde f(s)\over s^\alpha}\,,\q \alpha >0\,,	\eqno(1.10)$$
which is the straightforward generalization
of the case with an $n$-fold repeated integral ($\alpha =n$).
For the proof it is sufficient to recall the convolution
theorem for Laplace transforms and note the pair
$\Phi_\alpha (t)  \div 1/s^\alpha \,,$ with $\alpha >0\,, $
see \eg Doetsch [25].
\vfill\eject
\line{1.3 {\it The Fractional Derivative}\hfill}
\vsp
After the notion of fractional integral,
that of fractional derivative of order $\alpha$
($\alpha >0$)
becomes a natural requirement and one is attempted to
substitute $\alpha $ with $-\alpha $ in the above formulas.
However, this generalization  needs some care  in order to
guarantee the convergence of  the integrals   and
preserve the
well known properties of the ordinary derivative of integer
order.
\vsp
 Denoting by $D^n\,$ with $ n\in \NN\,, $
the operator of the derivative of order $n\,,$	we first note that
$$ D^n \, J^n = I\,, \q   J^n \, D^n \ne I\,,\q n\in \NN \,,
\eqno(1.11)$$
\ie $D^n$ is left-inverse (and not right-inverse) to
the corresponding integral operator $J^n\,. $
In fact we easily recognize from (1.1) that
$$  J^n \, D^n \, f(t) = f(t) - \sum_{k=0}^{n-1}
	f^{(k)}(0^+) \, {t^k\over k!}\,, \q t>0\,. \eqno(1.12)$$
\vsp
As a consequence we expect that $D^\alpha $ is defined as left-inverse
to $J^\alpha $.  For this purpose, introducing the positive
integer $m$ such that $m-1 <\alpha \le m\,, $
one defines the
 \underbar{{\it Fractional Derivative of order} $\alpha >0 $} :
$ D^\alpha \,f(t) := D^m \, J^{m-\alpha} \, f(t) \,,$ namely
$$
 D^\alpha \,f(t) :=
\cases{
  {\ds {d^m\over dt^m}}\,\l[
  {\ds \rec{\Gamma(m-\alpha)}\,\int_0^t
    {f(\tau )\over (t-\tau )^{\alpha +1-m}} \,d\tau}\r] \,,
 & $\; m-1 <\alpha < m\,,$ \cr\cr
     {\ds {d^m\over dt^m}} f(t)\,,
    & $\; \alpha =m\,. $\cr\cr }
   \eqno(1.13) $$
Defining for complementation $D^0 = J^0 =I\,, $ then
we easily recognize that
$$ D^\alpha \, J^\alpha = I \,,  \q \alpha \ge 0\,, \eqno(1.14)$$
and
$$ D^{\alpha}\, t^{\gamma}=
   {\Gamma(\gamma +1)\over\Gamma(\gamma +1-\alpha)}\,
     t^{\gamma-\alpha}\,,
 \q \alpha >0\,,
  \q \gamma >-1\,, \q t>0\,.
\eqno (1.15)
$$
Of course, the properties (1.14-15) are a natural generalization
of those known when the order is a positive integer.
Since in (1.15) the argument of the Gamma function in the denominator
can be negative,
we need to consider the analytical continuation of $\Gamma(z)$
in (1.5) to the left half-plane, see \eg Henrici [26].
\vsp
Note the remarkable fact that the fractional derivative $D^\alpha\, f$
is not	zero
for the constant function $f(t)\equiv 1$ if $\alpha \not \in {\NN}\,. $
In fact, (1.15) with $\gamma =0$ teaches us that
$$
D^\alpha 1 = {t^{-\alpha}\over \Gamma(1-\alpha)}\,,\q \alpha\ge 0\,,
\q t>0\,.  \eqno (1.16)
$$
This, of course, is $\equiv 0$ for $\alpha \in{\NN}$, due to the
poles of the gamma function in the points $0,-1,-2,\dots$.
\vfill\eject
\vsp
We now observe that an alternative definition
of fractional derivative, originally introduced by Caputo
 [19], [27]
in the late sixties and
adopted by Caputo and Mainardi [28]
in the framework  of the theory of {\it Linear Viscoelasticity}
 (see a review in [24]), is
the so-called
\underbar{{\it Caputo Fractional Derivative of order} $\alpha >0$} :
$D_*^\alpha  \, f(t) :=   J^{m-\alpha}\, D^{m} \, f(t)$
with $m-1 <\alpha \le m\,, $ namely
$$
 D_*^\alpha \,f(t) :=
\cases{
    {\ds \rec{\Gamma(m-\alpha)}}\,{\ds\int_0^t
 {\ds {f^{(m)}(\tau)\over (t-\tau )^{\alpha +1-m}}} \,d\tau} \,,
  & $\; m-1<\alpha <m\,, $\cr\cr
     {\ds {d^m\over dt^m}} f(t)\,,
    & $\; \alpha =m\,. $\cr\cr }
   \eqno(1.17) $$
This definition is of course more restrictive than (1.13), in that
requires the absolute integrability of the  derivative of order $m$.
Whenever we use the operator   $D_*^\alpha$   we (tacitly) assume that
     this condition is met.
 We  easily recognize that in general
$$  D^\alpha\, f(t) := D^{m} \, J^{m-\alpha} \, f(t)
 \ne J^{m-\alpha}\, D^{m} \, f(t):= D_*^\alpha \, f(t)\,,
 \eqno(1.18)
 $$
 unless   the function	$f(t)$ along with its first $m-1$ derivatives
 vanishes at $t=0^+$.
In fact, assuming that
the passage of the $m$-derivative under
the integral is legitimate, one     
recognizes that,  for $ m-1 <\alpha  < m \,$  and $t>0\,, $
$$
    D^\alpha \, f(t) =
  D_*^\alpha   \, f(t) +
  \sum_{k=0}^{m-1}   {t^{k-\alpha}\over\Gamma(k-\alpha +1)}
    \, f^{(k)}(0^+) \,, \eqno(1.19)    $$
 and therefore, recalling the fractional derivative of the power
functions (1.15),
$$
   D^\alpha \l( f(t) -
 \sum_{k=0}^{m-1} {t^k \over k!} \, f^{(k)} (0^+)\r)
     =	D_*^\alpha  \, f(t)  \,.\eqno(1.20)  $$
The alternative definition (1.17) for the
fractional derivative  thus incorporates the initial values
of the function and of its integer derivatives of lower order.
The subtraction of the Taylor polynomial of degree $m-1$ at $t=0^+$
from $f(t)$ means  a sort of
regularization	of the fractional derivative.
In particular, according to this definition,
the relevant property for which the fractional derivative
of a constant is still zero, \ie
$$ D_*^\alpha  1 \equiv 0\,,\q	 \alpha >0\,. \eqno(1.21)$$
can be easily recognized.
   \vsp
We now explore the most relevant differences between the two
fractional derivatives (1.13) and (1.17). We agree to
denote (1.17) as the {\it Caputo fractional derivative}
to distinguish it from the standard Riemann-Liouville fractional
derivative (1.13).
We observe, again by looking at ({1.15}), that
$$
D^\alpha t^{\alpha -1} \equiv 0\,, \q \alpha>0\,, \q t>0\,.\eqno(1.22)
$$
 \vfill\eject
From (1.22) and (1.21)	we thus recognize
the following statements about functions
which  for $t>0\, $   admit the same fractional derivative
of    order $\alpha \,, $
with $m-1 <\alpha \le m\,,$ $\; m \in \NN\,, $
$$    D^\alpha \, f(t) = D^\alpha  \, g(t)
   \,  \Longleftrightarrow  \,
  f(t) = g(t) + \sum_{j=1}^m c_j\, t^{\alpha-j} \,,
    \eqno(1.23) $$
$$    D_* ^\alpha \, f(t) = D_*^\alpha	\, g(t)
   \,  \Longleftrightarrow  \,
  f(t) = g(t) +  \sum_{j=1}^m c_j\, t^{m-j} \,.
    \eqno(1.24) $$
In these formulas the coefficients $c_j$ are arbitrary constants.
\vsp
Incidentally, we note that  (1.22) provides an instructive example to
show how  
$D^\alpha$   is not right-inverse
to $J^\alpha\,, $ since
$$
J^\alpha D^\alpha t^{\alpha -1} \equiv 0, \q {\rm but}\q
D^\alpha J^\alpha
t^{\alpha -1} =t^{\alpha -1}\,,
    \q \alpha>0\,, \q  t>0,.	\eqno(1.25)
$$
\vsp
For the two definitions we also note a difference
with respect to the {\it formal} $\,$ limit  as
 $\alpha \to {(m-1)}^+$. From (1.13) and (1.17) we obtain
respectively,
$$ \alpha \to (m-1)^{+}\,\Longrightarrow\,
 D^\alpha \,f(t) \to	D^m\,  J\, f(t) = D^{m-1}\, f(t)
   \,; \eqno(1.26) $$
$$\alpha \to {(m-1)}^{+} \,\Longrightarrow\,
 D_*^\alpha \, f(t) \to J\, D^m\, f(t) =
       D^{m-1}\, f(t) - f^{(m-1)} (0^+)\,. \eqno(1.27) $$
\vsp
We now consider the {\it Laplace transform} of the two fractional
derivatives.
For the standard fractional derivative $D^\alpha $
the Laplace transform,	assumed to exist,  requires the knowledge of the
(bounded) initial values of the fractional integral $J^{m-\alpha }$
and of its integer  derivatives of order $k =1,2, m-1\,, $
as we learn from [2], [5], [10].
The corresponding rule reads, in our notation,
$$ D^\alpha \, f(t) \div
      s^\alpha\,  \widetilde f(s)
   -\sum_{k=0}^{m-1}  D^k\, J^{(m-\alpha)}\,f(0^+) \, s^{m -1-k}\,,
  \q m-1<\alpha \le m \,. \eqno(1.28)$$
\vsp
The {\it Caputo fractional derivative} appears more suitable to
be treated by the Laplace transform technique in that it requires
the knowledge of the (bounded)
initial values of the function
and of its integer  derivatives of order $k =1,2, m-1\,, $
in analogy with the case when $\alpha =m\,. $
In fact,
by using (1.10) and noting that
$$ J^\alpha  \, D_*^\alpha \, f(t) =
    J^\alpha\, J^{m-\alpha }\, D^m \, f(t) =
     J^m\, D^m \, f(t) = f(t) -
  \sum_{k=0}^{m-1} {f^{(k)}(0^+)}\, {t^k \over k!}
  \,. \eqno(1.29)$$
we easily prove  the following rule for the Laplace transform,
$$ D_*^\alpha \, f(t) \div
      s^\alpha\,  \widetilde f(s)
   -\sum_{k=0}^{m-1}  f^{(k)}(0^+) \, s^{\alpha -1-k}\,,
  \q m-1<\alpha \le m \,, \eqno(1.30)$$
Indeed, the  result (1.30), first stated by Caputo [19] by using the
Fubini-Tonelli theorem, appears  as the most "natural"
generalization of the corresponding result well known for $\alpha =m\,. $
\vfill\eject
\vsp
We  now show how both the  definitions (1.13) and (1.17)
for the fractional
derivative of $f(t)$ can be derived,
at least {\it formally}, by  the convolution  of $\Phi_{-\alpha}(t)$
with $f(t)\,, $
in a sort of analogy with (1.8) for the fractional integral.
For this purpose we need to recall from the treatise  on generalized
functions by	Gel'fand  and  Shilov [16]  that
(with proper interpretation  of the quotient as a limit if $t=0$)
$$    \Phi_{-n} (t) := {\;t_+^{-n-1} \over \Gamma(-n)} =
  \delta ^{(n)}(t)\,, \q n=0\,,\,1\,,\ldots  \eqno(1.31)$$
where $\delta ^{(n)}(t)$ denotes the generalized derivative of
order $n$ of the Dirac delta distribution.
Here, we assume that the reader has  some minimal knowledge
concerning these distributions, sufficient for	handling
classical problems in physics and engineering.
\vsp
The  equation (1.31) provides  an interesting (not so well known)
representation of $   \delta ^{(n)}(t)\,, $
which is useful in  our
following treatment  of  fractional derivatives.
In fact, we  note that
the derivative of order $n$ of a causal function $f(t)$
can be obtained  {\it formally}
by the (generalized) convolution between  $\Phi_{- n}$ and  $f\,, $
$$   {d^n\over dt^n}\, f(t) = f^{(n)} (t) =
       \Phi_{-n}(t)\,*\, f(t)  =
  \int_{0^-}^{t^+} \!\! f(\tau )\, \delta ^{(n)}(t-\tau )\, d\tau
  \,, \q t>0\,,
    \eqno(1.32)$$
based on the  well known properties
   $$  \int_{0^-}^{t^+} \!\! f(\tau )\, \delta ^{(n)}(\tau-t )\, d\tau
   = (-1)^n \, f^{(n)}(t) \,, \q
  \delta^{(n)} (t-\tau)  =  (-1)^n \, \delta^{(n)} (\tau-t)
    \,.  \eqno(1.33)$$
According  to a usual convention,
in (1.32-33) the limits of integration are extended to take into account
for the possibility of impulse functions centred at the extremes.
\vsp
Then, a formal definition of the fractional derivative
of order $\alpha $ could be
$$ \Phi_{-\alpha } (t) \,* \, f(t)
 = \rec{\Gamma(-\alpha )}\,
 \int_{0^-}^{t^+}\!\!  {f(\tau)\over (t-\tau )^{1+\alpha}}\,d\tau
\,, \q \alpha \in \RR^+\,.
$$
The formal character is evident
in that   the kernel $\Phi_{-\alpha} (t)$ turns out to
be not locally absolutely integrable  and consequently
the integral is in general divergent.
In order to obtain  a  definition
that is still
valid for classical functions,
we need to {\it regularize} the divergent integral  in some way.
For this purpose
let us	consider the integer $ m \in \NN$
such that $ m-1 < \alpha < m$ and  write
$ -\alpha = -m +(m-\alpha)$  or
$ -\alpha = (m-\alpha) -m $. We then obtain
$$
     \l[\Phi_{-m}(t) \,*\, \Phi_{m-\alpha}(t)\r]\,*\, f(t) =
  \Phi_{-m}(t) \,*\, \l[\Phi_{m-\alpha}(t)\,*\, f(t) \r] =
     D^m\, J^{m-\alpha}\, f(t)	 \,,   \eqno(1.34)  $$
or
$$
   \l[\Phi_{m-\alpha}(t) \,*\,\Phi_{-m}(t)\r] \,*\, f(t) =
  \Phi_{m-\alpha}(t) \,*\,\l[\Phi_{-m}(t) \,*\, f(t)\r]=
     J^{m-\alpha}\, D^m\, f(t) \,. \eqno(1.35)	     $$
As a consequence we derive
two alternative  
definitions
for the fractional derivative, corresponding
to (1.13) and (1.17), respectively.
The singular behaviour of $\Phi_{-m} (t)$ is reflected in the
non-commutativity of convolution in these formulas.
\vfill\eject
\line{1.4 {\it Other Definitions and Notations}\hfill}
\vsp
Up to now we have considered the approach to fractional calculus
usually referred to  Riemann and Liouville. However, while
Riemann (1847) had generalized the integral Cauchy formula with starting
point $t=0\,$ as reported in (1.1),
 originally Liouville (1832) had chosen $t=-\infty\,. $
In this case we define
$$J_{-\infty}^\alpha  \,f(t) :=
   \rec{\Gamma(\alpha )}\,
 \int_{-\infty}^t (t-\tau )^{\alpha -1}\, f(\tau )\,d\tau \,,
    \q \alpha  \in \RR^+
 \,,  \eqno(1.36)  $$
and consequently, for
$m-1 <\alpha \le m\,,\; m\in \NN\,, $
$ D_{-\infty}^\alpha \,f(t) :=
  D^m \, J_{-\infty}^{m-a} \, f(t)\,, $
namely
$$
D_{-\infty}^\alpha \,f(t) := \cases{
 {\ds{d^m\over dt^m}}\,\l[
   {\ds \rec{\Gamma(m-\alpha)}}\,{\ds \int_{-\infty}^t} \!\!
    {\ds {f(\tau )\, d\tau \over (t-\tau)^{\alpha +1-m}}}\r] \,,
 & $\; m-1 <\alpha <m\,,$  \cr\cr
{\ds{d^m\over dt^m}} f(t)\,,
 & $\; \alpha =m\,. $\cr}
  \eqno(1.37) $$
In this case, assuming $f(t)$  to  vanish
as $t \to -\infty$ along with its first $m-1$ derivatives, we have the
identity
$$ D^{m} \, J_{-\infty}^{m-\alpha} \, f(t)
 = J_{-\infty}^{m-\alpha}\, D^{m} \, f(t)\,, \eqno(1.38)$$
in contrast with (1.18).
\vsp
While for the fractional integral (1.2) a sufficient condition that
the integral  converge is that
$$ f\l( {t}\r) = O \l(t^{\epsilon-1}\r)\,, \q \epsilon >0\,,
  \q t \to 0^+\,, \eqno(1.39)$$
a sufficient condition that (1.36) converge is that
$$ f\l( {t}\r) = O \l(|t|^{-\alpha -\epsilon}\r)\,, \q \epsilon >0\,,
  \q t \to -\infty\,. \eqno(1.40)$$
Integrable functions satisfying the properties (1.39) and (1.40) are
sometimes referred to as functions of Riemann class and Liouville
class, respectively [10].
For example power functions $t^\gamma $  with $\gamma >-1$  and $t>0$
(and hence also constants) are of Riemann class,
while $|t|^{-\delta }$	with $\delta >\alpha >0$ and $t<0$
and ${\rm exp}\, (ct)$ with $c>0$
are of Liouville class. For the above functions we obtain
(as real versions of the formulas given in [10])
$$
   J_{-\infty}^\alpha  \,|t|^{-\delta} =
   {\Gamma(\delta -\alpha)\over \Gamma(\delta)}\,
  |t|^{-\delta+\alpha}\,,  \q
  D_{-\infty}^\alpha  \,|t|^{-\delta} =
   {\Gamma(\delta +\alpha )\over \Gamma(\delta )}\,
    |t|^{-\delta-\alpha}\,,  \eqno(1.41)$$
and
$$
J_{-\infty}^\alpha  \,\e^{\, ct} =
  c^{\, -\alpha} \, \e^{\,  ct}\,,
\q
  D_{-\infty}^\alpha  \,\e^{\, ct} =
  c^{\, \alpha} \, \e^{\,  ct}\,.\eqno(1.42)$$
\vfill\eject
\vsp
Causal functions can be considered in the above integrals with
the due care. In fact, in view of the possible jump discontinuities of
the integrands at $t=0\,, $ in this case it is worthwhile to write
	$$ \int_{-\infty}^t (\dots)\, d\tau =
      \int_{0^-}^t (\dots)\, d\tau \,. $$
As an example we consider for $0<\alpha <1$  the  chain
of identities
$$ \eqalign{
\rec{\Gamma(1-\alpha)}\,  \int_{0^-}^t \,
 {f'(\tau) \over (t-\tau)^\alpha }\, d\tau &=
   {t^{-\alpha }\over \Gamma (1-\alpha)}\, f(0^+)  +
   \rec{\Gamma(1-\alpha)}\,  \int_{0}^t \,
 {f'(\tau) \over (t-\tau)^\alpha }\, d\tau \cr
  &=   {t^{-\alpha }\over\Gamma(1-\alpha)}\,f(0^+) + D_*^\alpha \, f(t)
   =   D^\alpha \, f(t) \,,\cr} \eqno(1.43)$$
where we have used (1.19) with $m=1\,. $
\vsp
In recent years it has become customary to use in place of
(1.36) the Weyl fractional integral
$$ W_\infty^\alpha	  \,f(t) :=
      \rec{\Gamma(\alpha )}\,
 \int_t^\infty (\tau-t)^{\alpha -1}\, f(\tau )\,d\tau \,,
    \q \alpha  \in \RR^+
 \,,  \eqno(1.44)$$
based on a definition of Weyl (1917).
For $t>0$ it is a sort of complementary integral with respect to the
usual Riemann-Liouville integral (1.2).
The relation between (1.36) and (1.44) can
be readily obtained by noting
that, see \eg  [10],
$$ \eqalign{J_{-\infty}^\alpha \,f(t) &=
   \rec{\Gamma(\alpha )}
 \int_{-\infty}^t\! \!\!(t-\tau)^{\alpha -1}\, f(\tau )\,d\tau
    = - \rec{\Gamma(\alpha)}
 \int_\infty^{-t}\!\!\!(t+\tau')^{\alpha -1}\, f(-\tau')\, d\tau' \cr
 &=  \rec{\Gamma(\alpha )}\,
 \int_{t'}^\infty (\tau'-t')^{\alpha -1}\, f(-\tau' )\,d\tau'
  = W_\infty^\alpha \, g(t') \,, \cr} \eqno(1.45)$$
with $g(t')=f(-t')\,, \, t'=-t\,. $
In the above passages we have carried out the changes of variable
$\tau \to \tau '=-\tau$ and $t \to t' = -t\,. $
\vsp
For convenience of the reader, let us recall that exhaustive tables of
Riemann-Liouville and Weyl fractional integrals are available in the
second volume of the Bateman Project collection of Integral
Transforms [16], in  the chapter $XIII$ devoted to {\it fractional
integrals}. \vsp
Last but not the least, let us consider the question of notation.
The present authors oppose to the use of the notation
$D^{-\alpha }$ for denoting the fractional integral, since it is
misleading,
even if it is used in distinguished treatises as  [2], [10], [15].
It is well known that derivation and integration operators are
not inverse to each other, even if their order is integer, and
therefore   such unification of symbols,
present only in the framework of the fractional calculus, appears not
justified. Furthermore, we have to keep in mind that for
fractional order
the derivative is yet an {\it integral} operator, so that, perhaps,
it would be less disturbing to denote our $D^\alpha $ as $J^{-\alpha }$,
than our $J^\alpha$ as $D^{-\alpha }\,. $
\vfill\eject
\line{1.5 {\it The Law of Exponents} \hfill}
\vsp
In the ordinary calculus the  properties
of  the    operators of integration and differentiation
with respect to the laws of commutation and additivity of their
(integer) exponents are well known.  Using our notation,
the (trivial) laws
$$  J^m\, J^n = J^n \, J^m = J^{m+n}\,, \qq
     D^m\, D^n = D^n \, D^m = D^{m+n}\,,   \eqno(1.46)$$
where $ m,n = 0,1,2, \dots \,,$
can be referred to as the {\it Law of Exponents}  for the
operators of integration and differentiation of integer order,
respectively.  Of course,  for any positive integer order,
the operators $D^m$ and $J^n$  do not  commute,
 see  (1.11-12).
\vsp
In the fractional calculus  the {\it Law of Exponents}
is known to be generally  true	for  the operators of {\it fractional
integration} thanks to	their {semigroup property} (1.3).
In general,  both the operators of fractional differentiation,
$D ^\alpha$ and $D_*^\alpha\,,$ do not satisfy	either the  semigroup
property,  or the (weaker)  commutative property.
To show how the {\it Law of Exponents} does not necessarily
hold for the standard fractional derivative,
we provide two simple examples (with power functions) for which
$$ \l\{\eqalign{
  {\rm (a)}& \q D^\alpha D^\beta\, f(t) = D^\beta  D^\alpha\, f(t)
		 \ne D^{\alpha +\beta }\, f(t)\,, \cr
  {\rm (b)}& \q D^\alpha D^\beta\, g(t) \ne D^\beta  D^\alpha\, g(t)
     = D^{\alpha +\beta }\,g(t)\,.\cr}\r.
	     \eqno(1.47)  $$
In the example
(a) let us take   $f(t) = t^{-1/2}$ and $\alpha =\beta =1/2\,. $
Then,  using (1.15), we get  $D^{1/2} \, f(t) \equiv 0\,, $
$D^{1/2}D^{1/2} \, f(t) \equiv 0\,, $ but
$D^{1/2+1/2} \, f(t)= D\, f(t) = -t^{-3/2}/2\,.  $
In the example
(b) let us take   $g(t) = t^{1/2}$ and $\alpha =1/2\,, \,\beta =3/2\,. $
Then,  again using (1.15), we get
 $D^{1/2} \, g(t) = \sqrt{\pi}/2\,, $
$D^{3/2} \, g(t) \equiv 0\,, $
 but
$D^{1/2}D^{3/2} \, g(t)\equiv 0\,,$
$\, D^{3/2}D^{1/2} \, g(t)= -t^{3/2}/4 \, $
and
$D^{1/2+3/2} \, g(t)= D^2\, g(t) = -t^{3/2}/4\,.  $
\vsp
Although modern mathematicians would seek the conditions
to justify the {\it Law of Exponents} when the order of
differentiation and integration are composed together,
we resist the temptation to dive into the delicate
details of the matter, but rather refer the interested reader
to \S IV.6 ("The Law of Exponents") in the book by Miller and
Ross [10].
Let us, however, extract (in our notation, writing
$J^\alpha  $ in place of $D^{-\alpha}$ for $\alpha >0$)
three important cases, contained in their  Theorem 3:
{\it If $f(t)= t^\lambda \, \eta (t)$  or
   $f(t)= t^\lambda \,{\rm ln}\, t\; \eta (t)\,, $
where $\lambda >-1$ and
  $\eta(t) = \sum_{n=0}^{\infty} a_n\, t^n$ having a positive
radius $R$ of convergence, then for
$0\le t <R\,, $   the following three formulas are valid:}
$$ \l\{\eqalign{
 \,& \mu \ge 0 \q {\rm and}\q	 0\le \nu \le \mu
  \, \Longrightarrow \, D^\nu J^\mu \, f(t) =  J^{\mu -\nu}\, f(t)\,, \cr
 \,& \mu \ge 0 \q {\rm and}\q	  \nu > \mu
  \, \Longrightarrow \, D^\nu J^\mu \, f(t) =  D^{\mu -\nu}\, f(t)\,, \cr
 \,& 0\le \mu <\lambda +1 \q {\rm and}\q     \nu \ge 0
  \, \Longrightarrow \, D^\nu D^\mu \, f(t) =  D^{\mu +\nu}\, f(t)\,. \cr
  }\r. \eqno(1.48)$$
At least
in the case of $f(t)$ without  the factor ${\rm ln}\, t\,,$
 the proof
of these formulas is straightforward. Use the definitions (1.2) and (1.13)
of fractional integration and differentiation, the semigroup property
(1.3) of fractional integration, and apply the formulas (1.4) and (1.15)
termwise to the infinite series you meet in the course of calculations.
Of course, the condition that the function $\eta (t)$ be analytic can be
considerably relaxed; it only need be "sufficiently" smooth."
\vsp
The lack  of commutativity and the non-validity of  the  law
of exponents
has   led to the notion of
{\it sequential fractional differentiation} in which the order
in which fractional differentiation operators
$\,D^{\alpha _1}\,,$
$\, D^{\alpha _2}\,,  \dots \,, D^{\alpha _k}\,$  are concatenated
is   crucial.
For this and the related field of
fractional differential equations we refer again to Miller and Ross [10].
Furthermore, Podlubny [29]  has also given formulas
for the Laplace transforms of sequential fractional derivatives.
\vsp
In order   to give an impression on the strange effects
to be	expected in use of sequential
fractional derivatives we consider for a function
$f(t)$ continuous for $t\ge 0$ and
for positive numbers $\alpha $ and  $\beta $ with $\alpha +\beta =1$
the three  problems $(a)$, $(b)$, $(c)$
with the respective general solutions
$u\,,v\,,w\,$
in the set of locally integrable functions,
$$\l\{ \eqalign{
 (a)& \; D^\alpha D^\beta \, u(t)  =f(t)\, \Rightarrow \,
   u(t) = J\, f(t) + a_1 + a_2 \, t^{\beta -1}\,,
 \cr
 (b)&\; D^\beta  D^\alpha \, v(t)  =f(t)\, \Rightarrow \,
 v(t) = J\, f(t) + b_1 + b_2 \, t^{\alpha  -1}\,,
\cr
 (c)& \; D\, u(t)  =f(t)\, \Rightarrow \, w(t) = J\, f(t) + c \,,
 \cr
   } \r. \eqno(1.49)$$
where
$a_1\,,\, a_2\,, b_1\,,\,b_2\,,c\,$ are arbitrary constants.
Whereas the result for $(c)$ is obvious, in order to obtain the
final results for $(a)$ [or $(b)$]
we need to apply first the operator  $J^\alpha$  [or $J^\beta$]
and then the operator $J^\beta$ [or $J^\alpha $].
The additional terms must be taken  into account because
$ D^\gamma  \, t^{\gamma -1} \equiv 0\,,\;
  J^{1-\gamma}\, t^{\gamma-1} =\Gamma(\gamma )\,, \;
  \gamma = \alpha ,\beta \,.  $
We observe that, whereas the general solution of $(c)$ contains
{\it one} arbitrary constant, that of $(a)$ and likewise of $(b)$
contains {\it two} arbitrary constants, even though $\alpha +\beta =1\,. $
In case $\alpha \ne \beta $ the singular behaviour of $u(t)$ at
$t=0^+$ is distinct from that of  $v(t)\,. $
\vsp
From above we can conclude in rough words: sufficiently fine
sequentialization increases the number of free
constants in the general solution of a fractional differential
equation, hence the number of conditions that must be
imposed to make the solution unique.
For an example see Bagley's treatment of a composite
fractional oscillation equation [30];  there the highest order
of derivative is 2, but four conditions are required to
achieve uniqueness.
\vsp
In the present lectures we shall avoid the above troubles
since we shall consider only differential equations
containing {\it single} fractional derivatives.
Furthermore  we shall adopt the {\it Caputo fractional derivative}
in order to meet the  usual physical requirements
for which the initial conditions
are expressed in terms of a given number of bounded values assumed by the
field variable and its derivatives of integer order, see
(1.24) and (1.30).
\vfill\eject
\vsp
\pageno=235
\line{2. FRACTIONAL INTEGRAL EQUATIONS \hfill} \vsp
In this section we shall consider the most simple integral
equations of fractional order, namely the Abel integral equations
of the first and the second kind.
The former investigations on such equations are due to
 Abel (1823-26), after whom they are named, for the
first kind,  and to Hille and Tamarkin (1930) for the second
kind. The interested reader is	referred to [5], [21]
and [31-33]
for historical notes and detailed analysis with applications.
Here we limit ourselves to put some emphasis on the method
of the Laplace transforms, that makes easier and more
comprehensible the treatment
of such fractional integral equations, and provide some
applications. 
\vs
\line{{\it 2.1 Abel integral equation of the first kind \hfill}}
\vsp
Let us consider the Abel integral  equation of the first kind
$$ \rec{\Gamma(\alpha )}\, \int_0^t \!\!
{u(\tau)\over (t-\tau )^{1-\alpha}} \, d\tau  = f(t)\,,
  \q 0< \alpha <1\,, \eqno(2.1)$$
where $f(t)$ is a given function.
We easily recognize that this equation can be expressed in terms
of a fractional integral, \ie
 $$    J^{\alpha} \, u(t) = f(t)  \,, \eqno(2.2)$$
 and consequently solved in
terms of a fractional derivative, according to
$$
   u(t) = D^{\alpha} \, f(t)  \,. \eqno(2.3) $$
To this end we need to recall the definition (1.2) and the property
(1.14) $ D^\alpha \, J^\alpha = I$.
\vsp
Let us now solve (2.1)	using  the Laplace transform.
Noting from (1.7-8) and (1.10)	that
$	      J^{\alpha} \, u(t)
  = \Phi_\alpha (t) \,*\, u(t)\,
 \, \div \, \bar u(s) /{s^\alpha }\,,$
we then obtain
$$   {\bar u(s)\over s^\alpha } = \bar f(s)
\, \Longrightarrow \,
      \bar u(s) = s^\alpha \, \bar f(s)\,.\eqno(2.4) $$
Now we can choose  two different ways to get the inverse Laplace
transform  from (2.4), according to the standard rules.
Writing  (2.4) as
$$ \bar u(s) = s \, \l[{\bar f(s)\over s^{1-\alpha }}\r]\,, \eqno(2.4a)$$
we obtain
$$ {u(t) =\rec{\Gamma(1-\alpha)} \,{d\over dt}\,
   \int_0^t \!\! {f(\tau )\over (t-\tau )^\alpha }\, d\tau} \,.
   \eqno(2.5a) $$
On the other hand,  writing  (2.4) as
$$  \bar u(s) = \rec{s^{1-\alpha}}\,
  [ s\, \bar f(s) - f(0^+)] + {f(0^+)\over s^{1-\alpha }}\,,
   \eqno(2.4b)$$
we obtain
$$   {u(t) =\rec{\Gamma(1-\alpha) } \,
   \int_0^t \!\! {f^\prime(\tau )\over (t-\tau )^\alpha }\, d\tau
  +  f(0^+) \, {t^{-\alpha }\over \Gamma(1-a)}} \,.
   \eqno(2.5b) $$
\vfill\eject
\noindent
Thus, the solutions (2.5a) and (2.5b)
are expressed in terms of the fractional derivatives
$D^\alpha $ and $D_*^\alpha\,,	$ respectively,
according to (1.13), (1.17) and (1.19) with $m=1\,. $
\vsp
The way $b)$ requires that $f(t)$ be differentiable with
 ${\cal{L}}$-transformable derivative;
consequently $0 \le |f(0^+)| < \infty \,. $
Then it turns out from (2.5b) that
$u(0^+)$ can be infinite   if $f(0^+) \ne 0\,, $ being
$u(t) = O(t^{-\alpha})\,, $ as $t\to 0^+\,. $
The way $a)$ requires weaker conditions in that the integral
at the R.H.S. of (2.5a) must  vanish as $t \to 0^+$;
consequently $f(0^+)$ could be infinite but with
$f(t) = O(t^{-\nu})\,,\;0<\nu <1-\alpha \, $ as $t\to 0^+\,. $
To this end keep in mind that $ \Phi_{1-\alpha }\,*\,\Phi_{1-\nu} =
  \Phi_{2-\alpha -\nu }$.
Then it turns out from (2.5a) that
$u(0^+)$ can be infinite   if $f(0^+)$ is infinite, being
$u(t) = O(t^{-(\alpha+\nu)})\,, $ as $t\to 0^+\,. $
\vsp
Finally, let us remark that we can analogously treat the case
of equation (2.1) with $0<\alpha <1$ replaced by $\alpha >0\,. $
If $m-1<\alpha \le m$ with $m\in\NN\,, $ then again we have
(2.2), now with $D^\alpha \, f(t)$ given by the formula (1.13)
which can also be obtained by the Laplace transform method.
\vvs
\line{{\it 2.2 Abel integral equation of the second kind \hfill}}
\vsp
Let us now consider the Abel equation of the  second kind
$$ {u(t)  + {\lambda \over \Gamma(\alpha )}\, \int_0^t \!\!
{u(\tau)\over (t-\tau )^{1-\alpha}} \, d\tau = f(t)\,,
  \q \alpha >0\,,\q \lambda \in \CC\,.	}\eqno(2.6)$$
In terms of the fractional integral operator such equation reads
$$
   \l( 1 +\lambda \, J^{\alpha}\r) \,u(t)=f(t)\,,\eqno(2.7)
$$
and consequently can be  formally solved as follows:
$$
   u(t)
  = \l(1 + \lambda J^{\alpha}\r)^{-1} \, f(t)
 = \l(1+ \sum_{n=1}^{\infty} (-\lambda) ^n \, J^{\alpha n}\r) \, f(t)
  \,. \eqno(2.8)$$
Noting by (1.7-8) that
$$ J^{\alpha n} \, f(t) = \Phi_{\alpha n}(t)\,*\, f(t)
 = {t_+^{\alpha n-1}\over \Gamma(\alpha n)}\,*\, f(t)$$
the formal solution reads
$$ u(t) = f(t) +  \l( \sum_{n=1}^{\infty} (-\lambda) ^n \,
   {t_+^{\alpha n-1}\over \Gamma(\alpha n)}\r)\,*\, f(t)\,.
  \eqno(2.9)$$
Recalling from the Appendix the definition of the 
function,
  $$ e_\alpha (t;\lambda ) := E_\alpha (-\lambda \, t^\alpha ) =
\sum_{n=0}^\infty {{\l(-\lambda \,t^{\alpha}\r)}^n
 \over \Gamma (\alpha n+1)} \,,
  \q t>0\,,  \q \alpha > 0\,, \q \lambda \in \CC\,,
     \eqno(2.10)$$
where
$E_\alpha\, $
denotes the  Mittag-Leffler function of order $\alpha\,,$
 we note that
$$
   \sum_{n=1}^{\infty} (-\lambda) ^n \,
    {t_+^{\alpha n-1} \over \Gamma (\alpha n)}
 = {d \over dt}  E_\alpha  (-\lambda t^\alpha )
 = e_\alpha ^\prime(t;\lambda )  \,, \q t>0\,.\eqno(2.11)  $$
\vfill\eject
\noindent
Finally, the solution  reads
$$u(t)= f(t)+ e_\alpha^\prime (t;\lambda)\,*\, f(t) \,. \eqno(2.12)
 $$
\vsp
Of course the above formal proof
can be made  rigorous.
Simply observe that because of the rapid growth of the gamma function the
infinite series in (2.9) and (2.11) are uniformly convergent in every
bounded interval of the variable $t$ so that term-wise integrations and
differentiations are allowed.
However,
we prefer to use the alternative technique of Laplace transforms,
which will allow us  to obtain
the solution in different forms, including  the result (2.12).
\vsp
Applying the Laplace transform to (2.6) we
obtain
$$ \l[1 +{\lambda \over s^\alpha}\r] \bar u(s) =  \bar f(s)
 \, \Longrightarrow \,
\bar u(s) = {s^\alpha \over s^\alpha +\lambda}\, \bar f(s)
\,. \eqno(2.13)$$
Now,  let us proceed to obtain the inverse Laplace transform
of (2.13) using    the following
Laplace transform pair	(see Appendix)
   $$ e_\alpha (t;\lambda) := E_\alpha (-\lambda \, t^\alpha)
 \, \div \,
	  {s^{\alpha -1}\over s^\alpha +\lambda}\,.
 \eqno(2.14) $$
\vsp
As for the Abel equation of the first kind, we can choose  two different
ways to get the
inverse Laplace transforms   from (2.13), according to the standard rules.
Writing  (2.13) as
$$ \bar u(s) = s \, \l[
 {s^{\alpha -1}\over s^\alpha+ \lambda}\, \bar f(s)\r]\,,
\eqno(2.13a)$$ we obtain
$$ {u(t) ={d\over dt}\,
   \int_0^t \!\! f(t-\tau ) \, e_\alpha (\tau;\lambda)
  \, d\tau} \,.      \eqno(2.15a) $$
If we write  (2.13) as
$$  \bar u(s) = {s^{\alpha -1}\over s^\alpha +\lambda }\,
  [ s\, \bar f(s) - f(0^+)] + f(0^+)\,
  {s^{\alpha -1}\over s^\alpha +\lambda }\,,	  \eqno(2.13b)$$
we obtain
$$ {u(t) =  \int_0^t
\!\! f^\prime(t-\tau)  \,    e_\alpha (\tau;\lambda)\,	d\tau
   +	f(0^+) \, e_\alpha(t;\lambda )} \,.
   \eqno(2.15b) $$
We also note that,  $e_\alpha (t;\lambda)$ being a
function differentiable with respect to $t$
with $e_\alpha (0^+;\lambda)= E_\alpha	(0^+)=1\,, $
there exists another possibility to re-write (2.13), namely
$$ \bar u(s) = \l[ s \,
 {s^{\alpha -1}\over s^\alpha+ \lambda} -1\r] \, \bar f(s) +
  \bar f(s)\,.
\eqno(2.13c)$$
Then  we obtain
$$u(t) =
   \int_0^t \!\!  f(t-\tau)\, e_\alpha^\prime (\tau;\lambda)\, d\tau
  + f(t) \,,	  \eqno(2.15c) $$
in agreement with (2.12).
We see that
the way $b)$ is  more restrictive than the ways $a)$ and $c)$
since it requires that	$f(t)$ be differentiable with
 ${\cal{L}}$-transformable derivative.
\vfill\eject
\line{{\it 2.3 Some applications of Abel integral equations \hfill}}
\vsp
     It is well known that Niels Henrik Abel was led to his famous
equation by the mechanical problem of the   {\it tautochrone},
that is by the problem of determining the shape of a curve in the vertical
plane such that the time required for a particle to slide down the curve
to its lowest point is equal to a given function of its initial height
(which is considered as a variable in an interval $[0, H]$). After
appropriate changes of variables he obtained his famous integral equation
of first kind with  $\alpha = 1/2\,. $
He did, however, solve the general case
$0<\alpha < 1\,.$
See Tamarkin's translation$\,\null^{1)}$ of and comments to
Abel's short paper$\,\null^{2)}\,.$
As a special case Abel discussed the problem of the
{\it isochrone}, in which it is required that the time of sliding
down is independent of the initial height. Already in his earlier
publication$\,\null^{3)}$
he recognized the solution as derivative of non-integer order.
\footnote{}{\baselineskip=12truept   \note 
\item{$\null^{1)}$\ }
  Abel, N. H.: Solution of a mechanical problem. Translated from the
   German. In: D. E. Smith, editor: A Source Book in Mathematics, pp.
   656-662. Dover Publications, New York, 1959.
\item{$\null^{2)}$\ }
  Abel, N. H.: Aufloesung einer mechanischen Aufgabe.
  Journal f\"ur die
  reine und angewandte Mathematik (Crelle), Vol. I (1826), pp. 153-157.
\item{$\null^{3)}$\ }
 Abel, N. H.: Solution de quelques probl\`emes \`a l'aide
   d'int\'egrales   d\'efinies.
   Translated from the Norwegian original,
   published in Magazin for Naturvidenskaberne. Aargang 1, Bind 2,
   Christiana 1823. French Translation in Oeuvres Compl\`etes, Vol I,
   pp. 11-18. Nouvelle \`edition par L. Sylow et S. Lie, 1881.
\vsh}
\vsp
We point out that integral equations of Abel type, including the simplest
(2.1) and (2.6), have found so many applications in diverse fields
that it is almost impossible to provide an exhaustive list of them.
\vsp
 Abel integral equations occur in many situations where physical
measurements are to be evaluated. In many of these the
independent variable is the radius of a circle or a sphere and only after
a change of variables the integral operator has
the form  $J^\alpha\,, $ usually
with  $\alpha	=  1/2 \,,$ and the equation is of first kind.
Applications
are, \eg, in evaluation of spectroscopic measurements of cylindrical gas
discharges, the study of the solar or a planetary atmosphere, the
investigation of star densities in a globular cluster, the inversion of
travel times of seismic waves for determination of terrestrial
sub-surface structure, spherical stereology. Descriptions and analysis of
several problems of this kind can be found in the books by Gorenflo and
Vessella [21] and by Craig and Brown [31], see also [32]. Equations of
first and of second kind, depending on the arrangement of the measurements,
arise in spherical stereology. See [33] where an analysis of the basic
problems and many references to previous literature are given.
\vfill\eject
\vsp
  Another field in which Abel integral equations or integral equations
with more general weakly singular kernels are important is that of
{\it inverse boundary value problems} in partial differential equations,
in particular parabolic ones in which naturally the independent variable
has the meaning of time.
We are going to describe in detail the occurrence of Abel
integral equations of first and of second kind in the problem of heating
(or cooling) of a semi-infinite rod by influx (or efflux) of heat across
the boundary into (or from) its interior.
Consider the {\it equation of heat flow}
$$ u_t - u_{xx} =0 \,, \q u=u(x,t)\,, \eqno(2.16)$$
in the semi-infinite intervals $0<x<\infty$
and $0<t<\infty$  of space and time, respectively. In this
dimensionless equation $u=u(x,t)$ means temperature.
Assume vanishing initial temperature, \ie $u(x,0) =0$
for $0<x<\infty$ and given influx across the boundary $x=0$
from $x<0$ to $x>0\,, $
$$ -u_x(0,t) = p(t)\,. \eqno(2.17)$$
Then, under  appropriate regularity conditions,  $u(x,t)$ is given
by the formula, see \eg [34],
$$ u(x,t) = \rec{\sqrt{\pi}}\, \int_0^t
     {p(\tau)\over \sqrt{t-\tau}}\, \e^{\, \ds-{x^2/ [4(t-\tau)]}} \,
    d\tau \,, \q x>0\,,\; t>0\,. \eqno(2.18)$$
We turn our special interest to the interior boundary temperature
$ \phi	(t):= u(0^+,t)\,,$  $t>0\,, $
which by (2.18) is represented as
$$   \rec{\sqrt{\pi}}\, \int_0^t
     {p(\tau)\over \sqrt{t-\tau}}\,
    d\tau  = J^{1/2}\, p(t) = \phi (t)
\,, \q t>0\,. \eqno(2.19)$$
We recognize (2.19) as an Abel integral equation of {\it first kind}
for {\it determination of an unknown influx $p(t)$
if the interior boundary temperature $\phi (t)$ is given}
by measurements, or intended to be achieved by controlling
the influx. Its solution is given by formula (1.13) with
$m=1\,, $ $\alpha =1/2\,, $ as
$$ p(t) = D^{1/2}\, \phi (t) = \rec{\sqrt{\pi}}\,
   {d \over dt}\, \int_0^t
     {\phi (\tau )\over \sqrt{t-\tau }}\, d\tau \,.
	  \eqno(2.20)$$
\vsp
It may be illuminating to consider the following special cases,
  $$\left\{ \eqalign{
   (i)& \q \phi (t)= t\, \Longrightarrow \,
  p(t) = {1\over 2}\, \sqrt{\pi \,t} \,, \cr
 (ii)& \q \phi (t)= 1\, \Longrightarrow \,
  p(t) = {1\over \sqrt{\pi \,t}} \,, \cr} \right.
   \eqno(2.21)$$
where we have used formula (1.15).
So, for linear increase of interior boundary temperature
the required influx is continuous and increasing from
$0$ towards $\infty$ (with unbounded derivative at $t=0^+$),
whereas
for instantaneous jump-like increase from $0$ to $1$
the required influx decreases from $\infty$
at $t=0^+\, $
to $0$ as $t\to \infty\,. $
\vfill\eject
We now modify our problem to obtain an Abel integral equation
of {\it second kind}. Assume that the rod $x>0$ is bordered at $x=0$
by a bath of liquid in $x<0$ with controlled exterior
boundary temperature $u(0^-,t) := \psi(t)\,. $
\vsp
Assuming Newton's radiation law we have an influx of heat
from $0^-$ to $0^+$ proportional to the difference of
exterior and interior temperature,
$$ p(t) = \lambda \, \l[\psi(t)-\phi (t)\r]\,, \q \lambda >0\,.
\eqno(2.22)$$
Inserting (2.22) into (2.19) we obtain
$$
 \phi (t)  = {\lambda \over \sqrt{\pi}}\,
    \int_0^t
     {\psi(\tau )-\phi (\tau )\over \sqrt{t-\tau }}\, d\tau \,,
$$
namely, in operator notation,
  $$
\l(1+\lambda \, J^{1/2}\r) \, \phi (t)= \lambda \, J^{1/2}\, \psi(t)
\,. \eqno(2.23)$$
If we {\it now assume the exterior boundary temperature $\psi(t)$ as
given and the evolution in time of the interior
boundary temperature $\phi (t)$ as unknown}, then (2.23)
is an Abel integral equation of {\it second kind}
for determination of $\phi (t)\,. $
\vsp
With $\alpha =1/2$ the equation (2.23) is of the form (2.7),
and by (2.8) its solution is
$$ \phi (t) = \lambda \,
\l(1+\lambda J^{1/2}\r)^{-1}\, J^{1/2}\, \psi(t) =
   - \sum_{m=0}^\infty (-\lambda )^{m+1}\, J^{(m+1)/2}\, \psi(t)\,.
\eqno(2.24)$$
\vsp
Let us investigate the very special case of constant exterior boundary
temperature
$$\psi(t) =1\,.\eqno(2.25)$$
Then, by (1.4) with $\gamma =0\,,$
$$  J^{(m+1)/2}\, \psi(t) = {t^{(m+1)/2}\over
   \Gamma \l[(m+1)/2 +1\r]}
 \,, $$
hence
$$ \phi (t) =
   - \sum_{m=0}^\infty (-\lambda )^{m+1}\,
  {t^{(m+1)/2}\over \Gamma \l[(m+1)/2 +1\r]}=
 1 - \sum_{n=0}^\infty (-\lambda )^{n}\,
  {t^{n/2}\over \Gamma \l(n/2 +1\r)} \,, $$
so that
$$ \phi (t) = 1- E_{1/2}\l(-\lambda t^{1/2}\r) =
  1- e_{1/2}(t;\lambda )\,. \eqno(2.26)$$
Observe that $\phi (t)$ is creep function, increasing strictly
monotonically from 0 towards 1 as $t$ runs from 0 to $\infty\,. $
\vsp
   For more or less distinct treatments of this problem of "Newtonian
heating" the reader may consult [21], and [35-37]. In [37] a
formulation in terms of fractional differential equations is derived
and, furthermore, the analogous problem
of "Newtonian cooling" is discussed.
\vfill\eject

\pageno=241	   
\line{3. FRACTIONAL DIFFERENTIAL EQUATIONS: 1-st PART \hfill} \vsp
We now analyse	 
the most simple differential equations of fractional order.
For this purpose, following  our  recent works [37-42],
we choose the  examples
which, by means of  fractional derivatives, generalize the
well-known ordinary differential equations related to
relaxation and oscillation phenomena.
In this section we  treat the simplest types,
which we refer to as
the {\it simple fractional relaxation and oscillation
equations}. In the next section  we  shall consider
the types, somewhat more cumbersome,
which we refer to as
the {\it composite fractional relaxation and oscillation
equations}.
\vs
\line{{\it 3.1 The simple fractional relaxation and oscillation
equations \hfill}}
\vsp
The classical phenomena of
relaxation and oscillations in their simplest form
are known to be
governed by  linear ordinary differential equations,
of order one and two  respectively, that
hereafter we  recall
with the corresponding solutions.
Let us denote by $u=u(t)$ the field variable
 and by $q(t)$ a given continuous
function, with $t\ge0\,. $
\vsp
The {\it relaxation} differential equation  reads
$$ u^{\prime}(t) = - u(t) +q(t)\,, \eqno(3.1)$$
whose solution, under the initial condition $u(0^+)=c_0\,, $ is
$$ u(t) = c_0 \, \e^{\ds-t} +
  \int_0^t \!\! q(t-\tau)\, \e^{\ds-\tau} \, d\tau \,. \eqno(3.2)$$
\vsp
The {\it oscillation} differential equation  reads
$$ u^{\prime\prime}(t) = - u(t) +q(t)\,, \eqno(3.3)$$
whose solution, under the initial conditions $u(0^+)=c_0$ and
$u^{\prime}(0^+) = c_1\,, $ is
$$ u(t) = c_0 \, \cos t+ c_1 \, \sin t +
\int_0^t \!\! q(t-\tau )\,\sin {\tau}\, d\tau \,. \eqno(3.4) $$
\vsp
From the point of view of the fractional calculus
a natural generalization of equations (3.1) and (3.3)
is obtained by replacing the ordinary
derivative with a fractional one of order  $\alpha\,.  $
In order to
preserve the type of initial conditions required in the classical
phenomena, we agree to replace the first and second derivative
in (3.1) and (3.3) with a Caputo fractional derivative of
order $\alpha $ with $0<\alpha <1$ and $1<\alpha <2\,, $
respectively. We agree to refer to the corresponding
equations   as the {\it simple fractional relaxation equation}
and the {\it simple fractional oscillation equation}.
\vfill\eject
Generally speaking, we consider
the following differential equation of fractional order
$\alpha >0\,, $
$$ D_*^\alpha \, u(t) =
D^\alpha  \l( u(t) - \sum_{k=0}^{m-1} {t^k \over k!} \,
u^{(k)} (0^+)\r)    = - u(t) + q(t) \,,  \q t>0\,. \eqno(3.5)$$
Here $m$ is a positive integer uniquely defined
by $m-1 <\alpha  \le m\,, $
which provides the number of the prescribed initial values
$u^{(k)}(0^+) = c_k\,, \; k=0,1,2, \dots , m-1\,. $
Implicit in the form of (3.5) is our desire to obtain solutions
$u(t)$ for which the $u^{(k)}(t)$ are continuous   for $t \ge 0,\
k=0,1,\dots,m-1.$
In particular, the cases of {\it fractional relaxation} and
{\it fractional oscillation}
are obtained for $m=1$ and $m=2\,, $   respectively
\vsp
We note that when $\alpha$ is the integer $m$
the equation (3.5) reduces to an ordinary differential equation
whose solution	can be expressed in terms of $m$ linearly
independent solutions of the {\it homogeneous} equation
and of one particular solution of the {\it inhomogeneous}
equation.
We summarize the well-known result
as follows
   $$ u(t) = \sum_{k=0}^{m-1} c_k u_k(t) +
 \int_0^t \!\! q(t-\tau )\,u_\delta (\tau)\, d\tau \,. \eqno(3.6) $$
    $$ u_k(t) = J^k \, u_0(t)\,,\q
    u_k^{(h)}(0^+)=\delta _{k\,h}\,, \q
   h,k =0,1,\dots,m-1\,,		 \eqno(3.7)$$
    $$ u_\delta (t) = - \, u_0^\prime(t)\,. \eqno(3.8)$$
Thus, the $m$ functions $u_k(t)$
represent the {\it fundamental solutions}
of the differential equation of order $m\,, $
namely
those linearly independent solutions
of the {\it homogeneous} equation which satisfy
the initial conditions in (3.7).
The function $u_\delta (t)\,, $  with which the free term $q(t)$ appears
convoluted, represents the so called
{\it impulse-response solution},
namely the particular solution of the {\it inhomogeneous}
equation with all
$ c_k \equiv 0 \,, \; k=0,1, \dots , m-1\,,$
and with  $q(t) = \delta (t)\,. $
In the cases of ordinary relaxation and oscillation
we  recognize that
$u_0(t)  = \e^{-t} = u_\delta (t)$ and
$u_0(t) = \cos \,t\,,$
$\, u_1(t) = J\, u_0(t) = \sin \,t =  \cos \,(t -\pi/2) = u_\delta (t)
\,, $
respectively.
\vsn
\underbar{Remark 1} :
 The more general equation
$$ D^\alpha  \l( u(t) -
\sum_{k=0}^{m-1} {t^k \over k!} \, u^{(k)} (0^+)\r)
   = -	{\rho}^\alpha \, u(t) + q(t) \,, \q \rho >0\,,	\q t>0\,,
 \eqno(3.5')$$
can be reduced to (3.5) by a change of scale
$t \to t/\rho \,. $
We prefer, for ease of notation, to discuss the "dimensionless"
form (3.5).
\vsp
Let us now solve (3.5) by the method of Laplace transforms.
For this purpose we can use directly the Caputo formula (1.30)
or, alternatively, reduce (3.5) with the prescribed initial
conditions as an equivalent (fractional) integral  equation
and then treat the integral equation by the Laplace transform
method.
Here we prefer to follow the second way.
Then, applying the operator $J^\alpha $ to both sides of (3.5) we obtain
\vfill\eject
$$ u(t) = \sum_{k=0}^{m-1} c_k \, {t^k \over k!}
     - J^\alpha \, u(t) + J^\alpha  \, q(t) \,. \eqno(3.9)$$
The application of the Laplace transform yields
$$  \bar u(s) = \sum_{k=0}^{m-1} {c_k  \over s^{k+1}}
     - \rec{s^\alpha }\, \bar u(s) + \rec{s^\alpha }\, \bar q(s)\,, $$
hence
$$ \bar u(s) = \sum_{k=0}^{m-1} c_k \,{s^{\alpha -k -1} \over s^\alpha +1}
       + \rec{s^\alpha+1 }\, \bar q(s)\,.  \eqno(3.10)$$
Introducing the Mittag-Leffler type functions 
$$ e_\alpha (t) \equiv	e_\alpha (t;1) :=E_\alpha (-t^\alpha )
  \,\div\,
{s^{\alpha  -1} \over s^\alpha +1}   \,,  \eqno(3.11)$$
$$ u_k(t) := J^k e_\alpha (t) \, \div \,
   {s^{\alpha -k -1} \over s^\alpha +1}  \,,\q
   k=0,1,\dots,m-1\,,  \eqno(3.12)$$
we find, from inversion of the Laplace transforms in (3.10),
$$ u(t) = \sum_{k=0}^{m-1} c_k \, u_k (t)
   - \int_0^t \!\! q(t-\tau )\, u_0 ^\prime(\tau)\, d\tau \,.
   \eqno (3.13)$$
For finding the last term in the R.H.S. of (3.13), we have used
the well-known rule for the Laplace transform of the derivative
noting that
$u_0(0^+)=e_\alpha (0^+) = 1\,, $ and
    $$	{1\over s^\alpha +1} =
 - \l( s\, {s^{\alpha -1}\over s^\alpha +1} -1\r)\, \div  \,
   - u_0 ^\prime(t)= - e_\alpha ^\prime(t)\,. \eqno(3.14)$$
The  formula (3.13)
encompasses the solutions (3.2) and (3.4)
found for $\alpha = 1\,,\,2\,,$ respectively.
\vsp
When $\alpha$ is not integer, namely for $m-1 <\alpha <m\,, $
we note that $m-1$ represents
the integer part of $\alpha$ (usually denoted by $[\alpha]$)
and $m$  the number  of initial  conditions
necessary and sufficient to ensure the uniqueness of the
solution $u(t)$. Thus the $m$ functions
$u_k(t) = J^k e_\alpha (t)$ with $k =0, 1, \dots, m-1 \, $
represent those particular solutions
of the {\it homogeneous} equation which satisfy
the initial conditions
$$ u_k^{(h)} (0^+)  = \delta _{k\,h} \,,
\q h,k =0, 1, \dots, m-1 \, ,
    \eqno(3.15)$$
and therefore they represent the {\it fundamental solutions}
of the fractional equation (3.5), in analogy with the case
$\alpha =m\,. $  Furthermore, the function $ u_\delta (t) =
 -e_\alpha ^\prime (t)$ represents the {\it impulse-response solution}.
Hereafter, we are going to compute  and exhibit
the {\it fundamental solutions} and the {\it impulse-response solution}
 for  the cases (a) $0<\alpha < 1$ and (b) $1<\alpha <2\,, $
pointing out the comparison with
the corresponding solutions obtained when $\alpha =1$ and $\alpha =2\,. $
\vfill\eject
\noindent
\underbar{Remark 2} :
The reader is invited to verify that
the solution (3.13) has continuous derivatives $u^{(k)} (t)$
for $k = 0,1,2, \dots, m-1\,, $ which fulfill the
$m$ initial conditions $u^{(k)}(0^+) =c_k\,. $
In fact, looking back at (3.9), one must recognize the smoothing
power of the operator $J^\alpha \,. $
However,  the so called
{\it impulse-response solution}
of our equation (3.5),	$u_\delta (t)\,, $
is expected to be  not so regular like the ordinary solution (3.13).
In fact, from (3.10) and (3.13-14), one  obtains
$$  u_\delta (t) = - u^\prime _0(t) \div   {1\over s^\alpha +1}\,,
   \eqno(3.16)$$
and  therefore, using the limiting theorem for Laplace transforms,
one  can recognize that, being $m-1 <\alpha < m\,, $
$$  u_\delta ^{(h)}(0^+) =0 \,, \; h=0,1,\dots,m-2\,;\q
 u_\delta ^{(m-1)}(0^+) = \infty\,.   \eqno(3.17)$$
\vsp
We now prefer to derive the relevant properties of the basic functions
$e_\alpha (t)$	directly   from their
representation as a Laplace inverse integral
$$ e_\alpha (t) = \rec{2\pi i}\,
   \int_{Br} \!\! \e^{\ds \,st} \, { s^{\alpha -1} \over s^\alpha +1}\, ds
  \,,\eqno(3.18)$$
in detail for $0<\alpha \le 2\,, $ without detouring on
the general theory of Mittag-Leffler functions
in the complex plane.
In (3.18) $Br$ denotes the Bromwich path, \ie a line
${\rm Re} \,\{ s\} = \sigma $ with a value $\sigma \ge 1\,,$
and ${\rm Im} \,\{ s\} $ running from $-\infty$ to $+\infty\,. $
\vsp
For transparency reasons, we  separately discuss the cases
$$ {\rm (a)}\q 0<\alpha <1 \q {\rm and} \q
   {\rm (b)}\q 1<\alpha <2\,,$$
recalling that in the limiting cases
$\alpha =1\,, \,2\,,$
we  know $e_\alpha (t)$ as elementary function,
 namely
   $ e_1(t) = \e^{\ds\,-t}\,$
and $	 e_2(t) = \cos \, t\,.$
For $\alpha $ not integer the power function  $s^\alpha $ is uniquely
defined as
 $ s^\alpha = |s|^\alpha \, \e ^{i\, {\rm arg}\,s}\,, $
with $-\pi < {\rm arg} \, s <\pi \,,$
that is in the complex
$s$-plane cut along the negative real axis.
\vsp
The essential step  consists in decomposing $e_\alpha (t)$ into two parts
according to
$e_\alpha(t)  = f_\alpha(t) + g_\alpha(t) \,, $
as indicated below.
In case (a) the function $f_\alpha(t)\,, $
in case (b) the function $-f_\alpha(t)\, $
is {\it completely monotone}; in both cases  $f_\alpha (t)$
tends to zero as $t$ tends to infinity,
from above in case (a), from below in case (b).
The other part, $g_\alpha (t)\,, $
is identically vanishing in case (a), but of
{\it oscillatory} character with exponentially	decreasing amplitude
in case (b).
\vsp
In order to obtain the desired decomposition of $e_\alpha $
we bend the Bromwich path of integration  $Br$ into the equivalent
Hankel path $Ha(1^+)$,
 a loop which starts from
$-\infty$ along the lower side of the negative real axis, encircles
the circular disc $|s | =1\,$ in the positive
sense and ends at $-\infty$ along the upper side of the negative real axis.
\vfill\eject
One obtains 
$$  e_\alpha(t)  = f_\alpha(t) + g_\alpha(t) \,, \q t\ge 0\,,
 \eqno(3.19)$$
with
$$  f_\alpha (t) :=
   \rec{2\pi i}\,
   \int_{Ha(\epsilon)} \!\! \e^{\ds \,st} \, { s^{\alpha -1} \over
  s^\alpha +1}\, ds   \,,\eqno(3.20)$$
where now the Hankel path $Ha(\epsilon) $ denotes a loop constituted
by a   small circle $ |s| = \epsilon $ with $\epsilon \to 0$
and by the two borders of the cut negative real axis,
and
$$ g_\alpha (t) := \sum_h   \e^{\,\ds s'_h \,t}\,
Res \,\l[ { s^{\alpha -1} \over s^\alpha +1} \r]_{s'_h} =
   \rec{\alpha }\, \sum_h   \e^{\,\ds s'_h \,t} \,,
  \eqno(3.21)$$
where $s'_h$ are the relevant poles  of
$ s^{\alpha -1} / (s^\alpha +1) $. In fact the poles turn out to
be $s_h = {\rm exp}\l[i(2h+1)\pi/\alpha\r]\,$ with unitary
modulus;
they are all simple but relevant are only those situated
in the main Riemann sheet, \ie the poles $s'_h$ with
argument such that
$ -\pi <\arg \,s'_h  <\pi \,. $
\vsp
If $\,0<\alpha <1\,,$ there are no such poles,
since for all integers $h$
we have $|\arg\, s_h| = |2h+1|\,\pi/\alpha >\pi\,; $ as a consequence,
$$ g_\alpha (t) \equiv 0 \,, \q {\rm hence} \q	e_\alpha (t)
    = f_\alpha (t) \,, \q {\rm if} \q 0<\alpha <1\,. \eqno(3.22)$$
\vsp
If $\,1<\alpha < 2\,, $ then there exist precisely two relevant poles,
namely $s'_0 = {\rm exp} (i\pi/\alpha)$ and
$s'_{-1}= {\rm exp} (-i\pi/\alpha) = {\overline{s_0}}'\,, $
which are located in the  left half plane. Then
one obtains
$$ g_\alpha (t) = {2\over \alpha }\,
     \e^{\ds  t\, \cos \,(\pi /\alpha )}\,
\cos\,\l[t\,\sin\,\l({\pi\over \alpha}\r) \r]
  \,, \q {\rm if}\q 1<\alpha<2 \,. \eqno(3.23)$$
We note that this function exhibits oscillations with
circular frequency $\omega  (\alpha ) = \sin\,(\pi/\alpha)$
and with an  exponentially decaying amplitude
with  rate
$ \lambda (\alpha ) = |\cos \,(\pi /\alpha )|\,. $
\vsn
\underbar{Remark 3} :
One easily recognizes that (3.23) is valid also for
$2\le \alpha <3\,. $  In the classical case $\alpha =2$
the two poles are purely imaginary (coinciding with $\pm i$)
 so that    we recover the sinusoidal behaviour
with unitary frequency.  In the case $2<\alpha <3\,, $
however, the two poles are located in the right half plane,
so providing amplified oscillations. This instability,
which is common to the case $\alpha =3\,, $ is the
reason why we limit ourselves to consider $\alpha $ in the range
$0<\alpha \le 2\,. $
\vsp
It is now an exercise  in complex analysis to show that the contribution
from the Hankel path $Ha(\epsilon)$ as $\epsilon \to 0$ is provided by
$$ f_\alpha (t) := \int_0^\infty \! \e^{\,\ds -rt}\,
     K_\alpha (r)\, dr\,, \eqno(3.24)$$
with
$$ K_\alpha (r) =
  -\,\rec{\pi}\,    {\rm Im}\,
  \l\{ \l.{s^{\alpha -1} \over s^\alpha +1}\r\vert_{s=r\, \e^{i\pi}} \r\}
 = \rec{\pi}\,
   { r^{\alpha -1}\, \sin \,(\alpha \pi)\over
    r^{2\alpha} + 2\, r^{\alpha} \, \cos \, (\alpha \pi) +1}\,.
      \eqno(3.25)$$
\vfill\eject
This function $K_\alpha(r)$ vanishes identically if $\alpha$ is
an integer, it is positive for all $r$ if $\,0<\alpha <1\,,$
negative for all $r$ if $1<\alpha <2\,. $
In fact in (3.25) the denominator  is, for $\alpha$ not integer, always
positive being	$> (r^\alpha -1)^2 \ge0\,. $
Hence $f_\alpha (t)$ has the aforementioned
monotonicity properties, decreasing towards zero in case (a),
increasing towards zero in case (b).
We also note that, in order to satisfy the initial condition
$e_\alpha (0^+) =1\,, $ we find
$\, \int_0^\infty K_\alpha (r)\, dr =1 $ if $0<\alpha <1\,, $
$\,\int_0^\infty K_\alpha (r)\, dr = 1- 2/\alpha  $ if
$1<\alpha <2\,. $
In Fig. 1 we show the plots of the
{\it spectral functions}
$K_{\alpha } (r) $ for some values of $\alpha $ in the intervals
(a) $\; 0<\alpha <1\,, $
(b) $\; 1<\alpha <2 \,. $
$\null$

 \cen{\epsfxsize=7.0truecm \epsfysize=7.0truecm \epsffile{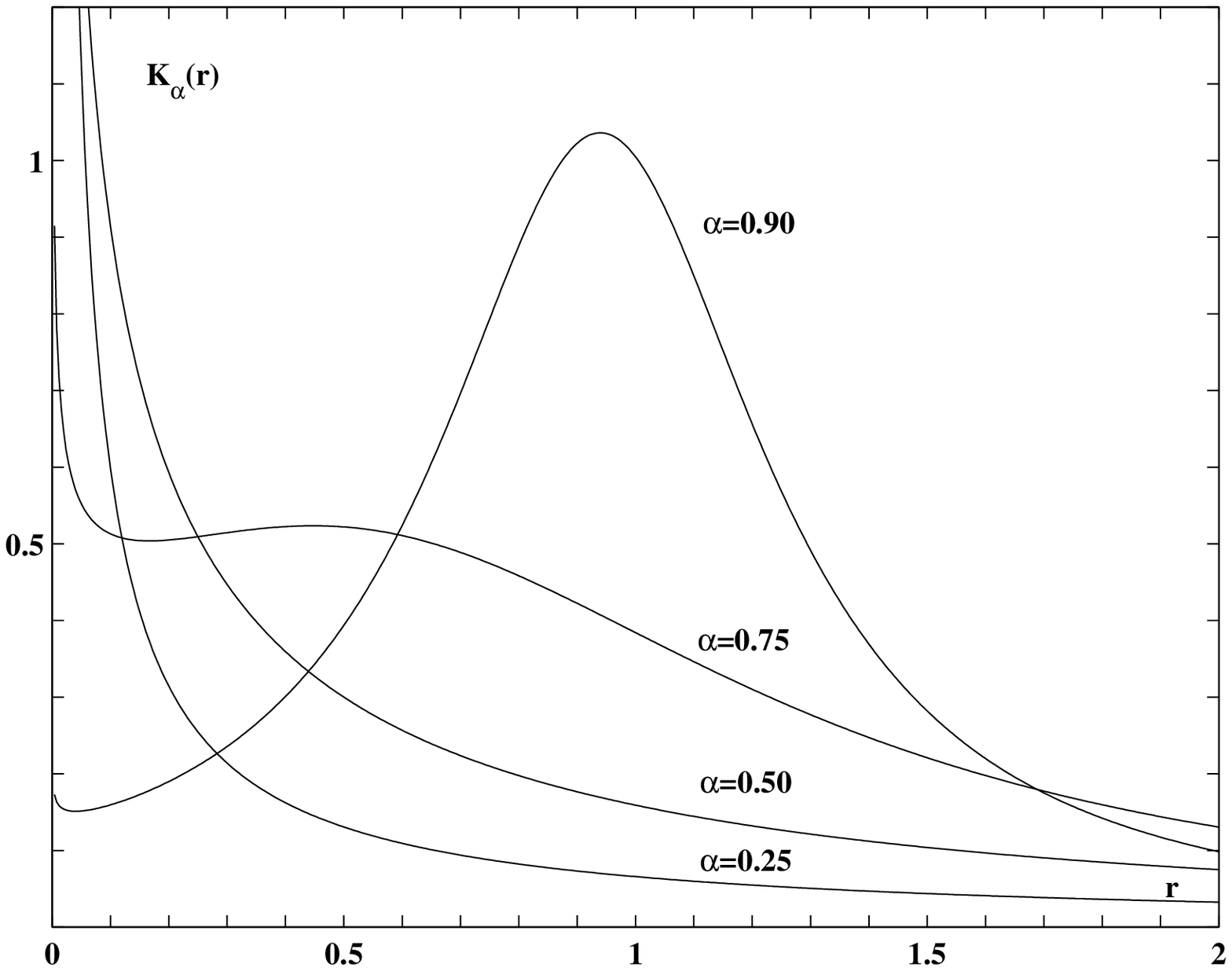}}

 \vskip 0.5 truecm
\centerline{{\bf Fig. 1a} -- Plots of the  {\it basic spectral function}
$K_\alpha(r)$  for  $\,0<\alpha <1$}
  \vskip 0.5 truecm

  \cen{\epsfxsize=7.0truecm \epsfysize=7.0truecm \epsffile{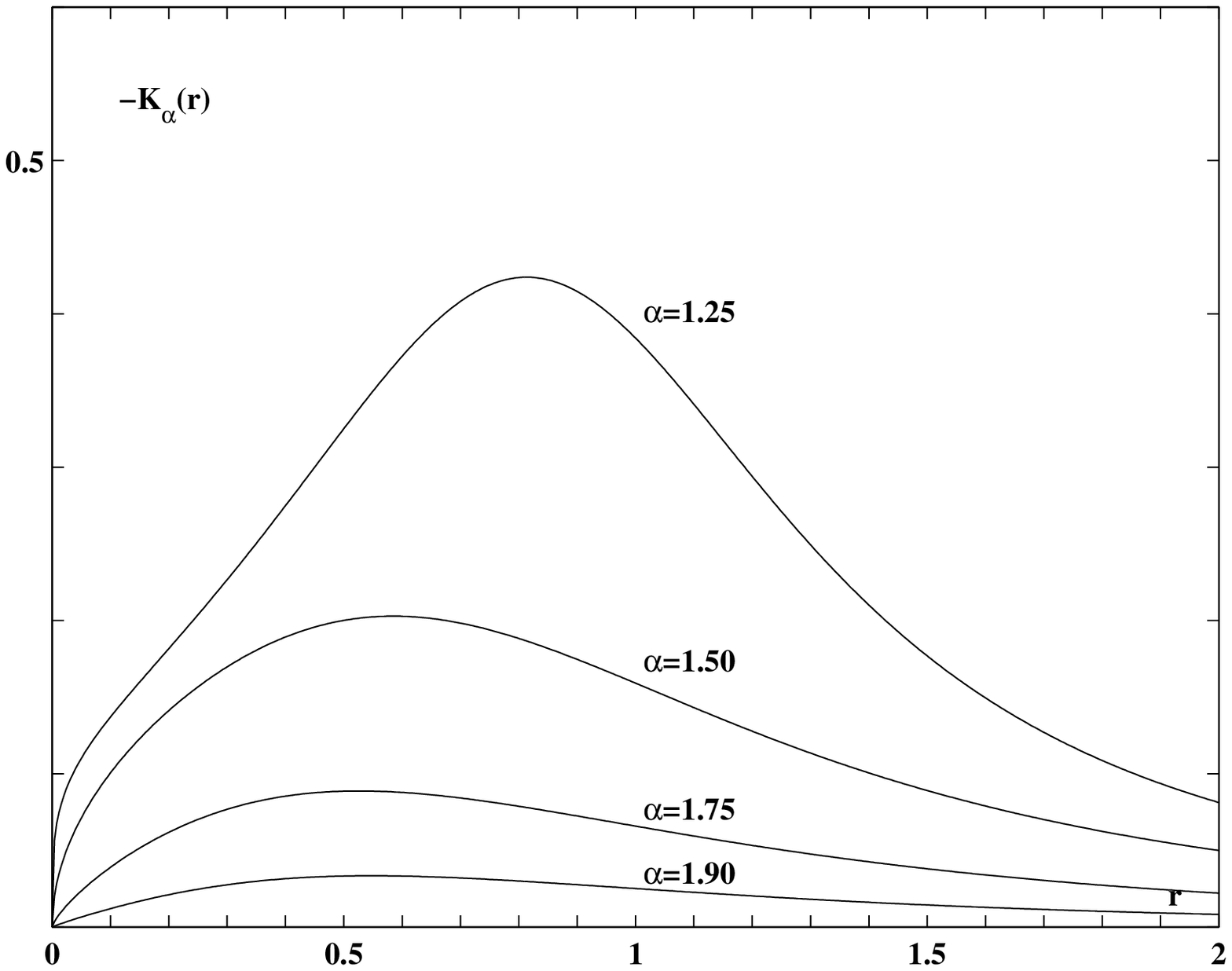}}
  
\vskip 0.5 truecm

\centerline{{\bf Fig. 1b} -- Plots of the  {\it basic spectral function}
$-K_\alpha(r)$ for   $\,1<\alpha <2$}

\vfill\eject
 \vsp
In addition to the   basic fundamental solutions, $u_0(t)= e_\alpha (t)$
we need to compute the impulse-response solutions
$u_\delta (t) = - D^1 \, e_\alpha (t)$ for cases (a) and (b)
and,  only in case (b), the second fundamental solution
$u_1(t) = J^1 \, e_\alpha (t)\,. $
For this purpose we note that in general  it turns out that
$$ J^k \, f_\alpha (t) =
    \int _0^\infty  \!\! \e^{\,\ds -r t} \, K_{\alpha ,k}(r)\, dr \,,
  \eqno(3.26)$$
with
  $$ K_{\alpha,k}(r) := (-1)^k \, r^{-k} \, K_\alpha(r) =
     {	(-1)^k \over\pi}\,
   { r^{\alpha -1-k}\, \sin \,(\alpha \pi)\over
    r^{2\alpha} + 2\, r^{\alpha} \, \cos \, (\alpha \pi) +1}\,,
      \eqno(3.27)$$
where $K_\alpha (r) = K_{\alpha ,0}(r)\,, $  and
$$ J^k g_\alpha (t) =
 {2\over \alpha }\,
     \e^{\ds  t\, \cos \,(\pi /\alpha )}\,
\cos\,\l[t\,\sin\,\l({\pi\over \alpha}\r)  - k {\pi\over \alpha}\r]\,.
    \eqno(3.27)$$
This can be done in direct analogy to the computation of
the functions $e_\alpha (t)$, the Laplace transform of
$J^k e_\alpha (t)$ being given by (3.12). For the impulse-response
solution we note that the effect of the differential operator $D^1$
is the same as that of the virtual operator $J^{-1}\,. $
\vsp
In conclusion we can resume  the solutions  for the
fractional relaxation and oscillation equations as follows:
\vsn  $ {\rm (a)} \q 0<\alpha <1\,,$
$$	  u(t) = c_0 \, u_0(t)
+ \int_0^t q(t-\tau)\, u_{\delta }(\tau)\, d\tau \,,\eqno(3.28a)$$
where
$$  \l\{\eqalign{
u_0(t) &= \int_0^\infty\!\!\e^{\,\ds -rt}\,K_{\alpha,0}(r)\,dr\,,\cr
u_\delta (t) &=
 -\int_0^\infty\!\!\e^{\,\ds -rt}\,K_{\alpha,-1}(r)\,dr\,,\cr} \r.
\eqno(3.29a)$$
  with
$  u_0(0^+) =1\,,\;  u_{\delta }(0^+) =\infty\,;$
\vsn  $ {\rm (b)} \q 1<\alpha <2\,,$
$$	  u(t) = c_0 \, u_0(t) + c_1 \, u_1(t)
    + \int_0^t q(t-\tau)\, u_{\delta}(\tau)\, d\tau\,,
\eqno(3.28b)$$
where
$$ \l\{ \eqalign{
 u_0(t) &= \int_0^\infty\!\! \e^{\,\ds -rt}\,K_{\alpha,0}(r)\, dr+
   {2\over \alpha }\, \e^{\ds  t\, \cos \,(\pi /\alpha )}\,
   \cos\,\l[t\,\sin\,\l({\pi\over\alpha}\r)\r]	 \,,   \cr
   u_1(t) &=  \int_0^\infty\!\! \e^{\,\ds -rt}\,K_{\alpha,1}(r)\,dr +
    {2\over \alpha }\, \e^{\ds	t\, \cos \,(\pi /\alpha )}\,
 \cos\,\l[t\,\sin\,\l({\pi\over\alpha}\r) -{\pi\over \alpha}\r] \,,\cr
   u_{\delta}(t)
  &= -\int_0^\infty\!\!\e^{\,\ds -rt}\,K_{\alpha,-1}(r)\,dr
   -  {2\over \alpha }\, \e^{\ds  t\, \cos \,(\pi /\alpha )}\,
 \cos\,\l[t\,\sin\,\l({\pi\over\alpha}\r) +{\pi\over \alpha}\r]\,,\cr
  }\r.\eqno(3.29b)$$
with
$  u_0(0^+) =1\,, \, u_0^\prime(0^+)=0\,, \,
  u_1(0^+) =0\,, \, u_1^\prime(0^+)=1\,,$
 $u_{\delta }(0^+) =0\,, \, u_{\delta }^\prime(0^+)=+\infty\,.$
\vsp
\vfill\eject
$\null$

 \cen{\epsfxsize=10truecm \epsfysize=8.0truecm \epsffile{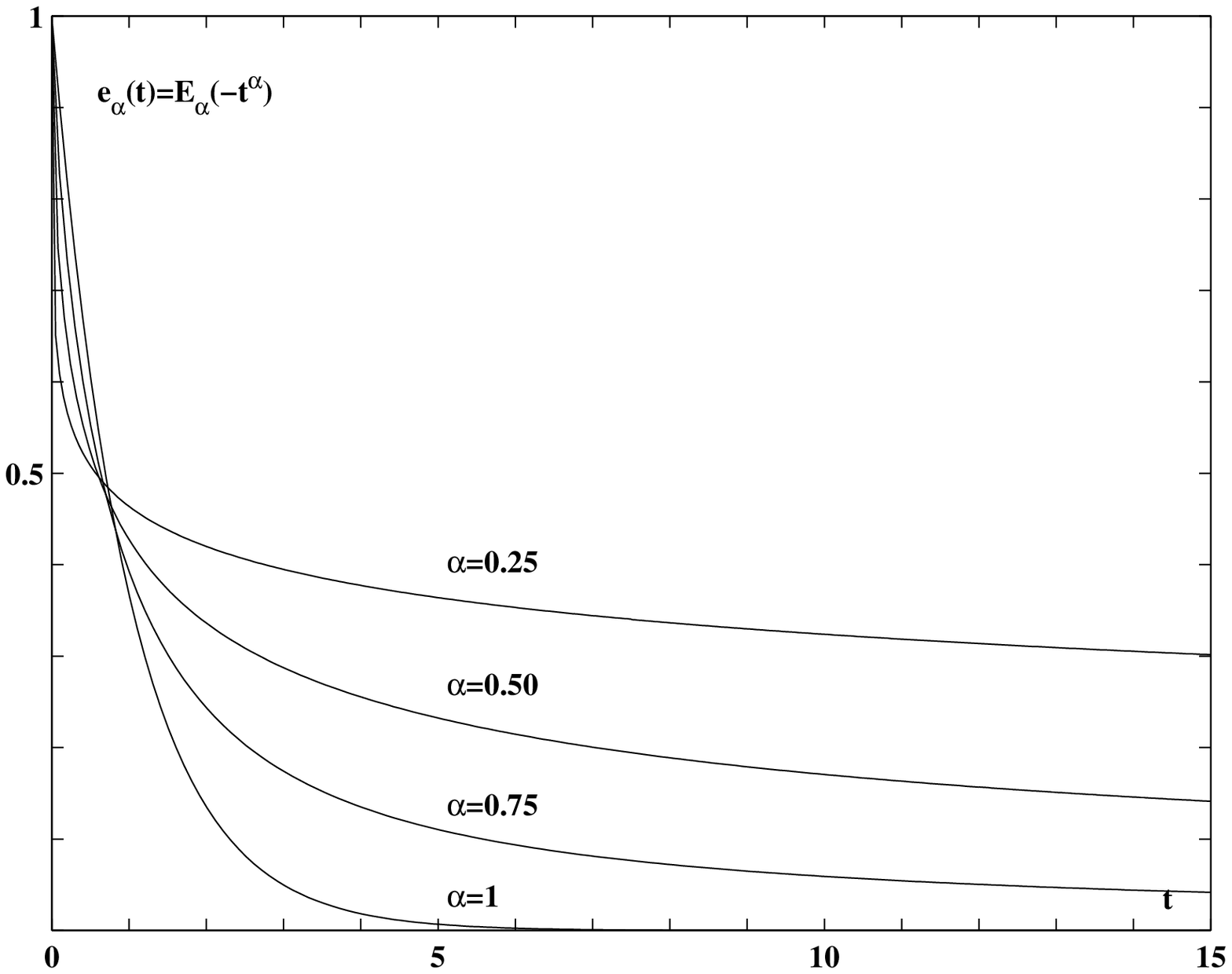}}

 \vskip 0.5 truecm

\centerline{{\bf Fig. 2a} -- Plots of the {\it basic fundamental solution}
 $u_0(t) = e_\alpha (t) $
   for $\, 0<\alpha \le 1$}


 \vskip 0.5 truecm
 \cen{\epsfxsize=10.0truecm \epsfysize=8truecm \epsffile{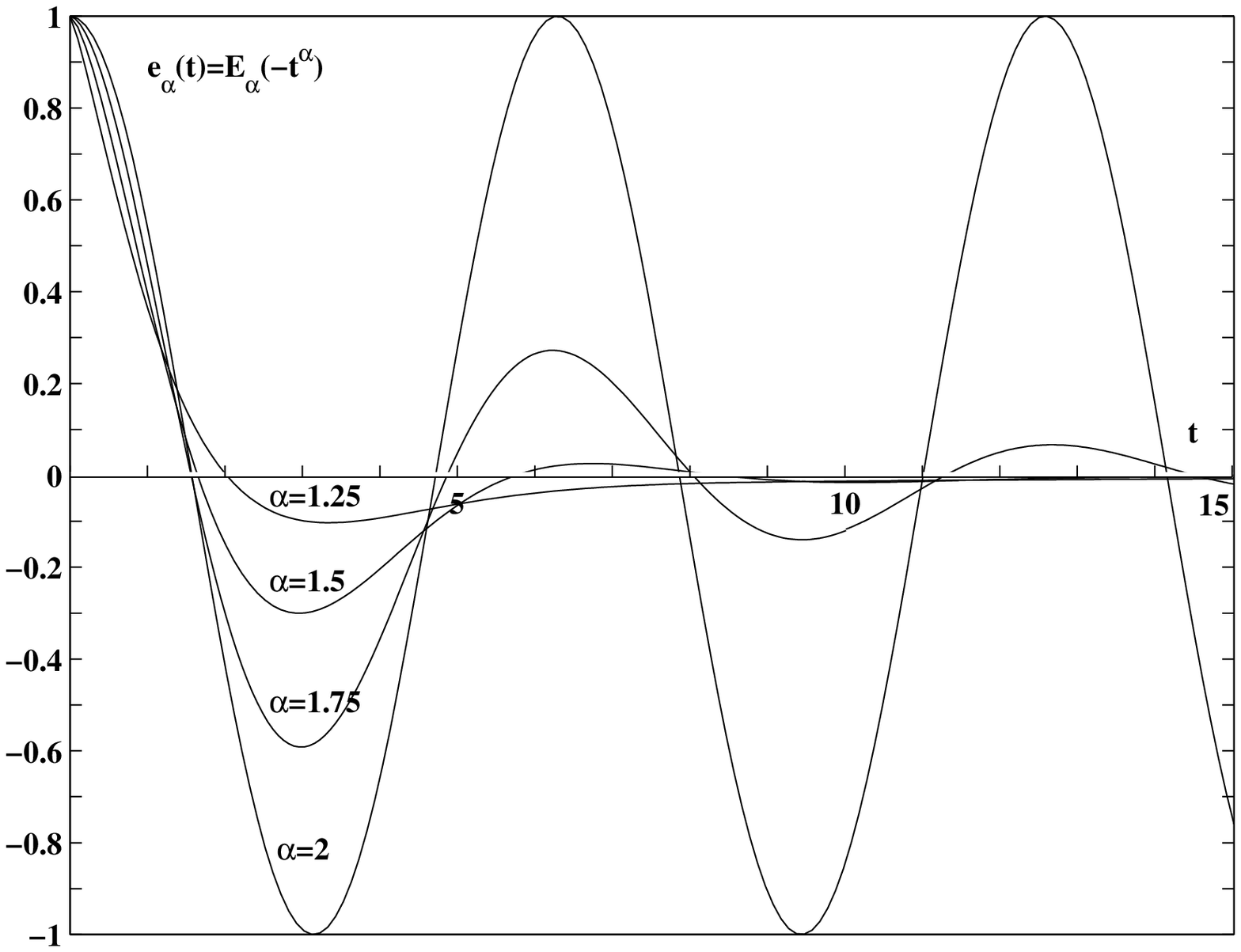}}

 \vskip 0.5 truecm

\centerline{{\bf Fig. 2b} -- Plots of the {\it basic fundamental solution}
 $u_0(t) = e_\alpha (t) $
   for $\, 1<\alpha \le 2$ :}


\vfill\eject
\vsp
In Fig. 2 we quote the plots of the basic fundamental solution	for the
following cases :
(a) $\; \alpha = 0.25\,, \, 0.50\,,\, 0.75\,, \, 1\,,$
and
(b) $\; \alpha = 1.25\,, \, 1.50\,,\, 1.75\,, \, 2\,,$
obtained from the first formula in  (3.29a) and (3.29b), respectively.
We have verified that our present results confirm  those obtained
by Blank [43] by a numerical treatment	and  those obtained
by Mainardi [39]
by an analytical treatment, valid when $\alpha $ is a rational
number, see  \S $A2\,$	of the Appendix.
Of particular interest is the case $\alpha =1/2$
where we recover a well-known formula of the Laplace
transform theory, see (A.34), 
\def\ss{{s}^{1/2}}
$$
   e_{1/2}(t) := E_{1/2} (-\sqrt{t}) =
    {\rm e}^{\ds \, t}\, {\rm erfc} (\sqrt{t})
    \,\div\,	    {1\over \ss\,(\ss +1 )}
\,, \eqno(3.30)
  $$
where $ \, {\rm erfc}\,$ denotes the
{\it complementary error} function.
 \vsp
We now desire to point out that in both the cases
(a) and  (b) (in which $\alpha $  is  just  not integer)
\ie for  {\it fractional relaxation}
and {\it fractional oscillation},
all the fundamental and impulse-response solutions
exhibit an {\it algebraic decay}   as $t \to \infty\,, $
as discussed below.
Let us start with the asymptotic behaviour of $u_0(t)\,. $
To this purpose we first derive an asymptotic series for the
function $f_\alpha (t)$, valid for $t \to \infty\,. $
Using the identity
$$ \rec{s^\alpha +1} = 1 - s ^\alpha  + s ^{2\alpha } - s^{3\alpha }
   + \dots + (-1)^{N-1}\, s^{(N-1)\alpha }
   + (-1)^N \, {s^{N\alpha }\over s^\alpha +1}\,,
$$
  in  formula (3.20) 
and the Hankel representation of the reciprocal
Gamma function, we (formally)  obtain  the   asymptotic
expansion (for $\alpha $ non integer)
$$ f_\alpha (t) = \sum_{n=1}^N (-1)^{n-1}
   \, { t^{-n\alpha } \over \Gamma(1-n\alpha )}
    + O\l(t^{-(N+1)\alpha}\r) \,, \q {\rm as}\q t \to \infty\,.
\eqno(3.31)
$$
The validity of this asymptotic expansion can be established rigorously
using the (generalized)  Watson lemma, see [44].
We also can start from the spectral representation (3.24-25) and
expand the spectral function for small $r\,. $	Then  the
(ordinary) Watson lemma  yields (3.31).
 We note that this asymptotic expansion coincides with that for
$u_0(t)= e_\alpha (t)$, having assumed
$0<\alpha <2$  ($\alpha  \ne 1$).
In fact   the contribution
of $g_\alpha (t)$ is identically zero if $ 0 <\alpha <1$ and
    exponentially small as $t \to \infty$
if $ 1 <\alpha <2 \,.$
\vsp
The asymptotic expansions of the solutions $u_1(t)$ and
$u_\delta (t)$	are obtained from (3.31) integrating   or differentiating
term by term with respect to $t\,. $
In particular,	taking	 the leading term in (3.31),   we obtain
the  asymptotic representations
$$
 u_0(t) \sim  {t^{-\alpha }\over\Gamma(1-\alpha )}\,, \q
 u_1(t) \sim  {t^{1-\alpha }\over\Gamma(2-\alpha )}\,,	\q
 u_\delta (t) \sim  - {t^{-\alpha-1 }\over\Gamma(-\alpha )}\,,
   \q {\rm as}\q t \to \infty\,,
 \eqno(3.32)$$
that point out the algebraic decay of the fundamental and impulse-response
solutions.
\vsp
In Fig. 3 we show  some plots of the {\it basic fundamental solution}
$ u_0(t) = e_\alpha (t)$
for   $\; \alpha = 1.25\,, \, 1.50\,,\, 1.75\,.$
Here the algebraic decay of the fractional oscillation
can be recognized and compared with the two contributions provided by
$f_\alpha $ (monotonic behaviour ) and $g_\alpha (t)$ (exponentially
damped oscillation).
\vfill\eject
$\null$

 \cen{\epsfxsize=7.0truecm \epsfysize=5.0truecm \epsffile{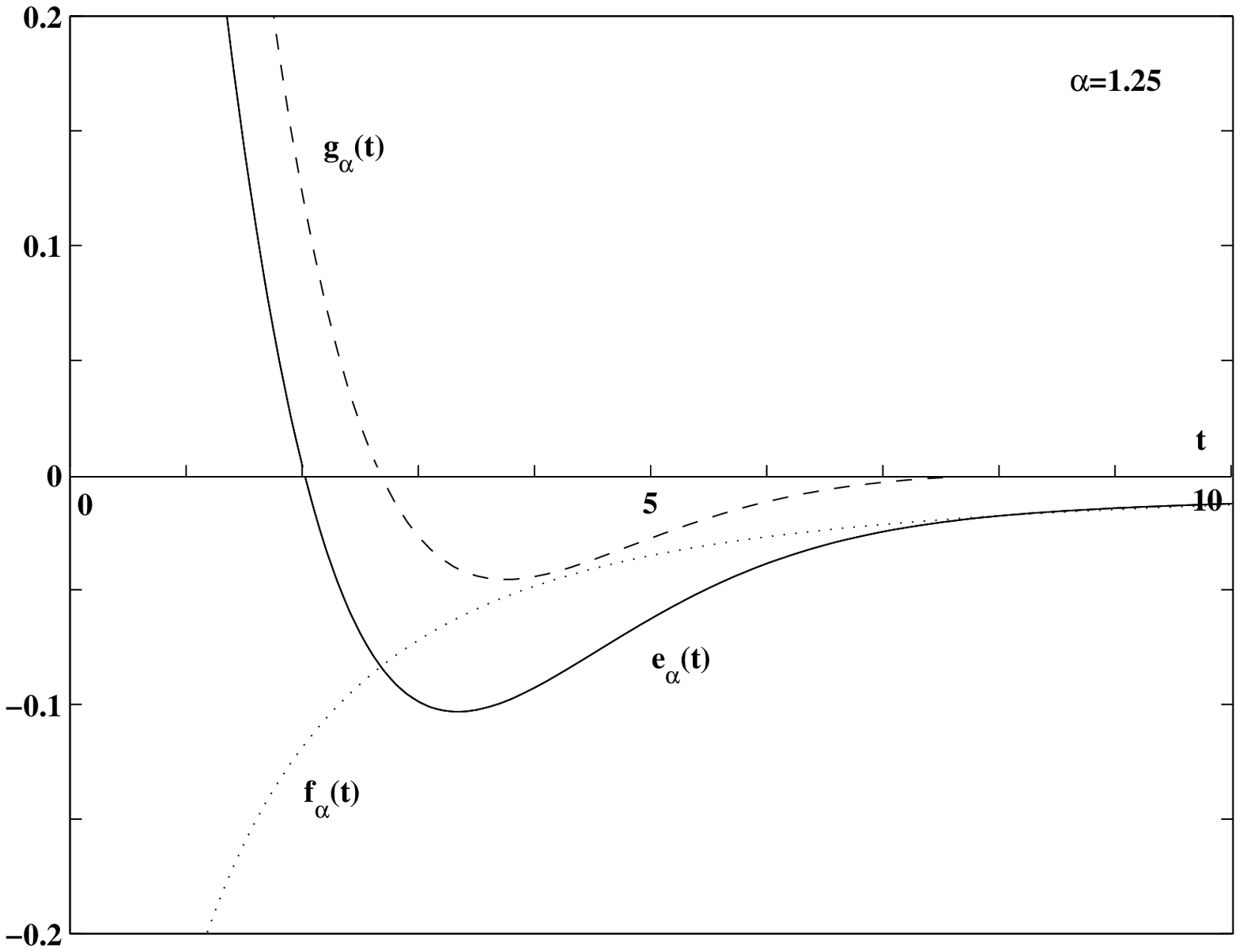}}

 \vskip 0.5 truecm

\centerline{{\bf Fig. 3a} --  Decay of the {\it basic fundamental
solution}  $u_0(t) = e_\alpha (t) $
for $\, \alpha =1.25$}
 \vskip 1.0 truecm

 \cen{\epsfxsize=7.0truecm \epsfysize=5.0truecm \epsffile{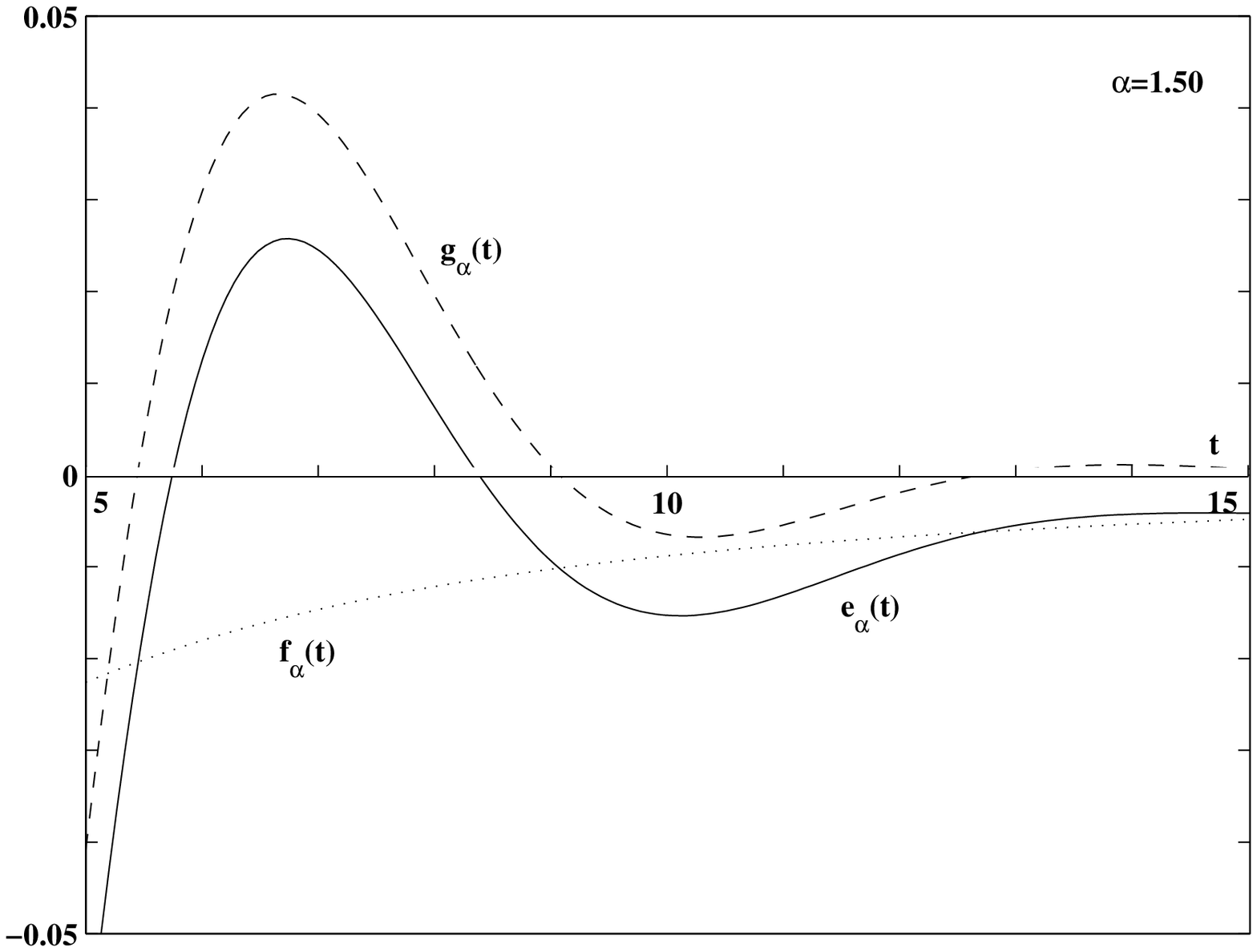}}

 \vskip 0.5 truecm

\centerline{{\bf Fig. 3b} --  Decay of the {\it basic fundamental
solution}  $u_0(t) = e_\alpha (t) $
for $\, \alpha =1.50$}
 \vskip 1.0 truecm

 \cen{\epsfxsize=7.0truecm \epsfysize=5.0truecm \epsffile{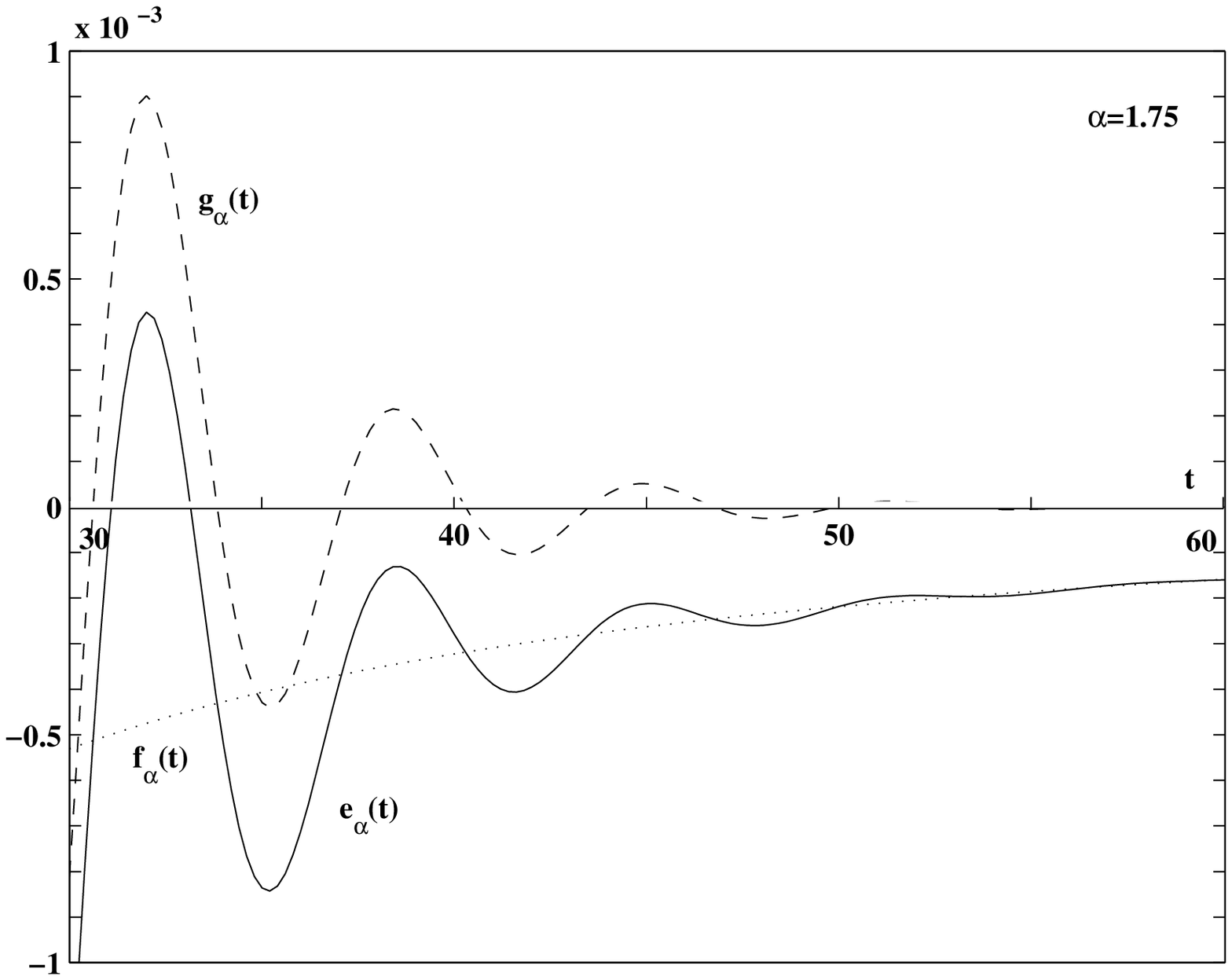}}
 
\vskip 0.5 truecm

\centerline{{\bf Fig. 3c} --  Decay of the {\it basic fundamental
solution}  $u_0(t) = e_\alpha (t) $
for $\, \alpha =1.75$}
\vfill\eject
\line{{\it 3.2 The zeros of the solutions of the fractional
oscillation equation \hfill}}
\vsp
Now we find it interesting to carry out
some investigations
about the zeros of the basic fundamental solution
$u_0(t) = e_\alpha (t)\,$   in the case (b) of fractional oscillations.
For the second fundamental solution
and the  impulse-response solution the analysis of the zeros
can be easily carried out analogously.
\vsp
Recalling the first equation in (3.29b),
the required zeros of $e_\alpha (t)$ are the solutions of
the  equation
  $$ e_\alpha (t) =  f_\alpha (t)  +
{2\over \alpha }\, \e^{\,\ds t\, \cos\, (\pi/\alpha )}\,
     \cos\,\l[t\,\sin\,\l({\pi\over \alpha}\r)\r]
=0\,. \eqno(3.33)$$
\vsp
We first note  that
the function $e_\alpha (t)$  exhibits an {\it odd} number
of zeros, in that  $e_\alpha (0)= 1\,,$
and,
for sufficiently large $t$,   $e_\alpha (t)$
turns out to be permanently negative, as shown in (3.32)
by the sign of $\Gamma(1-\alpha )\,. $
The smallest zero lies in the first positivity
interval of $\cos \, [t \sin \,(\pi/\alpha)]\,, $
hence in the interval $0<t< \pi/[2 \, \sin\,(\pi/\alpha)]\,; $
all other zeros can only lie in the succeeding
positivity intervals   of $\cos\, [t\,\sin\,(\pi/\alpha)]\,, $
in each of these two zeros are present as long as
$$    {2\over \alpha }\, \e^{\,\ds t \, \cos\, (\pi/\alpha )}
     \ge |f_\alpha (t)|\,. \eqno(3.34)$$
When $t$ is sufficiently large the zeros are expected
to be found approximately from the equation
$$ {2\over \alpha }\, \e^{\,\ds t\, \cos\, (\pi/\alpha )}
   \approx {t^{-\alpha}\over |\Gamma(1-\alpha)|}
\,,\eqno(3.35) $$
obtained from (3.33)
by ignoring the oscillation factor of  $g_\alpha (t)$
[see  (3.23)]
and taking the first  term in the asymptotic  expansion
of  $f_\alpha (t)$ [see (3.31-32)].
As we have shown in [40], such 
approximate equation
turns out to be
useful when $\alpha \to 1^+$ and $\alpha \to 2^-\,. $
\vsp
For $\alpha \to 1^+\,, $  only one zero is present,
which is expected to be very far from the origin in view of the
large period of the function   $\cos\, [t\, \sin\,(\pi/\alpha)]\,. $
In fact, since there is no zero for $\alpha =1$, and by
increasing $\alpha $ more and more zeros arise, we are sure
that only one zero exists  for $\alpha $ sufficiently close to 1.
Putting $\alpha = 1+\epsilon $
the asymptotic position $T_{*} $ of this zero
can be found from the relation (3.35) in the limit
$\epsilon \to 0^+\,. $
Assuming in this  limit 
the first-order approximation,	we get
$$ T_{*} \sim  \log \l({2\over \epsilon}\r)\,, \eqno(3.36)$$
which shows that $T_*$ tends to infinity slower than
$1/\epsilon \,, $  as $\epsilon  \to 0\,. $
For details see [40].
\vfill\eject
\noindent
For $\alpha \to 2^-$, there is an increasing number of zeros
up to infinity
since $\,e_2(t) = \cos \, t\,$ has infinitely many zeros
[in $t^*_n = (n+1/2)\pi\,,\, n = 0,1,\dots$].
Putting now $\alpha =2-\delta $ the asymptotic position $T_*$ for
the largest zero
can be found again   from (3.35)
in the limit $\delta  \to 0^+\,. $
Assuming in this  limit 
the first-order approximation,	we get
$$ T_* \sim
{12\over \pi\, \delta }\,  \log \l( {1 \over \delta}\r)\,. \eqno(3.37)
$$
For details see [40].
Now, for $\delta \to 0^+$ the  length of the positivity
intervals of $g_\alpha (t)$ tends to $\pi\, $
and, as long as $t \le T_{*}\,, $ there are two zeros in
each positivity interval. Hence, in the limit $\delta \to 0^+\,,$
there is in average one zero per interval of  length $\pi\,,$
so we expect that
$      N_{*} \sim   T_{*}/\pi \,. $
\vsn
\underbar{Remark 4} : For the above considerations we got inspiration
from an interesting paper by Wiman [45] who
at the beginning of our century,
 after having treated
the Mittag-Leffler function  in the complex plane,
considered the	position of the zeros
of the function  on the negative real axis
(without providing  any detail).
Our expressions  of  $T_* $  are in disagreement with those
by Wiman for  numerical factors; however,
the results of our numerical studies carried out in [40]
confirm and  illustrate the validity of our  analysis.
\vsp
Here, 
we find it interesting to analyse
the phenomenon of the transition
of the	(odd) number of  zeros as
$1.4 \le \alpha \le 1.8\,. $
For this purpose, in Table I we report	the intervals
of amplitude $\Delta \alpha  = 0.01$  where
these transitions occur, and the
location $T_*$ (evaluated  within a relative
error of $0.1\%\,$)  of the  largest zeros found at the
two extreme  values  of the above intervals.
We recognize that the transition from 1 to 3 zeros occurs as
$ 1.40 \le \alpha \le 1.41$,   that one from  3 to 5 zeros occurs
as $1.56 \le \alpha \le 1.57$, and so on. The last transition
in the considered range of   $\alpha\,$  is
from 15 to 17 zeros, and it just occurs
as $1.79 \le \alpha \le  1.80\,. $
$$  \vcenter{
\vbox{
\offinterlineskip
\halign{
&
\vrule#
&
\strut
\quad
\hfil
#
\quad
\cr
\noalign{\hrule}
      height 2 pt & \omit && \omit && \omit & \cr
 & $N_*\;\;$ && $\alpha\quad \quad$ && $T_*\quad\quad$ & \cr
      height 2 pt & \omit && \omit && \omit & \cr
\noalign{\hrule}
   height 2 pt & \omit && \omit && \omit & \cr
 & $ 1\div 3$  && $ 1.40 \div 1.41 $ &&  $ 1.730\div 5.726 $  & \cr
      height 2 pt & \omit && \omit && \omit & \cr
\noalign{\hrule}
     & $ 3\div 5$
  && $ 1.56\div 1.57 $	 &&  $ 8.366\div 13.48 $  & \cr
      height 2 pt & \omit && \omit && \omit & \cr
\noalign{\hrule}
   height 2 pt & \omit && \omit && \omit & \cr
 & $ 5\div 7$  && $ 1.64 \div 1.65 $   && $ 14.61\div 20.00 $  & \cr
      height 2 pt & \omit && \omit && \omit & \cr
\noalign{\hrule}
   height 2 pt & \omit && \omit && \omit & \cr
 & $ 7\div 9$  && $ 1.69 \div 1.70 $   &&  $ 20.80\div 26.33 $	& \cr
      height 2 pt & \omit && \omit && \omit & \cr
\noalign{\hrule}
   height 2 pt & \omit && \omit && \omit & \cr
 & $ 9\div 11$	&& $ 1.72 \div 1.73 $	&&  $ 27.03\div 32.83 $  & \cr
      height 2 pt & \omit && \omit && \omit & \cr
\noalign{\hrule}
   height 2 pt & \omit && \omit && \omit & \cr
 & $ 11\div 13$  && $ 1.75 \div 1.76 $	 &&  $ 33.11\div 38.81 $  & \cr
      height 2 pt & \omit && \omit && \omit & \cr
\noalign{\hrule}
   height 2 pt & \omit && \omit && \omit & \cr
 & $ 13\div 15$  && $ 1.78 \div 1.79 $	 &&  $ 39.49\div 45.51 $  & \cr
      height 2 pt & \omit && \omit && \omit & \cr
\noalign{\hrule}
   height 2 pt & \omit && \omit && \omit & \cr
 & $ 15\div 17$  && $ 1.79 \div 1.80 $	 &&  $ 45.51\div 51.46 $  & \cr
      height 2 pt & \omit && \omit && \omit & \cr
\noalign{\hrule}
\noalign{\hrule}
\noalign{\medskip}
\multispan 7 \hfil {\bf Table I} \hfil	\cr
  }
  }
  }
$$
\centerline{{$N_*$ = number of zeros, $\, \alpha$ = fractional order,
 $\, T_*$ location of the largest zero.}}
\vfill\eject

\pageno=253 
\line{4. FRACTIONAL DIFFERENTIAL EQUATIONS: 2-nd PART \hfill} \vsp
In this section we shall consider the following
fractional differential equations for $t\ge 0\,, $
equipped with the
necessary initial conditions,
$$  {du\over dt} + a\, {d^\alpha u\over dt^\alpha}  + u(t) = q(t) \,,
  \q u(0^+) = c_0 \,, \q 0<\alpha <1\,,   \eqno(4.1)$$
$$  {d^2 v\over dt^2} + a\, {d^\alpha v\over dt^\alpha} + v(t) = q(t) \,,
 \q v(0^+) = c_0 \,, \q v^\prime (0^+) = c_1\,, \q 0<\alpha <2\,,
\eqno(4.2)$$
where $a$ is a positive constant.
The unknown  functions $u(t)$ and $v(t)$  (the field variables)
are required to  be
sufficiently well behaved
to be treated with their derivatives $u^\prime(t) $ and
$v^\prime(t)\,, \, v^{\prime\prime}(t)$
by the technique
of Laplace transform.  The given function $q(t)$ is supposed
to be continuous.
In the above equations the fractional derivative of order
$\alpha $ is assumed to be provided  by the operator
$D_*^\alpha $, the {\it Caputo derivative}, see (1.17),
in agreement with our choice followed in the previous section.
Note that in (4.2) we must distinguish the cases
$\,{\rm (a)}\;\,  0<\alpha <1\,,$  $\,{\rm (b)}\;\,  1<\alpha <2\,$
and $\alpha =1\,. $
\vsp
The equations (4.1) and(4.2)
will be referred to as the
{\it composite fractional relaxation equation}
 and the {\it composite fractional oscillation equation}, respectively,
 to be distinguished from the corresponding {\it simple} fractional
equations treated in \S 3. 
\vsp
The fractional differential equation in (4.1) with  $\alpha = 1/2$
corresponds to the 
{\it Basset problem},
a classical  problem in fluid dynamics
concerning the unsteady motion of a particle accelerating in a viscous
fluid under the action of the gravity, see  [24]. 
\vsp
The  fractional differential equation in (4.2) with $0 < \alpha  < 2 $
models an oscillation process with fractional damping term.
It was formerly treated by Caputo [19], who provided a
preliminary analysis  by  the Laplace transform.
The special cases
$\alpha  = 1/2$  and  $\alpha  = 3/2\,, $
but with  the standard
 definition $D^\alpha $ for the fractional derivative,
have been discussed  by Bagley [30].
Recently,  Beyer and Kempfle [46] discussed (4.2)
for $-\infty <t<+\infty$ to investigate the uniqueness and {\it causality}
of the solutions. As they let $t$ running in all of $\RR\,, $ they
used Fourier transforms and characterized the fractional derivative
$D^\alpha $ by its properties in frequency space, thereby requiring that
for non-integer $\alpha $ the principal branch of $(i\omega )^\alpha $
should be taken. Under the global condition that the
solution is square summable, they showed that the system described by (4.2)
is {\it causal} {\it iff} $a > 0\,. $
\vsp
Also here we shall apply the method of Laplace transform
to solve  the  fractional differential equations  and get some insight
into their {\it fundamental} and {\it impulse-response solutions}.
However, in contrast with the previous section,
we now find it more convenient to
apply directly the formula (1.30) for the Laplace transform of
fractional and integer derivatives,
than reduce the equations with the prescribed initial
conditions as  equivalent (fractional) integral  equations
to be treated by the Laplace transform.
\vfill\eject
\line{{\it 4.1 The composite fractional relaxation equation \hfill}}
\vsp
Let us apply the Laplace transform to the  fractional relaxation equation
(4.1).
Using the rule (1.30) we are led to the transformed
algebraic equation
 $$\u(s) = c_0 { 1 + a\, s^{\alpha -1}\over w_1(s)} + {\ql(s)\over w_1(s)}
\,,\q 0<\alpha <1\,,  \eqno(4.3)$$
where
$$    w_1(s) := s+ a\, s^\alpha +1\,, \eqno(4.4)$$
and $a>0\,.$
Putting
$$ u_0(t) \div \u_0(s) :=  { 1 +  a\, s^{\alpha -1}\over w_1(s)} \,,
   \q u_\delta(t) \div \u_\delta  (s) :=  { 1 \over w_1(s)} \,,
 \eqno(4.5)$$
and recognizing that
$$ u_0(0^+) = \lim_{s\to \infty} s\, \u_0(s) =1\,,\q
  \u_\delta (s)  = - \, \l[s\, \u_0(s) - 1\r] \,,\eqno(4.6)
$$
we can conclude that
$$     u(t) = c_0 \, u_0(t) + \int _0^t \!\!q(t-\tau )\,u_\delta (\tau )
       \,d\tau\,, \q u_\delta(t) = - \, u_0^\prime(t)
 \,. \eqno(4.7)$$
We thus recognize that $u_0(t)$ and
$ u_\delta (t)	$
are  the {\it fundamental solution} and
{\it impulse-response solution} for the equation (4.1),
respectively.
\vsp
Let us first consider the problem to get $u_0(t)$ as
the inverse Laplace transform of $\u_0(s)\,. $
We easily  see that the function  $w_{1}(s)$ has no zero in the main
sheet of the Riemann surface including its boundaries on the cut (simply
show that ${\rm Im}\,\l\{ w_{1}(s)\r\}$ does not vanish
if $s$ is not a real positive number), so that the inversion of
the Laplace transform $\u_0(s)$ can be carried
out by deforming the original Bromwich path into
 the Hankel path $Ha(\epsilon )$ introduced in the previous
section, \ie
into the loop constituted
by a   small circle $ |s| = \epsilon $ with $\epsilon \to 0$
and by the two borders of the cut negative real axis.
As a consequence we write
$$  u_0 (t) =
   \rec{2\pi i}\,
   \int_{Ha(\epsilon)} \!\! \e^{\ds \,st} \, { 1+ as^{\alpha -1} \over
 s+ a\,s^\alpha +1}\, ds   \,.\eqno(4.8)$$
It is now an exercise  in complex analysis to show that the contribution
from the Hankel path $Ha(\epsilon)$ as $\epsilon \to 0$ is provided by
$$ u_0 (t) = \int_0^\infty \! \e^{\,\ds -rt}\,
     H^{(1)}_{\alpha,0} (r;a)\, dr\,, \eqno(4.9)$$
with
$$ \eqalign{
H^{(1)}_{\alpha,0} (r;a) &=  - \,\rec{\pi}\,	{\rm Im}\,
  \l\{ \l.{1+ as^{\alpha -1} \over w_1(s)}\r\vert_{s=r\, \e^{i\pi}} \r\}\cr
   &= \rec{\pi}\,
   { a\,r^{\alpha -1}\, \sin \,(\alpha \pi)\over
 (1-r)^2 + a^2\, r^{2\alpha} + 2\, (1-r)\,a\,r^{\alpha} \, \cos \,
(\alpha \pi)}\,.\cr}	   \eqno(4.10)$$
For $a>0$ and $0<\alpha <1$ the function $H^{(1)}_{\alpha,0} (r;a)$
is positive for all $r>0$
since it has the sign of the numerator;
in fact in (4.10) the denominator  is strictly
positive being	equal to $| w_1(s)|^2 $ as $s = r\, \e^{\pm i\pi}\,. $
Hence,	the {\it fundamental solution} $u_0 (t)$ has the peculiar property
to be  {\it completely monotone}, and
 $H^{(1)}_{\alpha ,0}(r;a)$ is its
{\it spectral function}.
\vsp
Now the  determination of $u_\delta (t) = -u_0^\prime(t)$ is
straightforward. We see that
also the {\it impulse-response solution} $u_\delta (t)$
is  {\it completely monotone} since it can be represented by
$$ u_\delta  (t) = \int_0^\infty \! \e^{\,\ds -rt}\,
     H^{(1)}_{\alpha,-1} (r;a)\, dr\,, \eqno(4.11)$$
with  {\it spectral function}
$$ H^{(1)}_{\alpha,-1 } (r;a) =  r \, H^{(1)}_{\alpha,0} (r;a) =
 \rec{\pi}\,
   { a\,r^{\alpha}\, \sin \,(\alpha \pi)\over
 (1-r)^2 + a^2\, r^{2\alpha} + 2\, (1-r)\,a\,r^{\alpha} \, \cos \,
(\alpha \pi)}\,.
   \eqno(4.12)$$
\vsp
We recognize that both the solutions $u_0(t)$ and $u_\delta (t)$ turn out
to be strictly decreasing from $1$ towards $0$ as
 $ t$ runs from $0$ to $\infty\,. $  Their behaviour as $t\to 0^+$ and
$t \to \infty$ can be inspected by means of a proper asymptotic analysis.
\vsp
The behaviour of the solutions as  $t\to 0^+$	can be determined
from the behaviour of their Laplace transforms as Re$\,\{s\} \to +\infty$
as well known from the theory of the Laplace transform, see \eg [25].
We obtain as  Re$\, \{s\} \to +\infty\,,$
$$ \u_0(s) = s^{-1} - s^{-2} + O\l(s^{-3+\alpha}\r) \,,
\q
  \u_\delta (s) = s^{-1} - a\,s^{-(2-\alpha)} + O\l(s^{-2}\r) \,,
\eqno(4.13)
  $$
so that
$$     u_0(t) = 1 - t + O\l(t^{2-\alpha}\r)\,,
      \q u_\delta (t) = 1 - a\,{t^{1-\alpha }\over \Gamma(2-\alpha )} +
O\l(t\r)\,, \q {\rm as}\q  t \to 0^+ \,.
 \eqno(4.14)$$
\vsp
The spectral representations (4.9) and (4.11) are suitable to obtain the
asymptotic behaviour of $u_0(t)$ and  $u_\delta (t)$
as $ t \to +\infty\,, $ by using the Watson lemma.
In fact, expanding the spectral functions for small $r$ and
taking the dominant
term in the corresponding asymptotic series,  we obtain
$$ u_0(t) \sim
     a\,  {t^{-\alpha} \over \Gamma(1-\alpha)} \,,\q
u_\delta (t) \sim
  - a\,  {t^{-\alpha-1} \over \Gamma(-\alpha)} \,,\q
{\rm as}\q t \to \infty\,.
 \eqno(4.15)$$
\vsp
We note that the limiting case $\alpha =1$ can be easily treated
extending the validity of eqs (4.3-7) to $\alpha =1\,, $
as it is legitimate.
In this case we obtain
$$ u_0(t) = \e^{\,\ds - t/(1+a)}\,, \q u_\delta (t) =
  \rec{1+a}\,	\e^{\,\ds - t/(1+a)}\,, \q \alpha =1\,. \eqno(4.16)$$
Of course, in the case $a\equiv 0$ we recover the standard solutions
$ u_0(t) = u_\delta (t) = \e^{\, -t}\,. $
\vfill\eject
We conclude this sub-section with some considerations on the
solutions when	the order $\alpha $ is just a rational
number. If we take $\alpha =p/q\,, $ where $p,q \in \NN$
are assumed (for convenience) to be relatively prime,
a factorization in (4.4)   is possible by using the
procedure indicated  by Miller and Ross [10]. In these cases
the solutions can be expressed in terms of a linear combination
of $q$ Mittag-Leffler functions of fractional order $1/q$, which,
on their turn  can be expressed
in terms of  incomplete gamma functions, see (A.14) of the Appendix.
\vsp
Here we shall illustrate   the factorization in the
simplest case $\alpha =1/2$ and provide the solutions $u_0(t)$
and $u_\delta (t)$ in terms of the functions
$e_\alpha (t;\lambda)$ (with $\alpha =1/2$), introduced in
the previous section.
In this case, in view of the application to the
{\it Basset problem}, see [24], the equation (4.1) deserves a particular
attention. 
For $\alpha =1/2$ we can write
$$ w_1(s) = s +a\,s^{1/2} +1
= (s^{1/2} -\lambda _+)\,(s^{1/2}-\lambda _-)\,,\q
  \lambda _{\pm} = -a/2 \pm (a^2/4-1)^{1/2} \,.
\eqno(4.17) $$
Here $\lambda _{\pm}$ denote the two roots (real or conjugate complex)
of the second degree polynomial with positive coefficients
$z^2 +az+1\,,$ which, in particular, satisfy the following binary
relations
$$ \lambda_+  \cdot \lambda_- =1\,, \q
  \lambda_+ +\lambda_- = -a\,,\q
  \lambda_+ -\lambda_- = 2(a^2/4-1)^{1/2}= (a^2-4)^{1/2}\,.
 \eqno(4.18)$$
\vsp
We recognize that
we must treat separately the following two cases
$$i)\q 0<a<2\,, \q{\rm or}\q  a>2\,, \q{\rm and}\q  ii)\q a=2\,, $$
which correspond to two distinct roots ($\lambda _+ \ne \lambda _-$),
or two coincident roots ($\lambda _+ \equiv \lambda _-= -1$),
respectively.
\def\ss{{s}^{1/2}}   
For this purpose,  using the notation introduced in [24],
we write
$$ \widetilde M(s) :=
 {1+a\, s^{-1/2}\over s+ a \,\ss +1}= \l\{
\eqalign{
i)\,&\, {A_-\over \ss\,(\ss -\lambda_+ )} +
    {A_+\over\ss (\ss -\lambda_- )} \,, \cr
ii)\,&\,{1 \over (\ss +1)^2}+{2 \over \ss (\ss +1)^2} \,, \cr }
   \r. \eqno(4.19)$$
and
$$ \widetilde N(s) :=
 {1\over s+ a  \,\ss +1}= \l\{
\eqalign{
i)\,&\, {A_+\over \ss\,(\ss -\lambda_+ )} +
    {A_-\over\ss (\ss -\lambda_- )} \,, \cr
ii)\,&\, {1 \over (\ss +1)^2}\,, \cr }
   \r. \eqno(4.20)$$
where
$$ A_{\pm} =
{\pm}\, { \lambda_\pm  \over \lambda_+	- \lambda_- }\,. \eqno(4.21)$$
Using (4.18) we note that
$$ A_+ + A_- =1\,, \q
   A_+\,\lambda_-  + A_-\, \lambda_+ =0\,, \q
   A_+\,\lambda_+  + A_-\, \lambda_- =-\,a\,.
\eqno(4.22)$$
\vfill\eject
\vsp
Recalling the Laplace transform pairs (A.34), (A.36) and (A.37)
in Appendix, we obtain
 $$u_0(t)= M(t) := \l\{ \eqalign{
i)\,&\;
A_-\,E_{1/2}
\,(\lambda_+\,\sqrt{t})+A_+\,E_{1/2}\,(\lambda_-\,\sqrt{t})\,,\cr
ii)\,&\;
    (1-2t)\, E_{1/2}\,(-\sqrt{t})  + 2\, \sqrt{t/\pi}\,, \cr}
\r.
   \eqno(4.23)$$
and
$$  u_\delta (t) = N(t) :=
  \l\{
\eqalign{
i)\,&\,   A_+\,  E_{1/2} (\lambda _+\,\sqrt{t})
   + A_-\,  E_{1/2}(\lambda _-\,\sqrt{t})
     \,, \cr
ii)\,&\,  (1 +2t)\,   E_{1/2} (-\sqrt{t})
       - 2\,\sqrt{t/\pi} \,.	    \cr }
   \r. \eqno(4.24)$$
We thus recognize in (4.23-24) the presence of the functions
$e_{1/2}(t;-\lambda _\pm) = E_{1/2}  (\lambda _\pm\,\sqrt{t})$
and $e_{1/2}(t) =  e_{1/2}(t;1) = E_{1/2}  (-\sqrt{t})\,. $
\vsp
In particular, the solution of the
{\it Basset problem} can be easily obtained from (4.7) with
$q(t) = q_0$ by using (4.23-24) and noting that
$\int_0^t N(\tau)\, d\tau  = 1- M(t)\,. $
Denoting this solution by $u_B(t)$ we get
   $$u_B(t) = q_0 - (q_0-c_0) \, M(t) \,.\eqno(4.25) $$
When $a \equiv 0$, \ie in the absence of term containing the fractional
derivative (due to  the Basset force),
we recover the classical Stokes solution, that we denote by
$u_S(t)\,, $
$$   u_S(t) = q_0 - (q_0-c_0) \, \e^{\ds \, -t} \,. $$
In the particular case $q_0 = c_0\,, $
we get	the steady-state solution
$u_B(t) = u_S(t) \equiv q_0\,. $
\vsp
For vanishing initial condition $c_0 =0\,, $ we have the creep-like
solutions
$$ u_B(t) = q_0\, \l[1 - M(t)\r]\,, \q
    u_S(t) = q_0\, \l[1 - \e^{\ds \, -t}\r]\,, $$
that we compare in the normalized plots of Fig. 5 of [24].
In this case
it is instructive to  compare the  behaviours of
the two   solutions as $t\to 0^+$ and $t\to \infty\,. $
Recalling the general asymptotic expressions of
$u_0(t) = M(t)$
in (4.14) and (4.15) with $\alpha =1/2\,, $
 we recognize that
$$u_B(t) = q_0\, \l[t + O\l(t^{3/2}\r)\r] \,,  \q
    u_S(t) = q_0\,\l[t + O\l(t^{2}\r)\r] \,, \q {\rm as} \q t \to 0^+\,, $$
and
$$
u_B(t) \sim q_0\, \l[  1- {a/ \sqrt{\pi\, t}}\,\r] \,,	\q
  u_S(t) \sim q_0\, \l[ 1- EST\,\r] \,, \q {\rm as} \q t \to \infty\,, $$
where $EST$ denotes {\it exponentially small terms}.
In particular we note  that the normalized plot of $u_B(t)/q_0$
remains under that  of $u_S(t)/q_0$ as $t$ runs from $0$ to
$\infty\,. $
\vsp
The reader is invited to convince himself of the following fact. In
the general case
$0 < \alpha  < 1\,$   the solution $u(t)$ has the particular property of
being  equal to 1 for all $t \ge 0$ if $q(t)$ has this property and
$ u(0^+) = 1\,,$
whereas $q(t) = 1$ for all $t \ge 0$ and $u(0^+) = 0$
implies that $u(t)$ is
a creep function tending to $1$ as  $t \to \infty\,. $
\vfill\eject
\line{{\it 4.2 The composite fractional oscillation equation \hfill}}
\vsp
Let us now apply the Laplace transform to the  fractional oscillation
equation (4.2).
Using the rule (1.30) we are led to the transformed
algebraic equations
 $${\rm (a)}\q \v(s) = c_0 { s + a\, s^{\alpha -1}\over w_2(s)} +
  c_1 {1\over w_2(s)} +
{\ql(s)\over w_2(s)}
\,, \q	0<\alpha < 1\,, \eqno(4.26a)$$
or
$${\rm (b)} \q \v(s) = c_0 { s + a\, s^{\alpha -1}\over w_2(s)} +
  c_1 {1+ a\, s^{\alpha -2} \over w_2(s)} +
{\ql(s)\over w_2(s)}
\,, \q 1<\alpha < 2\,, \eqno(4.26b)$$
where
$$    w_2(s) := s^2+ a\, s^\alpha +1\,, \eqno(4.27)$$
and $a>0\,. $
Putting
$$ \v_0(s) :=  { s +  a\, s^{\alpha -1}\over w_2(s)} \,,
 \q 0<\alpha <2\,,   \eqno(4.28)$$
we recognize that
$$ v_0(0^+) = \lim_{s\to \infty} s\, \v_0(s) =1\,,\q
   { 1 \over w_2(s)}  = - \, \l[s\, \v_0(s) - 1\r] \,
 \div \, - v_0^\prime (t) \,,
\eqno (4.29)$$
and
$$  {1 +  a\, s^{\alpha -2}\over w_2(s)} = {\v_0(s)\over s} \,\div \,
   \int_0^t\! v_0(\tau )\, d\tau   \,. \eqno(4.30)$$
Thus we can conclude that
$$ {\rm (a)}\q	  v(t) = c_0 \, v_0(t) - c_1\, v_0^\prime(t)
 -\int _0^t \!\!q(t-\tau )\,
      v_0^\prime(\tau ) \,d\tau \,,
    \q 0<\alpha < 1 \,,
 \eqno(4.31a)$$
or
$$ {\rm (b)}\q v(t) = c_0 \, v_0(t) + c_1\,\int_0^t\! v_0(\tau )\, d\tau
   -\int _0^t \!\!q(t-\tau )\,
      v_0^\prime(\tau ) \,d\tau \,,
    \q	1<\alpha < 2 \,.
 \eqno(4.31b)$$
In both of the above equations	the term
 $\, -v_0^\prime (t)\,$
represents the
{\it impulse-response solution}  $\,v_\delta (t)\,$
for the {\it composite fractional oscillation	equation} (4.2),
namely the particular solution of the inhomogeneous
equation with $ c_0 = c_1 = 0 $
and with  $q(t) = \delta (t)\,. $
For the {\it fundamental solutions} of (4.2) we recognize
from eqs (4.31)
that we have two distinct couples of solutions according
to the case (a) and (b) which read
\def\a{{\rm a}}   \def\b{{\rm b}}
$$
{\rm (a)}\q \{ v_0(t)\,,\, v_{1\,\a}(t) = - v_0^\prime(t)\}\,,
    \qq
{\rm (b)}\q \{ v_0(t)\,,\, v_{1\,\b}(t) = \int_0^t v_0(\tau)\, d\tau\}\,.
 \eqno(4.32)$$
\vsp
We first consider the particular
case $\alpha =1$ for which
the fundamental and impulse response
solutions
are known in terms of elementary functions.
This limiting case   can also be
treated by extending the validity of eqs (4.26a) and (4.31a)
to $\alpha =1\,, $
as it is legitimate.
From
$$ \v_0(s) = {s + a\over s^2 +a\,s +1} =
     {s +a/2\over (s+a/2)^2 + (1-a^2/4)} -
      {a/2\over (s+a/2)^2 + (1-a^2/4)} \,, \eqno(4.33)$$
we obtain  the {\it basic fundamental solution}
$$ v_0(t) = \l\{\eqalign{\,&
    \e^{\ds -at/2}\,\l[
 \cos (\omega t) + {a\over 2\omega }\, \sin (\omega t) \r]
\q {\rm if}\q 0<a<2\,, \cr
   \,& \e^{\ds -t}\, (1-t)  \qq{\rm if}\q a=2\,, \cr
\,&  \e^{\ds -at/2}\, \l[\cosh (\chi t)+{a\over 2\chi }\,\sinh (\chi t)\r]
\q {\rm if}\q a>2\,, \cr}\r.
  \eqno(4.34)$$
where
$$ \omega  = \sqrt{1-a^2/4}\,, \q \chi =\sqrt{a^2/4 -1}\,. \eqno(4.35)$$
By a  differentiation of (4.34) we easily obtain the {\it second
fundamental solution} $v_{1\,a}(t)$ and the {\it impulse-response solution}
$v_\delta (t)$
since $v_{1\,\a}(t)= v_\delta (t)= - v_0^\prime(t)\,. $
We point out that all the solutions exhibit an
{\it exponential decay}  as $t\to \infty\,. $
\vsp
Let us now consider the problem to get $v_0(t)$ as
the inverse Laplace transform of $\v_0(s)\,, $ as
given by (4.26-27),
$$	v_0(t)	=
  \rec{2\pi i}\,
   \int_{Br} \!\! \e^{\ds \,st} \, { s+ a\,s^{\alpha -1} \over w_2(s)}
       \, ds   \,,\eqno(4.36)$$
where $Br$ denotes the usual Bromwich path.
Using a result by Beyer and Kempfle [46]
we know  that  the function $w_2(s)$  (for $a>0$ and $0<\alpha <2\,, $
$\alpha \ne 1\, $) has exactly {\it two simple, conjugate complex
zeros on the principal branch in the open left half-plane}, cut along
the negative real axis, say $s_+= \rho \,\e^{\,+i\gamma }$
and $s_- = \rho \,\e^{\,-i\gamma }$   with $\rho >0\,$ and
$ \pi/2<\gamma <\pi\,. $
This enables us to repeat the considerations carried out for the
simple fractional oscillation equation to decompose the basic
fundamental solution  $v_0(t)$ into two parts according
to $v_0(t) = f_\alpha (t;a) + g_\alpha (t;a)\,. $
In fact,  the
evaluation of the Bromwich integral (4.36) can be achieved
by adding the contribution $f_\alpha (t;a)$
from the  Hankel path $Ha(\epsilon )$ as $\epsilon \to 0\,, $
to the residual contribution  $g_\alpha (t;a)$ from the two
poles $s_\pm\,. $
\vsp
As an exercise	in complex analysis we	obtain
$$ f_\alpha  (t;a) = \int_0^\infty \! \e^{\,\ds -rt}\,
    H^{(2)} _{\alpha,0} (r;a)\, dr\,, \eqno(4.37)$$
with   {\it spectral function}
$$ \eqalign{
H^{(2)}_{\alpha,0} (r;a) &= -\, \rec{\pi}\,    {\rm Im}\,
  \l\{ \l.{s+ as^{\alpha -1} \over w_2(s)}\r\vert_{s=r\, \e^{i\pi}} \r\}\cr
   &= \rec{\pi}\,
   { a\,r^{\alpha -1}\, \sin \,(\alpha \pi)\over
 (r^2+1)^2 + a^2\, r^{2\alpha} + 2\, (r^2+1)\,a\,r^{\alpha} \, \cos \,
(\alpha \pi)}\,.\cr}	   \eqno(4.38)$$
Since
in (4.38) the denominator  is strictly
positive being	equal to $| w_2(s)|^2 $ as $s = r\, \e^{\pm i\pi}\,, $
the {\it spectral function} $H^{(2)}_{\alpha,0} (r;a)$
turns out to be positive for all $r>0$
for $0<\alpha <1$ and
negative for all $r>0$
for $1<\alpha <2\,. $
Hence,
in case (a) the function $f_\alpha(t)\,, $
in case (b) the function $-f_\alpha(t)\, $
is {\it completely monotone}; in both cases  $f_\alpha (t)$
tends to zero as $t\to \infty\,, $
from above in case (a), from below in case (b), according to the
asymptotic behaviour
$$ f_\alpha (t;a) \sim a\,{t^{-\alpha}\over \Gamma(1-\alpha)}\,,
\q {\rm as} \q t\to \infty\,, \q 0<\alpha <1\,, \q
   1<\alpha <2\,, \eqno(4.39)$$
as derived  by applying the Watson lemma in (4.37) and
considering (4.38).
\vsp
The other part, $g_\alpha (t;a)\,, $
is obtained as
$$ \eqalign{
 g_\alpha (t;a) &=     \e^{\,\ds s_+ \,t}\,
Res \,\l[ {s+a\, s^{\alpha -1} \over w_2(s)} \r]_{s_+} +
  \q  {\rm conjugate} \;\;{\rm complex} \cr
  &=  2\, {\rm Re}\, \l \{
   {s_+ +a\, s_+^{\alpha -1}\over 2\,s_+ + a\,\alpha\,s_+^{\alpha -1}}\,
     \e^{\,\ds s_+ \,t} \r\}\,.\cr} \eqno(4.40)$$
Thus this term exhibits an
{\it oscillatory} character with exponentially	decreasing amplitude
like ${\rm exp} \l(-\rho \,t \, |\cos \, \gamma |\r)\,. $
\vsp
Then we recognize that
the basic fundamental solution $v_0(t)$
exhibits a {\it finite} number of zeros and that, for sufficiently
large $t\,, $ it turns out to be permanently positive if
$0<\alpha <1$ and
permanently negative if
$1<\alpha <2$	with an {\it algebraic decay}
provided by (4.39).
\vsp
For the second fundamental solutions $v_{1\,{\rm a}}(t)\,,\,
v_{1\,{\rm b}}(t)$
and for the impulse-response
solution $v_\delta (t)\,, $
the corresponding analysis is straightforward in view
of their  connection with $v_0(t)$, pointed out in (4.31-32).
The {\it algebraic decay} of all the solutions
as $t \to \infty\,,$ for $0<\alpha <1\,$ and
   $1<\alpha <2\,,$
is henceforth resumed in the relations
$$ v_{0}(t) \sim a\, {t^{-\alpha}\over \Gamma(1-\alpha)}  \,, \q
   v_{1\,{\rm a}}(t)  = v_{\delta }(t)
	     \sim -a\, {t^{-\alpha-1}\over \Gamma(-\alpha)}\,, \q
   v_{1\,{\rm b}}(t) \sim a\,{t^{1-\alpha}\over \Gamma(2-\alpha)}\,.
 \eqno(4.41)$$
\vsp
In conclusion, except in the particular case $\alpha =1\,, $
all the present solutions of the composite fractional
oscillation equation
exhibit similar characteristics
with the corresponding solutions of the simple fractional
oscillation equation, namely a {\it finite number of damped oscillations}
followed by a {\it monotonic algebraic decay} as $t\to \infty\,. $
\vfill\eject

\pageno=261 
\line{5. CONCLUSIONS\hfill} \vsp
    Starting from the classical {\it Riemann-Liouville} definitions
of the
fractional integration operator and its left-inverse, the fractional
differentiation operator, and using the powerful tool of the Laplace
transform method, we have described the basic {\it analytical} theory of
fractional integral and differential equations.
For a {\it numerical} treatment
we refer to Gorenflo [23] and to the references there
quoted.
\vsp
For the {\it fractional integral equations} we have considered the
basic examples provided by the linear {\it Abel  equations
 of first and second kind}.
  For both the kinds we have given
 the solution in different forms and
 discussed an  interesting  application
 to inverse heat conduction problems.
 \vsp
  Then we have analyzed in detail the
 scale of {\it fractional ordinary differential equations} ($FODE$),
see (3.5),
 $   D_{*}^\alpha \, u(t) + u(t) = q(t), \; t > 0\,, \;
    0 < \alpha \le 2\,, $
 with a modified fractional differentiation $D_{*}^\alpha\,,$
 the {\it Caputo fractional derivative},
 that takes account of given initial values $u(0^+)$
 if  $0 < \alpha < 1\,,$ the case of {\it fractional relaxation},
 $u(0^+)$ and $u'(0^+)$ if $1 < \alpha < 2\,,$ the case of {\it
 fractional oscillation}.
\vsp
We have investigated in depth the properties of the {\it fundamental}
and the {\it impulse-response solutions}.
All these solutions can be explicitly
written down in terms of {\it Mittag-Leffler functions}.
They tend to zero
like powers  $t^{- \beta}$ (with appropriate choices of $ \beta$),
monotonically if $0 < \alpha < 1\,,$ but exhibiting finitely many
oscillations
around zero if $1 < \alpha < 2$ (the more of these
the nearer $\alpha$ is to
the limiting value 2).
If $1 < \alpha < 2$ these equations are able to model
processes intermediate between exponential decay ($\alpha = 1$) and pure
sinusoidal oscillation ($\alpha = 2$). We have found these qualitative
properties essentially by bending the Bromwich integration path of
the Laplace inversion formula into the Hankel path, thus for each of these
functions obtaining an integral representation as the Laplace transform of
a function that nowhere changes its sign,
augmented if $1 < \alpha < 2$ by
an oscillatory contribution resulting from a pair of conjugate complex
poles lying in the left half-plane.
\vsp
   By  quite
analogous methods we have studied the {\it composite equations},
see (4.1-2),
$     (D + a D_{*}^\alpha + 1)\,u(t) = q(t)\,,
   \;	 0 < \alpha < 1\,,$
 and
$  (D^{2} + a D_{*}^\alpha + 1)\, v(t) = q(t)\,,
  \;   0<\alpha <2\,,	$
with $a>0\,,$ which model
processes of relaxation and of oscillation, respectively.
We have obtained similar properties of the {\it fundamental} and
{\it impulse-response solutions} with respect to monotonicity and
oscillatory behaviour.
 \vsp
 Let us stress the fact that our adoption of the {\it Caputo fractional
derivative} $D_*^\alpha $ with $m-1<\alpha \le m\,, \; m\in \NN\,, $
and the consequent prescription of the initial
values in analogy with the ordinary differential equations of integer
order $m\,,$
stands in contrast to the majority  of	the
treatments of fractional differential equations, where the
standard fractional derivative $D^\alpha $  is used, see \eg [5], [10].
As pointed in \S 1.3 the adoption of $D^\alpha $
requires  the prescription  of
certain fractional integrals  as $t\to 0^+\,. $
\vfill\eject
\vsp
In our opinion, the different  prescription of the
the initial data points out the major difference
between the two definitions for the fractional derivative.
The analogy with the  cases of integer order
would induce one to adopt the {\it Caputo derivative}
in the treatment of differential equations of fractional
order for {\it physical applications}.
In fact, in physical problems,	the initial conditions are usually
expressed in terms of a given number of bounded values assumed by the
field variable and its derivatives of integer order,
no matter if
the governing evolution equation may be a generic integro-differential
equation and therefore, in particular,	a  fractional differential
equation.
 \vsp
The liveliness of the field of fractional integral and differential
equations,
both in applications and in pure theory, is underlined by several papers
and some books that have appeared recently.
We would like to conclude the present lectures with
 brief hints to recent investigations, which have not
been quoted  explicitly  up to now since not
strictly related to our results,  but have attracted
our attention.
Naturally, this listing
is far from exhaustive. The interested reader can find
more on problems and aspects in several papers recently published or
in press in some conference
volumes and specialized magazines.
\vsp
We first like to quote the  most recent book
by Rubin [14],
who starting from
one-dimensional fractional calculus develops the theory of
multidimensional weakly singular integral equations of first kind, making
heavy use of the Marchaud approach.
We thus recognize that all existing books  on fractional calculus
vary widely from each
other in their character concerning problems treated and methods applied.
We quote also the Ph.D. thesis by Michalski [47], who treats
linear and nonlinear  problems of fractional calculus
(in one and in several dimensions) in a very elegant way.
\vsp
The importance	of using fractional methods in physics for describing
{\it slow decay processes} and
{\it processes intermediate between relaxation and oscillation}
was stressed by Nigmatullin [48] in 1984.
Nonnenmacher and associates 
published a series of papers (of
which we quote [49-50]) discussing various physical aspects of
{\it fractional relaxation}.
Fractional relaxation is overall a peculiarity of a class of
{\it viscoelastic bodies} which are extensively treated by
Mainardi [24], to which we refer for details and additional bibliography.
The fractional calculus finds important applications in different
areas of applied science including
{\it electrochemistry}, see \eg [51-54], 
{\it electromagnetism}, see \eg [55-56], 
{\it radiation physics}, see \eg  [57-59], 
and {\it control theory}, see \eg [60-64].
\vsp
Yet another field of applications of fractional calculus
is that of {\it fractional partial differential equations}
($FPDE$), including
certain equations of {\it fractional diffusion}, introduced
to explain the phenomena of anomalous diffusion  in complex or fractal
systems.
We refer again to Mainardi [24] for a mathematical
treatment of a relevant $FPDE$, referred to as the
{\it time fractional diffusion-wave equation},
with  some applications and related references.
\vfill\eject

\pageno=263	 
\line{APPENDIX: THE MITTAG-LEFFLER TYPE FUNCTIONS
\hfill} \vsp
In this Appendix we  shall consider  the Mittag-Leffler function
and some of the related   functions  which are relevant for
their connection with fractional calculus.
It is our purpose  to provide a review of the
main properties of these functions including their
Laplace transforms.
\vvs
\line{{\it A.1 The Mittag-Leffler functions $E_\alpha (z)\,, \,
  E_{\alpha,\beta }(z)$  \hfill}}
\vsp
    The Mittag-Leffler function $E_\alpha (z)$ with $\alpha >0$
is so named from the great Swedish mathematician
who introduced it at the beginning of
this century in a sequence of five notes, see  [65-69].  
The function is defined by the following series representation,
valid in the whole complex plane,
$$
E_\alpha (z) := \sum_{n=0}^\infty {z^n\over \Gamma (\alpha n+1)}
\,,\q \alpha > 0\,, \q z\in \CC\,. \eqno(A.1)$$
It turns out that
$E_\alpha (z) $ is an {\it entire function}
 of order $\rho =1/\alpha \,$ and type $1\,. $
This property is still valid but with
$\rho= 1/{\rm Re}\{\alpha \}\,, $
if  $\alpha \in \CC$ with {\it positive real part},
as formerly noted by Mittag-Leffler himself in [68].
\vsp
In the limit for $\alpha \to 0^+$ the analyticity
in the whole complex plane is lost   since
$$ E_0(z) := \sum_{n=0}^\infty z^n
= {1\over 1- z}\,,\quad |z|<1\,.  \eqno(A.2)$$
\vsp
The  Mittag-Leffler function provides a simple generalization of the
  exponential function because of the substitution of
$n! = \Gamma (n+1)$ with $(\alpha n)!=\Gamma (\alpha n+1)\,.$
Particular cases of (A.1), from which elementary functions
are   recovered,  
are
$$ E_2\l(+z^2\r) =  \cosh \, z\,, \qq
   E_2\l(-z^2\r) =  \cos \, z\,, \qq z\in \CC \,,   \eqno(A.3)$$
and
$$ E_{1/2} (\pm  z^{1/2}) =
     \e^{\ds z} \, \l[ 1+{\rm erf}\, (\pm z^{1/2})\r ] =
   \e^{\ds z} \, {\rm erfc} \, (\mp z^{1/2})\,,\q z\in \CC \,,
 \eqno(A.4)$$
where erf (erfc) denotes the (complementary) error function
defined as
$$  {\rm erf} \,(z) := {2\over \sqrt{\pi}}\,\int_0^z \e^{\ds -u^2}\,du \,,
 \q   {\rm erfc} \,(z) :=  1 - {\rm erf}\, (z)\,,
 \q z\in \CC\,. $$
In (A.4) for $z^{1/2}$ we mean the principal value of the
square root of $z$  in the complex plane cut along the the
negative real axis. With this choice $\pm z^{1/2}$
turns out to be positive/negative  for $z \in \RR^+$.
\vsp
Since the identities in (A.3) are  trivial, we present the proof only for
(A.4). Avoiding the inessential polidromy
with the substitution $\pm z^{1/2} \to z\,, $
we write
 $$  E_{1/2}(z)  =
  \sum_{m=0}^\infty {z^{2m}\over \Gamma (m+1)} +
 \sum_{m=0}^\infty {z^{2m+1}\over \Gamma (m+3/2)}
= u(z) + v(z)\,.
\eqno(A.5)$$
\vfill\eject
\noindent
Whereas  the even part is easily recognized to be
$u(z) = {\rm exp} (z^2)\,,$
only after some manipulations
the odd part
can be proved to be
$v(z) = {\rm exp} (z^2)\, {\rm erf} (z)\,.$
To this end we need to recall
the following series representation for the error function,
see \eg the handbook of the Bateman Project [70] or
that by Abramowitz and Stegun [71],
$$ {\rm erf} (z) =
     {2\over \sqrt{\pi}}\, {\rm e}^{\ds -z^2}\,
 \sum_{m=0}^\infty  {2^{\ds m}\over (2m+1)!! }\, z^{\ds 2m+1}  \,,
   \q z\in \CC $$
and note that	
$ (2m+1)!! := 1 \cdot 3 \cdot 5 \dots \cdot (2m+1)=
   2^{m+1} \,	\Gamma(m+3/2)/\sqrt{\pi} \,. $
An alternative proof  
is obtained by recognizing, after
a term-wise differentiation of the series representation in (A.5), that
$v(z)$ satisfies the differential equation  in $\CC\,, $
$$ v'(z) = 2 \l[ \rec{\sqrt{\pi}} + z\, v(z)\r]\,, \q v(0)=0\,, $$
whose solution can immediately be checked to be
$$ v(z) = {2\over \sqrt{\pi}} \,
    {\rm e}^{\ds z^2}
    \, \int _0^z {\rm e}^{\ds -u^2} \, du =
    {\rm e}^{\ds z^2} \, {\rm erf}(z)\,. $$
\vsp
A straightforward generalization of the  Mittag-Leffler function,
originally due to Agarwal in 1953 based on a note by Humbert,
see [72-74],
is obtained  by replacing the additive constant $1$
in the argument of the Gamma function in (A.1) by an
arbitrary complex parameter $\beta \,. $
Later, when we shall deal with Laplace transform pairs,
the parameter $\beta $ is required to be positive as $\alpha \,. $
For the new function we agree to use the following notation
$$ {
E_{\alpha, \beta } (z ) := \sum_{n=0}^\infty {z^n\over
 \Gamma(\alpha n+ \beta  )}\,,\quad \alpha >0\,,\;\, \beta \in \CC\,,
   \quad z \in\CC
  }\,.	\eqno(A.6)
$$
Particular simple cases are
$$
  E_{1,2} (z) = {\e^{\ds z} -1 \over z}\,, \quad
  E_{2,2} (z) = {\sinh \, (z^{1/2}) \over z^{1/2}}\,.  \eqno(A.7)
$$
We note that
$E_{\alpha,\beta}  (z)$ is still
an entire function of order $\rho =1/\alpha $
and type $1\,. $
\vsp
In  these lectures
we have preferred to use only
the original Mittag-Leffler function (A.1) since our problems  depend on
only a single parameter $\alpha $, the order of fractional integration
of differentiation.
However, for completeness, we list hereafter
the general functional relations for the generalized Mittag-Leffler
function (A.6), which involve both the two parameters $\alpha \,,\,
\beta  \,, $ see [18] and [70], 
$$
{ E_{\alpha,\beta}   (z)
= {1\over \Gamma(\beta )} + z \,E_{\alpha,\beta+\alpha}  ( z)}\,,
\eqno(A.8)
$$
$$
 {
E_{\alpha,\beta}   (z) =
\beta  E_{\alpha,\beta +1} (z) + \alpha z \, {d \over dz} \,
E_{\alpha, \beta +1} (z)}\,,
\eqno(A.9)
$$
$$
{
\left( {d\over dz} \right)^p \left [
   z^{\beta  -1} E_{\alpha, \beta }  (z^{\alpha})\right ] =
z^{\beta  - p -1} E_{\alpha,\beta -p}  (z^{\alpha})\,,
\quad p \in \NN }\,.  \eqno(A.10)
$$
\vfill\eject
\line{{\it A2. The Mittag-Leffler functions of rational order}\hfill}
\vsp
Let us now consider  the  Mittag-Leffler functions of rational
order $\alpha  = p/q\,$ with $ p\,,\, q\in \NN\,$ relatively  prime.
The relevant functional relations, that we quote from [18], [70],
turn out to be
 $$ {
\left( {d\over dz }\right)^p \, E_{p}\,(z^{p})
=E_{p}\,(z^{p}) }\,,  \eqno(A.11)
$$
 $$ { {d^p\over dz^p}\, E_{p/q} \l(z^{p/q} \r)
= E_{p/q} \l(z^{p/q} \r)
  + \sum_{k=1}^{q-1}{z^{-k\,p/q}\over \Gamma(1-k\,p/q)}} \,,
  \q q= 2, 3, \dots\,, \eqno(A.12)$$
$$ {
E_{p/q}\,(z) =
{1\over p}\sum_{h=0}^{p-1}E_{1/q} \l(z^{1/p} e^{i 2\pi h/p}\r)}
\,,
\eqno(A.13)  $$
 and
$$ {
E_{1/q} \l(z^{1/q} \r) = \e^{\ds z}\, \l[
   1+\sum_{k=1}^{q-1}{\gamma (1-k/q,z)\over \Gamma(1-k/q)}\r] \,,
     \q q= 2, 3, \dots\,,} \eqno(A.14)$$
where $\gamma(a,z) := \int_0^z \e^{\, -u}\, u^{\, a-1} \, du$
denotes the {\it incomplete gamma function}.
Let us now sketch the proof for the  above functional relations.
\vsp
One easily  recognizes that the relations (A.11) and (A.12) are immediate
consequences of the definition (A.1).
 \vsp
In order to prove the relation (A.13) we need to recall the identity
$$
\sum_{h=0}^{p-1} \e ^{\, \ds i 2\pi hk/p} =\left\{
 \eqalign{ \, p \quad &{\rm if}\;  k\equiv 0 \pmod{p}\,, \cr
	   \, 0 \quad &{\rm if}\;  k\not\equiv 0 \pmod{p}\,. \cr}
   \right.\eqno(A.15)
$$
In fact, using this identity and the definition (A.1),
 we have
 $$
\sum_{h=0}^{p-1}E_{\alpha}(z \,\e^{i 2\pi h/p}) =
 p\,E_{\alpha p}(z^p)\,,
 \quad p \in \NN\,. \eqno(A.16)
$$
Substituting in the above relation $\alpha/p $
instead of $\alpha $ and $z^{1/p}$ instead of $z\,,$ we obtain
$$
E_{\alpha}(z) =
{1\over p}\sum_{h=0}^{p-1}E_{\alpha/p}\l(z^{1/p} e^{i 2\pi h/p}\r)\,,
  \quad p \in \NN \,.
\eqno(A.17)  $$
Setting above $\alpha =p/q\,, $ we finally obtain (A.13).
\vsp
To prove the relation (A.14) we consider (A.12) for $p=1\,.$
Multiplying both sides by  $\e^{-z}\,, $ we obtain
$$ {d\over dz} \l[ \e^{\ds -z} \, E_{1/q} \l(z^{1/q}\r)\r] =
  \e^{\ds -z} \,  \sum_{k=1}^{q-1}{z^{-k/q}\over \Gamma(1-k/q)}\,.
\eqno(A.18) $$
Then, upon integration of this and recalling the definition of the
incomplete gamma function,
we arrive at (A.14).
\vfill\eject
\vsp
The relation (A.14) shows how the Mittag-Leffler functions
of rational order can be expressed in terms of exponentials
and incomplete gamma functions. In particular, taking in (A.14) $q=2\,, $
we now can verify again the relation (A.4).
In fact, from (A.14) we obtain
   $$ E_{1/2} (z^{1/2}) = \e^{z} \,\l[ 1 + \rec{\sqrt{\pi}}\,
    \gamma (1/2\,, \, z )\r]\,, \eqno(A.19)$$
which is equivalent to (A.4) if we use	the relation
$ {\rm erf}\, (z) =  \gamma (1/2, z^2)/\sqrt{\pi}\,,$
see \eg [70-71].
 \vvs
\line{{\it A3. The Mittag-Leffler integral representation
 and asymptotic expansions}\hfill}\vsp
Many of the most important properties of $E_\alpha (z)$
follow from  Mittag-Leffler's {\it integral representation}
$$ { E_\alpha (z) =
  \rec{2\pi i}\,
 \int_{Ha} {\zeta ^{\alpha -1}
 \, \e^{\,\zeta} \over \zeta ^\alpha -z}\,
	d\zeta \,,  \quad  \alpha  >0\,,
 \quad z\in \CC
 }\,, \eqno(A.20)$$
where the path of integration $Ha\,$ (the {\it Hankel path})
is a loop which
starts and ends at $-\infty$ and encircles the circular disk
$|\zeta | \le |z|^{1/\alpha }$ in the positive sense:
$-\pi  \le \arg \zeta \le \pi $ on $Ha\,.$
To prove (A.20), expand the integrand in powers of $\zeta$,
integrate term-by-term, and use Hankel's integral
for the reciprocal of the Gamma function.
\vsp
The integrand in (A.20) has a branch-point at $\zeta =0$.
The complex $\zeta$-plane is cut along the negative real axis,
and in the cut plane the integrand is single-valued:
the principal branch of $\zeta^\alpha $ is taken in the cut plane.
The integrand has poles at the points
$\zeta _m = z^{1/\alpha } \, \e^{2\pi\, i\, m/\alpha }	\,,
   \; m\;$ integer,
but only those of the poles lie in the cut plane for
which
$ -\alpha \,\pi  < \arg z + 2\pi \, m < \alpha \, \pi\,.$
Thus, the number of the poles inside $Ha\,$ is either $[\alpha ]$
or $[\alpha +1]$, according to the value of $\arg \, z$.
\vsp
The integral representation of the generalized
Mittag-Leffler function turns out to be
$$ { E_{\alpha,\beta}  (z) =
  \rec{2\pi i}\,
 \int_{Ha} {\zeta ^{\alpha -\beta  }\, \e^{\,\zeta} \over \zeta ^\alpha -z}\,
	d\zeta\,,
 \quad \alpha \,,\, \beta >0\,,
   \quad z \in\CC
  }\,. \eqno(A.21)$$
\vsp
The most interesting properties of the Mittag-Leffler function
are associated with its asymptotic developments
as $z \to \infty$ in various sectors of the complex plane.
These properties can be summarized as follows.
\pni For the case $0 < \alpha <2\,$ we have  
$$ E_\alpha (z) \sim
   \rec{\alpha }\, \exp (z^{1/\alpha})
   - \sum_{k=1}^\infty {z^{-k}\over \Gamma(1-\alpha k)} \,,
   \qq |z| \to \infty\,, \q
|\arg  \,z | <\alpha \pi/2\,, \eqno(A.22)$$
 $$ E_\alpha (z) \sim
      - \sum_{k=1}^\infty {z^{-k}\over \Gamma(1-\alpha k)} \,,
   \qq |z| \to \infty\,,
\q \alpha \pi /2 < \arg \,z < 2\pi -\alpha \pi/2\,. \eqno(A.23)$$
\vfill\eject
\pni For the case  $\alpha  \ge 2\,$ we have
 $$ E_\alpha (z) \sim	 \rec{\alpha }\,
\sum_{m} \exp \l(z^{1/\alpha}\e^{2\pi \,i\,m /\alpha } \r)
  - \sum_{k=1}^\infty {z^{-k}\over \Gamma(1-\alpha k)} \,,
   \qq |z| \to \infty\,, \eqno(A.24)$$
where $m$ takes all integer values 
such that
   $- \alpha \pi/2 <	\arg \, z + 2\pi \,m
	  < \alpha \pi /2\,, $
and $\arg \, z$ can assume any value between $-\pi $ and $+\pi $
inclusive.
 \vsp
From the asymptotic properties (A.22-24) and the definition
of the order of an entire function, we infer that
 the Mittag-Leffler function  is an {\it entire function of order}
$1/\alpha $ for $\alpha >0$; in a certain sense
each $E_\alpha (z)$ is the simplest entire function of its
order, see Phragm\'en [75].
The Mittag-Leffler function  also furnishes examples and counter-examples
for the growth and other properties of entire functions of finite order,
see Buhl [76].
 \vvs
\line{{\it A4. The  Laplace transform pairs related to
 the Mittag-Leffler functions}\hfill}
\vsp
     The Mittag-Leffler functions  are connected to the
Laplace integral through the equation
$$ {
\int_0^\infty \!\!\! \e^{-u} \, E_\alpha \l(u^\alpha \,z\r) \, du
   = \rec{1-z}=
\int_0^\infty \!\!\! \e^{-u}\, u^{\beta  -1} \,
  E_{\alpha,\beta } \l(u^\alpha \,z \r)\, du  \,,
  \q \alpha \,, \,\beta  >  0
   }\,.\eqno(A.25)$$
The integral at the L.H.S. was evaluated by Mittag-Leffler who showed that
the region of its convergence contains the
unit circle and is bounded by the line
${\rm Re}\, z^{1/\alpha } =1$.
\vsp
The above integral is fundamental in the  evaluation of
the {Laplace transform} of 
$E_\alpha \l(-\lambda \, t^\alpha \r)$
and $E_{\alpha,\beta } \l(-\lambda \, t^\alpha\r)$
with $\alpha \,,\, \beta  >0$ and $\lambda  \in \CC\,. $
  Since
these functions turn out to
play a key role in problems of fractional calculus,
we shall introduce a special notation  for them.
\vsp
Putting  in (A.25)
$u=st$ and $u^\alpha \, z = -\lambda \, t^\alpha $  with
$t \ge 0\, $ and $\lambda \in \CC \,, $  and
using the sign $\div$ for the juxtaposition of
a function depending on $t$ with its Laplace transform depending on $s$,
we get the following Laplace
transform pairs
$$ {
e_\alpha (t;\lambda ) := E_\alpha \l(-\lambda \, t^\alpha \r)
   \div {s^{\alpha -1}\over s^\alpha +\lambda} \,,
 \q Re \, s > |\lambda |^{1/\alpha }
}\,, \eqno(A.26)$$
and
$$ {
e_{\alpha,\beta}  (t;\lambda):=
   t^{\beta  -1}\,  E_{\alpha,\beta } \l(-\lambda \, t^\alpha\r)
   \div {s^{\alpha -\beta }\over s^\alpha +\lambda}
    \,,   \q Re \, s > |\lambda |^{1/\alpha }
	   }\,. \eqno(A.27)$$
We note that the results
(A.26-27), but with a different notation,
were used by Humbert and Agarwal [72-74] to obtain a
number of functional relations satisfied by $E_\alpha (z)$ and
$E_{\alpha,\beta}  (z)\,. $
Of course the results (A.26-27) can also be obtained formally by
 Laplace transforming term by term  the series (A.1)  and (A.6)
with $z= -\lambda \, t^\alpha\,, $
respectively,  and summing the resulting series.
\vfill\eject
We find worthwhile  to list the following relations for the
functions $\, e_{\alpha ,\beta }\,$ easily
obtained from (A.8-9) :
$$ { e_{\alpha,\beta }(t;\lambda )=
   {t^{\beta  -1}\over \Gamma(\alpha )} -
 \lambda \,  e_{\alpha, \beta+\alpha} (t;\lambda )\,,
    } \eqno(A.28)
$$
and
$$ { {d\over dt}\, e_{\alpha,\beta +1}(t; \lambda)
 = e_{\alpha,\beta } (t;\lambda)}\,.
\eqno(A.29)
  $$
\vsp
A remarkable property satisfied by the functions
$e_\alpha(t;\lambda ) \,,\,  e_{\alpha ,\beta }(t;\lambda)\,$  when
$\lambda\,$ {\it is positive} and $0<\alpha \le 1\,, $
$\,0 <\alpha \le \beta \le 1 \,, \,$ respectively,
is to be   {\it completely monotone}  for $t>0\,. $
\vsp
We recall that a function $f(t)$ is told to be {\it completely monotone}
for $t>0$ if $(-1)^n\, f^{(n)}(t) \ge 0$
for all $n=0,1,2,\dots $ and for all $t>0\,, $
and that a sufficient condition for this is the
existence of a nonnegative locally integrable  function
$K(r)\,,$ $\, r>0\,,$
referred to as the {\it spectral function},
 with which
$f(t)= \int_0^\infty \e^{\,-rt} \,K(r)\, dr\,. $
For more details see \eg the book by
Berg \& Forst [77].
\vsp
Excluding the trivial case $\alpha =\beta =1$ for which
$e_1(t;\lambda ) = e_{1,1} (t;\lambda )  = \e^{\,-\lambda\,t}\,, $
we can prove  the existence of the corresponding spectral functions
using the complex Bromwich formula to invert the Laplace
transform in (A.26-27) and bending the Bromwich path into the
Hankel path, as we have already shown in the special case
$e_\alpha (t) := e_\alpha (t;1)$   in \S 3.
As an exercise	in complex analysis (we kindly invite the
reader to carry it out) we obtain
the  integral representations [A.30-33],
$$ {
 e_\alpha (t;\lambda) := \int_0^\infty \! \e^{\,\ds -rt}\,
     K_\alpha (r;\lambda )\, dr \,,\q 0<\alpha < 1\,, \q \lambda >0
}\,, \eqno(A.30)$$
with {\it spectral function}
$$ {
 K_\alpha (r;\lambda) = {1 \over \pi}\,
   { \lambda \,r^{\alpha -1}\, \sin \,(\alpha \pi)\over
r^{2\alpha} + 2\,\lambda\,r^{\alpha}\,\cos\,(\alpha\pi)+\lambda^2}
  \, \ge 0    }\,,	    \eqno(A.31)$$
and
$$ {
e_{\alpha,\beta} (t;\lambda)
 := \int_0^\infty \! \e^{\,\ds -rt}\, K_{\alpha,\beta} (r;\lambda)\, dr
   \,,\q 0<\alpha \le\beta  < 1\,, \q \lambda >0} \,,  \eqno(A.32)$$
with  {\it spectral function}
$${  K_{\alpha,\beta} (r;\lambda) =  {1\over \pi}\,
   {\lambda\,\sin \,[(\beta -\alpha)\pi]  + r^\alpha \, \sin\,(\beta \pi)
   \over
r^{2\alpha} + 2\,\lambda\,r^{\alpha}\,\cos\,(\alpha\pi)+\lambda^2}
   \,r^{\, \ds \alpha -\beta}  \,    \ge 0 }\,.\eqno(A.33)$$
\vfill\eject
\vsp
Historically, the complete monotonicity of the Mittag-Leffler function
in the negative real axis, \ie of
$E_\alpha (-x)\,, $ for $x \in \RR^+$, when $0<\alpha < 1\,, $
was first conjectured 
by Feller using probabilistic  methods and rigorously
proved by Pollard in 1948 [78].
\vsp
Only recently, Schneider  [79]
has proved a theorem for  the complete monotonicity
of the generalized Mittag-Leffler function
in the negative real axis. He proved that
$E_{\alpha,\beta } (-x)\,, $ for $x \in \RR^+$,
is {completely	monotone}
{\it iff}
 $0<\alpha \le 1\, $
and $\beta \ge \alpha \,. $
Our conditions for $e_{\alpha,\beta} (t, \lambda)$ to be
 completely monotone appear more restrictive than those by
 Schneider for $E_{\alpha,\beta}(-x)\,;$ however, we must note
 that in our case (A.27) the factor  $t^{\beta - 1}\,$ precedes
 the generalized Mittag-Leffler function.
\vsp
We note that, up to our knowledge, in the handbooks containing tables for
the Laplace transforms, the Mittag-Leffler
function is ignored so that the transform pairs (A.26-27)
do not appear
if not in the special cases $\alpha =1/2$
and $\beta  = 1\,,\,1/2\,, $
written however in terms
of the error and complementary error functions,
see \eg [71].
In fact, in these cases
we can use (A.4) and (A.28) and recover from (A.26-27)
the two Laplace transform pairs
\def\ss{{s}^{1/2}}   
$$ {
{1\over \ss\,(\ss \pm\lambda)} \,\div \, e_{1/2}\,(t;\pm\lambda ) =
    {\rm e}^{\ds \lambda^2\,t}\, {\rm erfc}\, (\pm\lambda \,\sqrt{t})}
   \,,\q \lambda \in \CC\,,   \eqno(A.34)
 $$
$$  {
 {1\over \ss \pm\lambda} \,\div \,  e_{1/2,1/2}\,(t;\pm \lambda ) =
     \rec{\sqrt{\pi\,t}}
 \mp \lambda \, e_{1/2}\,(t;\pm\lambda )  }
   \,,\q \lambda \in \CC   \,.	\eqno(A.35)
$$
We also obtain the  related pairs
$$  {
     {1\over \ss\, (\ss \pm\lambda )^2}
 \,\div \,  2\,\sqrt{{t\over \pi}}
 \mp 2\,\lambda \, t\, e_{1/2}(t;\pm \lambda)
     }\,,\q \lambda \in \CC \,,  \eqno(A.36)
 $$
$$
     {1\over  (\ss \pm\lambda )^2}
\,\div \, \mp 2\, \lambda \,\sqrt{{t\over \pi}}
+(1+ 2\,\lambda^2\,t)\, e_{1/2}(t;\pm \lambda)
   \,,\q \lambda \in \CC  \,,  \eqno(A.37)
 $$
In the	pair (A.36) we have used the properties
$$ {1\over \ss\, (\ss \pm\lambda )^2} =
  -2\, {d\over ds}\,\l({1\over \ss \pm\lambda } \r)\,,	\qq
  {d^n\over ds^n}\, \bar f(s) \div (-t)^n \, f(t)\,. 
$$
The pair (A.37)  is easily obtained by noting that
 $${1\over  (\ss \pm\lambda )^2}   = {1\over \ss\, (\ss \pm\lambda)}
     \mp {\lambda \over \ss\, (\ss \pm\lambda )^2}    \,. 
$$
\vfill\eject
\line{{\it A.5 Additional references for the
  Mittag-Leffler type functions}
 \hfill}
\vsp
We note that the   Mittag-Leffler type functions
are  unknown to the majority of scientists, because
they are ignored in the  common books
on special functions.
Thanks to  our suggestion the new
{\it 2000 Mathematics Subject Classification}
has included these functions, see the item
 {\it 33E12:  Mittag-Leffler functions and generalizations}.
\vsp
A description of the most important properties of these
functions with relevant references
can be found in the third volume of the  Bateman Project  [70], in the
chapter $XVIII$ devoted to {\it miscellaneous functions}.
The specialized treatises  where great attention is
devoted to the Mittag-Leffler type functions are those by Dzherbashyan
[18], [22].
For the interested readers we also recommend
the classical treatise on complex functions by
Sansone \& Gerretsen [80], where
a sufficiently detailed  treatment of the  original  Mittag-Leffler
function is given.
Since the times of Mittag-Leffler
several scientists
 have recognized the importance of the Mittag-Leffler type
 functions,
providing  interesting	results and applications,
which unfortunately are not much known.
As pioneering works of mathematical nature in the field of fractional
integral and differential equations, we like
to quote  those by Hille \& Tamarkin  and by Barret.
In 1930 Hille \& Tamarkin [81] have provided the solution
of the Abel integral equation of the second kind
in terms of a Mittag-Leffler  function,
whereas in 1956 Barret [82] has expressed the	general
solution of the linear	fractional differential equation with constant
coefficients in terms of Mittag-Leffler functions.
\vsp
As former applications in physics we like to quote the contributions
by K.S. Cole (1933), quoted by H.T. Davis [15, p. 287] in connection
with nerve conduction, and by F.M. de Oliveira Castro (1939) [83]
and B. Gross  (1947) [84]
in connection with dielectrical and mechanical relaxation, respectively.
Subsequently, in 1971, Caputo \& Mainardi [28]	have  proved
that the Mittag-Leffler function is present whenever
 derivatives of fractional order are
introduced in the constitutive equations of a linear viscoelastic body.
Since then, several other authors
have pointed out  the relevance of the Mittag-Leffler function for
fractional viscoelastic models, see Mainardi [24].
\vsp
In recent times  the attention of mathematicians
towards the Mittag-Leffler type functions
 has increased
from both  the analytical and  numerical
point of view, overall because of their relation with the fractional
calculus. In addition to the books and papers already quoted in the text,
here we would like to draw the reader's attention
to the most recent papers  on the Mittag-Leffler type
functions, \eg Al Saqabi \& Tuan [85], Kilbas \& Saigo [86],
Gorenflo, Luchko \& Rogozin [87]
and Mainardi \& Gorenflo [88].
Since the fractional calculus	has actually  recalled a wide interest
for its applications in different areas of physics and engineering,
we expect that soon the Mittag-Leffler function
will  exit  from its isolated life as {\it Cinderella} 
(using the term coined	by   F.G. Tricomi
 in the 1950s for the {\it incomplete Gamma} function).
\vfill\eject

\pageno=271 
\line{REFERENCES \hfill} \vsn
\rf{1}	 Ross, B.  (Editor):
  {\it Fractional Calculus and its Applications},
  Lecture Notes in Mathematics \# 457, Springer Verlag, Berlin 1975.
  [Proc. Int. Conf. held at Univ. of New Haven, USA, 1974]
\vsp
\rf{2}	Oldham, K.B. and J. Spanier: {\it The Fractional Calculus},
   Academic Press, New	York 1974.
\vsp
\rf{3}	McBride, A.C.: {\it Fractional Calculus and Integral Transforms
  of Generalized Functions}, Pitman Research Notes in Mathematics
 \# 31, Pitman, London 1979.
\vsp
\rf{4}	McBride, A.C. and G.F. Roach (Editors):
  {\it Fractional  Calculus},
   Pitman   Research   Notes  in Mathematics  \# 138,
  Pitman, London 1985.
  [Proc. Int. Workshop. held at Univ. of Strathclyde, UK, 1984]
\vsp
\rf{5}	Samko S.G.,  Kilbas, A.A.  and	 O.I. Marichev:
  {\it Fractional Integrals and Derivatives, Theory and Applications},
   Gordon and Breach, Amsterdam 1993.
   [Engl. Transl. from Russian, {\it Integrals and Derivatives of
    Fractional	 Order and Some of  Their Applications},
    Nauka i Tekhnika, Minsk 1987]
\vsp
\rf{6} H.M. Srivastava and S. Owa (Editors):
  {\it Univalent Functions, Fractional Calculus, and their Applications},
  Ellis Horwood, Chichester 1989.
\vsp
\rf{7}	 Nishimoto, K. (Editor):
  {\it	Fractional Calculus and its Applications},
  Nihon University, Tokyo 1990.
  [Proc. Int. Conf. held at Nihon Univ., Tokyo	1989]
\vsp
\rf{8}	 Nishimoto, K.:
  {\it An Essence of Nishimoto's Fractional Calculus},
  Descartes Press, Koriyama 1991.
\vsp
\rf{9}	 Kalia, R.N. (Editor):
  {\it	Recent Advances in Fractional Calculus},
  Global Publ., Sauk Rapids, Minnesota	1993.
\vsp
 \rf{10}  Miller, K.S. and  B. Ross:
 {\it An Introduction to the Fractional
  Calculus and Fractional Differential Equations},
  Wiley, New York 1993.
\vsp
\rf{11}  Kiryakova, V.:
   {\it Generalized Fractional Calculus and Applications},
   Pitman Research Notes in Mathematics \# 301,
     Longman, Harlow 1994.
\vsp
\rf{12} Rusev, P., Dimovski, I	and V. Kiryakova (Editors):
  {\it Transform Methods  and Special Functions, Sofia 1994},
  Science Culture Technology, Singapore 1995.
  [Proc. Int. Workshop, Sofia, Bulgaria, 12-17 August 1994]
\vsp
\rf{13} Kilbas, A.A. (Editor):
 {\it Boundary Value Problems, Special Functions and Fractional
 Calculus}, Byelorussian State University, Minsk 1996.
 (ISBN 985-6144-40-X)
 [Proc. Int. Conf., 90-th Birth Anniversary of Academician
   F.D. Gakhov, Minsk, Byelorussia, 16-20 February 1996]
\vsp
\rf{14}  Rubin, B.:
 {\it Fractional Integrals and Potentials}, Pitman Monographs and
  Surveys in Pure and Applied Mathematics \#82, Addison Wesley Longman,
  Harlow 1996.
\vsp
\rf{15} Davis, H.T.:{\it The Theory of Linear Operators},
  The Principia Press, Bloomington, Indiana 1936.
\vsp
\rf{16} Erd\'elyi, A. (Editor):
      {\it Tables of Integral Transforms}, 
 Bateman Project,  Vols. 1-2, McGraw-Hill, New	York 1953-1954.
 \vsp
\rf{17}
Gel'fand, I.M. and G.E. Shilov: {\it Generalized Functions}, Vol. 1,
 Academic Press, New York 1964.
\vsp
\rf{18}  Dzherbashian, M.M.: {\it Integral Transforms and
  Representations of Functions in the Complex Plane}, Nauka, Moscow 1966.
 [in Russian]
\vsp
\rf{19}   Caputo, M.:  {\it Elasticit\`a e Dissipazione},
     Zanichelli, Bologna 1969. [in Italian]
\vsp
\rf{20} Babenko, Yu.I.: {\it Heat and Mass Transfer},
   Chimia, Leningrad 1986. [in Russian]
\vsp
\rf{21}  Gorenflo, R. and S. Vessella: {\it Abel Integral Equations:
  Analysis and Applications},
  Lecture Notes in Mathematics \# 1461,
  Springer-Verlag, Berlin 1991.
 \vsp
\rf{22} Dzherbashian, M.M.: {\it Harmonic Analysis and Boundary
  Value Problems in the Complex Domain}, Birkh\"auser Verlag, Basel
  1993.
\vsp
\rf{23}   Gorenflo, R.:
  Fractional calculus: some numerical methods,
   in: {\it Fractals and Fractional Calculus in Continuum Mechanics}
  (Eds. A. Carpinteri and F. Mainardi), Springer Verlag, Wien 1997,
  277-290.  (this  book)
\vsp
\rf{24}
   Mainardi, F.: Fractional calculus:
  some basic problems in continuum and statistical mechanics,
  in: {\it Fractals and Fractional Calculus in Continuum Mechanics}
  (Eds. A. Carpinteri and F. Mai\-nardi), Springer Verlag, Wien 1997,
  291-348. 
\vsp
\rf{25}   Doetsch, G.:
  {\it Introduction to the Theory and Application of the Laplace
  Transformation}, Springer Verlag, Berlin 1974.
\vsp
\rf{26}   Henrici, P.:
  {\it Applied and Computational Complex Analysis},
   Vol. 2,   
  Wiley, New York 1977.
\vsp
\rf{27}  Caputo, M.:
  Linear models of dissipation whose Q is almost frequency
  independent,	 Part II.,
  {\it Geophys. J. R. Astr. Soc.}, {\bf 13}  (1967), 529-539.
\vsp
\rf{28} Caputo, M. and	F. Mainardi:
  Linear models of dissipation in  anelastic solids,
  {\it Riv. Nuovo Cimento} (Ser. II), {\bf 1} (1971), 161-198.
\vsp
\rf{29} Podlubny, I.: Solutions of linear fractional differential
 equations with constant coefficients,
 in: {\it Transform Methods  and Special Functions, Sofia 1994}
  (Eds. P. Rusev, I. Dimovski  and V. Kiryakova),
  Science Culture Technology, Singapore 1995, 227-237.
  [Proc. Int. Workshop, Sofia, Bulgaria, 12-17 August 1994]
\vsp
\rf{30} Bagley, R. L.: On the fractional order initial value problem and
 its engineering applications, in:
 {\it Fractional Calculus and Its Applications} (Ed. K. Nishimoto),
  College of Engineering, Nihon University, Tokyo 1990, pp. 12-20.
  [Proc. Int. Conf. held at Nihon Univ., Tokyo,  1989]
\vsp
\rf{31}   Craig, J. D. and J. C. Brown:
 {\it Inverse Problems in Astronomy}
  Adam	   Hilger Ltd, Bristol 1986.
\vsp
\rf{32}
Gorenflo, R.: {\it Abel Integral Equations with special Emphasis on
  Applications},
Lectures in Mathematical Sciences  Vol. 13, The University
of Tokyo, Graduate School of Mathematical Sciences, 1996.
(ISSN 0919-8180)
\vsp
\rf{33} Gorenflo, R.: The tomato salad problem in spherical stereology,
 in  {\it Transform Methods  and Special Functions, Varna 1996}
 (Eds. Rusev, P., Dimovski, I  and V. Kiryakova),
  Science Culture Technology, Singapore 1997 ({\it in press}).
  [Proc. Int. Workshop, Varna, Bulgaria, 23-30 August 1996]
An extended version is available as  Pre-print	A-25/96, Fachbereich
Mathematik und Informatik, Freie Universit\"at, Berlin 1996,
via Internet:
    $\,<\,$http://www.math.fu-berlin.de/publ/index.html$\,>$
\vsp
\rf{34}  Duff, G. F. D. and D. Naylor:
  {\it Differential Equations of Applied  Mathematics},
  Wiley \& Sons, New York 1966.
\vsp
\rf{35}  Mann, W. R. and F. Wolf:
  {\it Heat transfer between solids and gases under
     nonlinear boundary conditions},
  Quart.  Appl. Math., {\bf 9}	(1951), 163 - 184.
\vsp
\rf{36}  Lebedev, N. N.:
  {\it Special Functions and Their Applications},
  Dover, New York 1965.
\vsp
\rf{37} Gorenflo, R. and R. Rutman: On ultraslow and intermediate
   processes, in: 
  {\it Transform Methods  and Special Functions, Sofia 1994}
  (Eds. Rusev, P., Dimovski, I	and V. Kiryakova),
  Science Culture Technology, Singapore 1995, 61-81.
  [Proc. Int. Workshop, Sofia, Bulgaria, 12-17 August 1994]
\vsp
\rf{38} Mainardi, F.: Fractional relaxation and fractional diffusion
  equations, mathematical aspects,  in:
  {\it Proceedings 12-th IMACS World Congress} (Ed. W.F. Ames),
   GeorgiaTech, Atlanta 1994, Vol. 1, 329-332.	
\vsp
\rf{39} Mainardi, F.: Fractional relaxation-oscillation and fractional
  diffusion-wave phenomena,
  {\it Chaos, Solitons \& Fractals}, {\bf 7} (1996), 1461-1477.
\vsp
\rf{40} Gorenflo, R. and F. Mainardi:
   Fractional oscillations and Mittag-Leffler func\-tions,
   Pre-print  A-14/96, Fachbereich Mathematik und Informatik,
   Freie Universit\"at, Berlin 1996,
Internet:
   $\,<\,$http://www.math.fu-berlin.de/publ/index.html$\,>$
\vsp
\rf{41} Mainardi, F. and R. Gorenflo:
 The Mittag-Leffler function in the Riemann-Liouville fractional
  calculus,   in: 
 {\it Boundary Value Problems, Special Functions and Fractional Calculus}
  (Ed. A.A.Kilbas), Byelorussian State University, Minsk 1996, 215-225.
  (ISBN 985-6144-40-X)
   [Proc. Int. Conf., 90-th Birth Anniversary of Academician
   F.D. Gakhov, Minsk, Byelorussia, 16-20 February 1996]
\vsp
\rf{42} Gorenflo, R. and F. Mainardi:
Fractional relaxation and oscillations in linear causal systems,
Pre-print, Department of Physics, University of Bologna, 1996.
\vsp
\rf{43}
Blank, L.: Numerical treatment of differential equations of fractional
order,
\rp MCCM Numerical Analysis Report No. 287,
   The University of Manchester 1996,
Internet:    $\,<\,$http://www.ma.man.ac.uk/MCCM/MCCM.html$\,>$
\vsp
\rf{44}  Bleistein, N. and R. A. Handelsman:
  {\it Asymptotic Expansions of Integrals}, Ch. 4, p. 162,
  Dover, New York  1986.
\vsp
\rf{45}  Wiman, A.:
   \"Uber die Nullstellen der Funktionen $E_\alpha (x)\,,\,$
   {\it Acta Math.}, {\bf 29} (1905), 217-234.
\vsp
\rf{46} Beyer, H. and S. Kempfle:
   Definition of physically consistent damping laws with
  fractional derivatives,
  {\it ZAMM}, {\bf 75} (1995), 623-635.
\vsp
\rf{47} Michalski, M.W.: {\it Derivatives of Noninteger Order
   and their Applications}, Dissertationes Mathematicae, Polska
   Akademia Nauk, Instytut Matematyczny, Warsaw (1993).
\vsp
\rf{48} Nigmatullin, R.:
   On the theory of relaxation with "remnant" memory,
  {\it Phys. Stat. Sol. B}, {\bf 124} (1984), 389-393.
   Translated from the Russian.
\vsp
\rf{49}  Nonnenmacher, T.F. and  W.G. Gl\"ockle:
 A fractional model for mechanical stress relaxation,
    {\it Phil. Mag. Lett.}, {\bf 64} (1991), 89-93.
\vsp
\rf{50}   Nonnenmacher, T.F. and R. Metzler:
 On the Riemann-Liouville  fractional calculus and some
recent applications,	 {\it Fractals}, {\bf 3} (1995), 557-566.
\vsp
\rf{51}  Oldham, K.B. and C.G. Zoski: Analogue instrumentation
 for processing polorographic data, {\it J. Electroanalytical Chemistry},
    {\bf 157} (1983), 27-51.  
\vsp
\rf{52}  Macdonald, J.R. and L.D. Potter Jr.:
   A flexible procedure for analyzing impedance spectroscopy  results;
   descriptions and illustrations,
 {\it Solid State Ionics},  {\bf 23} (1987), 61-79.
 \vsp
\rf{53}  Mulder, W.H.  and J.H. Sluyters:
   An  explanation of depressed semi-circular arcs in impedance plots
  for irreversible electrode reactions,
 {\it Electrochemica Acta},  {\bf 33} (1988), 303-310.
 \vsp
\rf{54}  Keddam, M.  and H. Takenouti:
   Impedance of fractional interfaces; new data on the
 Von Koch model,
 {\it Electrochemica Acta},  {\bf 33} (1988), 3445-448.
 \vsp
\rf{55}  Engheta, N.:
   On the role of non-integral (fractional) calculus in electrodynamics,
{\it Digest of the 1992 IEEE AP-S/URSI International Symposium},
Chicago, July 17-20, 1992, Vol. URSI Digest, 163-175.
 \vsp
\rf{56}  Engheta, N.:
   Fractional differintegrals and electrostatic fields and potentials
   near sharp conducting edges, Lectures presented at the
International Summer School: {\it Fractals and Hyperbolic Geometries,
 Fractional and Fractal Derivatives in Engineering, Applied Physics and
Economics} (Eds. LeM\'ehaut\'e and A. Oustaloup), Bordeaux, July 3-8, 1994,
23 pages.
 \vsp
\rf{57}  Kalla, S.L.,  Al Saqabi, B. and S. Conde:
  Some results related to radiation-field problems,
   {\it Hadronic Journal}, {\bf 10} (1987), 221-230.
\vsp
\rf{58}  Gabutti, B., Kalla, S.L. and J.H. Hubbell:
  Some expansions related to the Hubbell rectangular-source integral,
   {\it J.  Comp. Appl. Maths}, {\bf 37} (1991), 273-285.
\vsp
\rf{59}   Kalla, S.L.:
   The Hubbell rectangular source integral and its generalizations,
   {\it Radiation Physics and Chemistry}, {\bf 41} (1993), 775-781.
\vsp
\rf{60} Rutman, R.S.: On physical interpretations of fractional
 integration and differentiation,
 {\it Theor. and Math. Physics}, {\bf 105} (1995), 1509-1519.
Translated from the Russian.
 \vsp
\rf{61} Podlubny, I.: {\it Fractional-order systems and fractional-order
  controllers}, Report	UEF-03-94, Slovak Academy of Sciences,
  Institute of Experimental Physics, Kosice, Slovakia, November 1994,
 18 pages.
 \vsp
\rf{62}  Oustaloup, A.:
{\it La d\'erivation non enti\`ere: th\'eorie, synth\`ese, applications},
s\'erie Automatique, \'Editions Herm\`es (1995).
\vsp
\rf{63} Matignon, D.:
 Stability results for fractional differential equations
 with applications to control processing,
 Proceedings {\it Computational Engineering in Systems and Application
  multiconference}, IMACS, IEEE-SMC, Lille, France, July 1966,
 pp. 963-968.
\vsp
\rf{64}  Fliess, M. and R. Hotzel:
 Sur les syst\`emes lin\'eaires \`a d\'erivation
non enti\`ere,
 {\it C. R. Acad. Sci. Paris}, XXX, Automatique,
Preprint 1996, 6 pages.
\vsp
\rf{65}  Mittag-Leffler, G.M.:
    Sur l'int\'egrale de Laplace-Abel,
   {\it C.R. Acad. Sci. Paris}, (ser. II) {\bf 136} (1902), 937-939.
\vsp
\rf{66}  Mittag-Leffler, G.M.:
    Une g\'en\'eralisation de l'int\'egrale de Laplace-Abel,
   {\it C.R. Acad. Sci. Paris}, (ser. II) {\bf 137} (1903), 537-539.
\vsp
\rf{67}  Mittag-Leffler, G.M.:
    Sur la nouvelle fonction $E_\alpha (x)$,
   {\it C.R. Acad. Sci. Paris}, (ser. II) {\bf 137} (1903), 554-558.
\vsp
\rf{68} Mittag-Leffler, G.M:
    Sopra la  funzione $E_\alpha (x)$,
   {\it R. Accad. Lincei, Rend.}, (ser. V) (1904) {\bf 13}, 3-5.
\vsp
\rf{69}  Mittag-Leffler, G.M.:
 Sur la repr\'esentation analytique d'une branche uniforme
  d'une fonction monog\`ene,
   {\it Acta Math.}, {\bf 29} (1905), 101-181.
\vsp
\rf{70}  Erd\'elyi, A. (Ed.): {\it Higher Transcendental Functions},
  Bateman Project,  Vols. 1-3, McGraw-Hill, New York 1953-1955.
\vsp
\rf{71}  Abramowitz, M. and I.A. Stegun:
 {\it  Handbook of Mathematical Functions}, Dover, New York 1965.
\vsp
\rf{72} Humbert, P.:
  Quelques r\'esultats relatifs \`a la fonction de Mittag-Leffler,
 {\it C.R. Acad. Sci. Paris}, {\bf 236} (1953), 1467-1468.
\vsp
\rf{73}  Agarwal, R.P.:
  A propos d'une note de M. Pierre Humbert,
   {\it C.R. Acad. Sci. Paris},  {\bf 236} (1953), 2031-2032.
\vsp
\rf{74} Humbert, P. and R.P. Agarwal:
  Sur la fonction de Mittag-Leffler et quelques-unes de ses
  g\'en\'eralisations,
 {\it Bull. Sci. Math} (Ser. II), {\bf 77} (1953), 180-185.
\vsp
\rf{75}  Phragm\'en, E.:
   Sur une extension d'un th\'eoreme classique de la
   th\'eorie des fonctions,
  {\it Acta Math.}, {\bf 28} (1904), 351-368.
\vsp
\rf{76}  Buhl, A. :  {\it S\'eries Analytiques. Sommabilit\'e},
 M\'emorial des Sciences Ma\-th\'e\-ma\-ti\-ques,
 Acad. Sci. Paris, Fasc. VII,
 Gauthier-Villars, Paris 1925, Ch. 3. 
\vsp
\rf{77} Berg, C. and G. Forst: {\it Potential Theory on Locally Compact
   Abelian Groups}, Springer Verlag, Berlin 1975, \S 9., pp. 61-72.
\vsp
\rf{78}  Pollard, H.:
    The completely monotonic character of the Mittag-Leffler function
     $E_\alpha (-x)\,, $
   {\it Bull. Amer. Math. Soc.}, {\bf 54} (1948), 1115-1116.
 \vsp
\rf{79}  Schneider, W.R.:
Complete monotone  generalized Mittag-Leffler functions,
   {\it Expositiones Mathematicae}, {\bf 14} (1996), 3-16.
\vsp
\rf{80}  Sansone, G. and J. Gerretsen:
{\it Lectures on the Theory of Functions of a Complex Variable},
 Vol. I. {\it Holomorphic Functions},
 Nordhoff, Groningen 1960, pp. 345-349.
\vsp
\rf{81} Hille E. and J. D. Tamarkin:
  On the theory of linear integral equations,
  {\it Ann. Math.},  {\bf 31} (1930), 479-528.
\vsp
\rf{82} Barret, J. H.: Differential equations of non-integer order,
  {\it Canad. J. Math.}, {\bf 6} (1954), 529-541.
\vsp

\rf{83} de Oliveira Castro, F. M.:
Zur Theorie der dielektrischen Nachwirkung,
{\it Zeits. f. Physik}, {\bf 114} (1939), 116--126.
\vsp
\rf{84} Gross, B:
  On creep and relaxation,
 {\it J. Appl. Phys.}, {\bf 18} (1947), 212--221.
\vsp
\rf{85}  Al Saqabi, B. N. and Vu Kim Tuan:
 Solution of a fractional  differintegral equation,
 {\it Integral Transforms and Special Functions},
 {\bf 4} (1996), 321-326.
\vsp
\rf{86}  Kilbas, A. A. and M. Saigo: On Mittag-Leffler type functions,
     fractional calculus operators and solution of integral equations,
 {\it Integral Transforms and Special Functions},
 {\bf 4} (1996), 355-370.
\vsp
\rf{87} Gorenflo, R.,  Luchko, Yu. and S. Rogozin:
   Mittag-Leffler type func\-tions : notes on growth properties
  and distribution of zeros,
   Pre-print  A-04/97,
 Fachbereich Mathematik und Informatik,
   Freie Universit\"at, Berlin 1997,
  available via
 Internet:
   $\,<\,$http://www.math.fu-berlin.de/publ/index.html$\,>$
\vsp
\rf{88}
Mainardi F. and R. Gorenflo:
     On Mittag-Leffler-type functions in fractional evolution processes,
     {\it J. Comput. and Appl. Mathematics},
  {\bf 118} No 1-2 (2000), 283-299.
\vfill\eject

\bye